\documentclass[10pt,twocolumn,twoside]{IEEEtran}
\pdfoutput=1 
\usepackage{amsmath}
\usepackage{graphicx}
\usepackage{amsfonts}
\usepackage{array}
\usepackage{amssymb}
\usepackage{helvet}
\usepackage{color}
\usepackage{float,algorithm,algorithmic}
\usepackage{epstopdf}
\usepackage{rawfonts,psfrag,flushend}
\usepackage{bm}
\usepackage[caption=false,font=footnotesize]{subfig}
\usepackage{multirow}
\usepackage{bigstrut}
\usepackage{cite}

\newcounter{row}
\makeatletter
\@addtoreset{subfigure}{row}
\makeatother

\begin{document}
\title{Ensemble and Random Collaborative Representation-Based Anomaly Detector for Hyperspectral Imagery}
\author{Rong Wang, Yihang Lu, Qianrong Zhang, Feiping Nie, Zhen Wang, and Xuelong Li,~\IEEEmembership{Fellow,~IEEE}
\IEEEcompsocitemizethanks{
\IEEEcompsocthanksitem R. Wang, Z. Wang and X. Li are with the School of Artificial Intelligence, Optics and Electronics (iOPEN), Northwestern Polytechnical University, Xi'an 710072, P. R. China. E-mail: wangrong07@tsinghua.org.cn.
\IEEEcompsocthanksitem Y. Lu, Q. Zhang and F. Nie are with the School of Computer Science and School of Artificial Intelligence, Optics and Electronics (iOPEN), Northwestern Polytechnical University, Xi'an 710072, P. R. China. E-mail: feipingnie@gmail.com.}}

\markboth{}%
{Wang \MakeLowercase{\textit{et al.}}: }

\IEEEtitleabstractindextext{
\begin{abstract}
In recent years, hyperspectral anomaly detection (HAD) has become an active topic and plays a significant role in military and civilian fields. As a classic HAD method, the collaboration representation-based detector (CRD) has attracted extensive attention and in-depth research. Despite the good performance of the CRD method, its computational cost mainly arising from the sliding dual window strategy is too high for wide applications. Moreover, it takes multiple repeated tests to determine the size of the dual window, which needs to be reset once the dataset changes and cannot be identified in advance with prior knowledge. To alleviate this problem, we proposed a novel ensemble and random collaborative representation-based detector (ERCRD) for HAD, which comprises two closely related stages. Firstly, we process the random sub-sampling on CRD (RCRD) to gain several detection results instead of the sliding dual window strategy, which significantly reduces the computational complexity and makes it more feasible in practical applications. Secondly, ensemble learning is employed to refine the multiple results of RCRD, which act as various ``experts" providing abundant complementary information to better target different anomalies. Such two stages form an organic and theoretical detector, which can not only improve the accuracy and stability of HAD methods but also enhance its generalization ability. Experiments on four real hyperspectral datasets exhibit the accuracy and efficiency of this proposed ERCRD method compared with ten state-of-the-art HAD methods.
\end{abstract}

}

\maketitle
\IEEEdisplaynontitleabstractindextext
\IEEEpeerreviewmaketitle

\section{Introduction}
Hyperspectral imagery (HSI) offers plentiful useful spectral and spatial information to monitor the earth's surface for the fine identification of various land cover materials~\cite{Shaw2002,Stein2002,Chang2002,Zhang2018a,Luo2019}, and thus HSI has been widely applied to various remote sensing fields, such as scene classification~\cite{Chang2002,Luo2019}, unmixing~\cite{Bioucas-Dias2012}, clustering~\cite{Wang2017a,Wang2019}, change detection~\cite{Liu2019} and target or anomaly detection~\cite{Stein2002,Yuan2016,Li2021,Nasrabadi2014}. Since the spectral signal of the anomalies is unknown in the task of anomaly detection, it is very challenging to detect anomalies in HSI. Hence, hyperspectral anomaly detection (HAD) has attracted more and more interest and in-depth research for its widespread applications in military reconnaissance, civilian search and rescue, environmental monitoring, mineral exploration and so on~\cite{Huber-Lerner2016,Wang2017}.

In essence, HAD is an unsupervised binary classification problem, which detects anomalies against the background without any prior knowledge of this scene.
Different from the background pixels, anomaly pixels have two discriminative features: 1) low occurrence probability; 2) distinct spectral-spatial characteristic different from the background. That is to say, anomaly pixels in HSI are ``few and different". These two features of the anomalies allow the researchers to distinguish the anomalies from the backgrounds. In decades, HAD has been studied and extended. Generally, there are two main kinds of existing HAD methods: statistics modeling HAD and representation-based HAD. In addition, HAD has been incorporated in the strength of support vector data description (SVDD)~\cite{Banerjee2006,Sakla2011}, morphological and attribute filters~\cite{Kang2017,Li2018,Taghipour2017}, tensor decomposition~\cite{Zhang2016a,Xu2018,Zhang2018,Xie2019}, and deep convolutional neural networks~\cite{Li2017,Zhao2017b,Ma2018}, which shows that the in-depth study of HAD has become a popular topic.

Statistics modeling HAD assumes a multivariate normal (Gaussian) background distribution, in which Reed-Xiaoli (RX) detector~\cite{Reed1990} is a well-known method that identifies anomalies based on the Mahalanobis distance between pixel and background. RX detector has two vital variants: the global RX (GRX) using the entire image to model the background, and the local RX (LRX) using the local dual window image. Then, numerous variants have been proposed~\cite{Kwon2005,Carlotto2005,Schaum2007,Molero2013,Liu2013,Guo2014}. For example, the kernel RX (KRX)~\cite{Kwon2005} uses the kernel theory to characterize the non-Gaussian distributions, which helps to solve the nonlinear problem but has high computational complexity. To speed it up, the cluster KRX (CKRX)~\cite{Zhou2016} detector applied the fast eigenvalue decomposition on clusters obtained from the entire image, which is inspired by the cluster-based anomaly detection (CBAD)~\cite{Carlotto2005} that groups the entire image into clusters before presenting the RX for better performance. Since the background in the real-world is complex, it is not enough to model the background only using the Gaussian distribution. The subspace RX (SSRX)~\cite{Schaum2007} detector is obtained by applying the RX to the low-variance principal components. Afterward, many adaptations of the RX detector have been proposed for improvement in different aspects\cite{Chang2018,Guo2014,Imani2017,Zhao2017a,Zhao2018,Zhou2016}. As another line of work, representation-based methods have attracted considerable attention for HAD, due to its diverse application scenarios and rich theoretical basis such as sparse representation~ \cite{Chen2011,Li2015,Zhang2017a,Zhao2017,Li2018a,Ling2019}, low-rank representation and 
so on~\cite{Sun2014,Zhang2016,Xu2016,Qu2018,Madathil2019,Cheng2020}. For example, Chen \emph{et al.}~\cite{Chen2011} proposed the sparse representation-based HAD method, which approximates the background by several atoms in the dictionary. Based on the low-rank and sparse matrix decomposition (LRaSMD) detector~\cite{Sun2014}, Zhang \emph{et al.}~\cite{Zhang2016} proposed the LRaSMD-based Mahalanobis distance (LSMAD) method, which employs the low-rank prior knowledge of the background first and then scores each pixel based on the Mahalanobis distance. 

More recently, HAD based on collaborative representation~\cite{Li2015b,Imani2018,Vafadar2018,Su2018,Ma2019} has attracted substantial attention and in-depth research, among which the collaborative representation-based detector (CRD) proposed by Li \emph{et al.} is one of the most representative methods~\cite{Li2015b}. CRD assumes that each background pixel can be approximated by its spatial neighborhood pixels in a sliding dual window centered at this pixel, while the anomalies cannot. Afterward, many variants have been proposed recently. To make better use of the spatial features from HSI, a morphology-based collaborative representation detector (MCRD)~\cite{Imani2018} was presented. Besides, principal component analysis (PCA) was first applied in the CRD method to extract the main background pixel information and remove outliers (PCAroCRD)~\cite{Su2018}. In order to decrease the complexity of computing, a recursive CRD (RCRD)~\cite{Ma2019} algorithm was proposed based on the matrix inversion lemma. All in all, CRD is the first method that applies collaborative representation to HAD. More importantly, CRD and its variants adopt brand new ideas for HAD, and they all have a closed-form solution, which is much simpler than a sparseness-constrained detector. However, most of the existing CRD methods adopt a sliding dual window strategy to approximate each pixel by its spatial neighborhood pixels, which gives rise to high computational complexity. What's worse, multiple repeated tests are required to determine the size of the dual window, which needs to be reset once the dataset changes, but we cannot identify whether a specific size is suitable for a particular dataset in advance.

The isolation forest (iForest) introduced by Liu \emph{et al.}~\cite{Liu2008,Liu2012} is a well-known anomaly detector, which assumes that anomalies are ``few and different". Such two features make anomalies more susceptible to isolation in a binary tree (iTree) via random sub-sampling, through which each iTree is specialized and contains a small set of anomalies or even no anomaly. More importantly, iForest is ensembled by multiple iTrees acting as various ``experts" providing abundant complementary information to target different anomalies, which can not only improve the accuracy and stability of the detector but also enhance its generalization ability. In 2020, Li \emph{et al.}~\cite{Li2020} first introduced iForest into HAD field, and we also established a hyperspectral anomaly detector combined with multiple features~\cite{wang2020}. Afterward, iForest has been studied and extended in HAD~\cite{Chang2020,Song2021,Zhang2021}, which shows that the research in this field is active but no one has ever incorporated iForest in CRD.

Due to the successful attempt of iForest on HAD and CRD being more and more valued and welcomed by researchers, we intend to introduce the idea of iForest to collaborative representation-based HAD methods, rather than simply apply it on HSI as an anomaly detector. In this paper, we come up with a novel ensemble and random collaborative representation-based detector (ERCRD) with random sub-sampling and ensemble learning, which is consisted of two closely related stages to form an organic and theoretical detector. 1) ERCRD repeatedly processes the random sub-sampling on the collaborative representation-based detector (RCRD) to gain a number of detection results. In specific, when we repeatedly adopt random sub-sampling on the whole image, several results with a small set of anomalies or even no anomaly can be regarded as the background, which meets the assumption of collaborative representation-based HAD methods. Moreover, ERCRD does not utilize the sliding dual window strategy which mainly produces the computational complexity of CRD and its variants. Thus, ERCRD significantly reduces the computational complexity and makes it more feasible in practical applications. 2) The multiple detection results of RCRD are further refined to the final detection result through ensemble learning. On the analogy of iForest being ensembled by multiple iTrees, ERCRD was ensembled by RCRD acting as various ``experts" providing abundant complementary information to better target different anomalies, which can not only improve the accuracy and stability of HAD methods but also enhance its generalization ability.

Last but not least, the proposed ERCRD shows its advantages over many classic and benchmark HAD methods, such as GRX~\cite{Reed1990}, LRX~\cite{Reed1990}, SSRX~\cite{Schaum2007}, CBAD~\cite{Carlotto2005}, and LSMAD~\cite{Zhang2016}. In comparison with the state-of-the-art collaborative representation-based HAD methods like CRD~\cite{Li2015b} and its variants~\cite{Su2018,Imani2018,Ma2019}, we also validate that our ERCRD method is able to attain considerable or better detection accuracy with even much less implementation time. The rest of this paper is arranged as follows. In Section~\ref{Sec.2}, we briefly review the traditional CRD and the recently proposed PCAroCRD. The proposed approach ERCRD is described in Section~\ref{Sec.3}. In Section \ref{Sec.4}, we conduct empirical studies on four real-world datasets to validate the accuracy and efficiency of our proposed ERCRD. Finally, we summarize this paper in Section~\ref{Sec.5}.

\section{Related Work} \label{Sec.2}
Let $\mathbb{X}\in\mathbb{R}^{d\times h \times w}$ denotes the hyperspectral imagery, where $d$ is the number of spectral bands, and $h$ and $w$ are the height and width of the background respectively. The three-dimensional (3-D) hyperspectral imagery $\mathbb{X}$ is transformed into a two-dimensional (2-D) matrix $\bm{X}=[\bm{x}_1,\bm{x}_2,\cdots,\bm{x}_n]\in\mathbb{R}^{d\times n}$, where $n=h\times w$ is the total number of pixels.
\subsection{CRD}
\begin{figure}[!htb]
  \centering
  \includegraphics[scale=0.28]{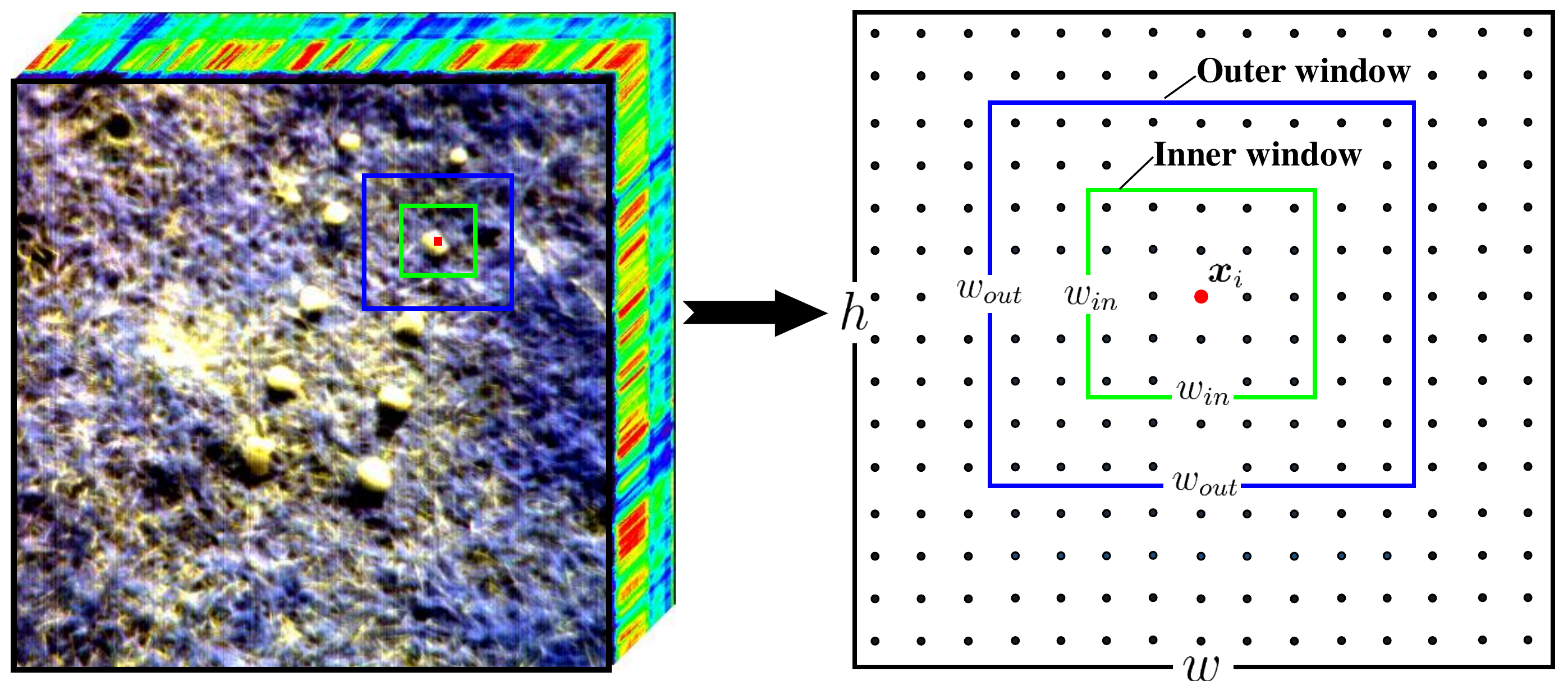}
  \caption{The sliding dual window of the CRD.}
  \label{Fig.1}
\end{figure}

In this subsection, we review the recently proposed collaborative representation-based detector (CRD)~\cite{Li2015b}. The CRD assumed that the background pixel can be well approximated by its spatial neighborhood pixels, whereas the anomalies cannot. For the pixel $\bm{x}_i\in \mathbb{R}^{d\times 1}$, its spatial neighborhood pixels are selected by a sliding dual window, as can be seen from Fig. \ref{Fig.1}. The spatial neighborhood pixels specifically refer to pixels outside the inner window and inside the outer window. The size of the outer window and the inner window is represented as $w_{out}$ and  $w_{in}$, respectively. Thus, the spatial neighborhood pixels are resized into a two-dimensional matrix $\bm{X}_s=[\bar{\bm{x}}_1,\bar{\bm{x}}_2,\cdots,\bar{\bm{x}}_s]\in\mathbb{R}^{d\times s}$, where $s$ denotes the number of spatial neighborhood pixels and $s=w_{out}\times w_{out}-w_{in}\times w_{in}$. The objective function of the CRD is defined as~\cite{Li2015b}
\begin{align}\label{Eq.1}
\min_{\bm{\alpha}_i}\|\bm{x}_i-\bm{X}_s\bm{\alpha}_i\|_2^2 +\lambda\|\bm{\alpha}_i\|_2^2,
\end{align}
where $\bm{\alpha}_i\in\mathbb{R}^{s\times 1}$ denotes the weight vector and $\lambda$ is the regularization parameter. Taking the derivative w.r.t $\bm{\alpha}_i$ and setting it to zero, so we have
\begin{align}\label{Eq.2}
  \hat{\bm{\alpha}}_i=(\bm{X}_s^{T}\bm{X}_s+\lambda\bm{I})^{-1}\bm{X}_s^T\bm{x}_i,
\end{align}
where $\bm{I}$ is an identity matrix. The reconstruction error $r_i$ for the pixel $\bm{x}_i$ is regarded as the anomaly score and can be computed by
\begin{align}\label{Eq.3}
  r_i = \|\bm{x}_i-\bm{X}_s\hat{\bm{\alpha}}_i\|_2.
\end{align}
If $r_i$ is greater than a threshold, then the pixel $\bm{x}_i$ is called anomalous pixel.

The computational complexity of this step is $O(ds^2+s^3+ds+d)$. Each pixel $\bm{x}_i$ needs to get the corresponding surrounding pixel matrix $\bm{X}_s$ on its sliding double window. Thus, the weight vector $\bm{\alpha}_i$ and the anomaly score $r_i$ need to calculate $n$ times, thereby the total computational complexity of the CRD being $O(nds^2+ns^3+nds+nd)$. This repeated process needs a high computational burden, reduces the speed of the anomaly detection, and limits the application of the CRD.
\subsection{PCAroCRD}
The PCAroCRD method was recently proposed by Su \emph{et al.}~\cite{Su2018}, which applied PCA to the CRD method for removing outliers. It has two versions: Global PCAroCRD and Local PCAroCRD.

The Global PCAroCRD first obtains the projection matrix $\bm{W}_g\in\mathbb{R}^{n\times n}$ by solving the standard PCA model:
\begin{align}\label{Eq.4}
   \max_{\bm{W}_g^T\bm{W}_g=\bm{I}}{\rm tr}(\bm{W}_g^T\bm{X}^T\bm{X}\bm{W}_g),
\end{align}
then the spatial-domain PCA transformation is represented as
\begin{align}\label{Eq.5}
   \hat{\bm{X}} = \bm{X}\bm{W}_g
\end{align}
where $\hat{\bm{X}}=[\hat{\bm{x}}_1,\hat{\bm{x}}_2,\cdots,\hat{\bm{x}}_n]\in\mathbb{R}^{d\times n}$ denotes the transformed data matrix. For the pixel $\bm{x}_i$, the objective function of the Global PCAroCRD can be written as follows~\cite{Su2018}
\begin{align}\label{Eq.6}
\min_{\bm{\alpha}_i}\|\bm{x}_i-\bm{X}_m\bm{\alpha}_i\|_2^2 +\lambda\|\bm{\Gamma}_i\bm{\alpha}_i\|_2^2.
\end{align}
where $\bm{X}_m=[\hat{\bm{x}}_1,\hat{\bm{x}}_2,\cdots,\hat{\bm{x}}_m]\in\mathbb{R}^{d\times m}$ denotes the first $m$ principal components of $\hat{\bm{X}}$ and contains the most information of $\bm{X}$ in the spatial domain. $\bm{\Gamma}_i^g$ denotes the Tikhonov regularization matrix and is defined as:
\begin{align}\label{Eq.7}
\bm{\Gamma}_{i}^g =
\left[
  \begin{array}{ccc}
    \|\bm{x}_i-\hat{\bm{x}}_1\|_2 & \cdots & 0\\
    \vdots & \ddots & \vdots \\
     0 & \cdots & \|\bm{x}_i-\hat{\bm{x}}_m\|_2
  \end{array}
\right].
\end{align}
Similarly, taking the derivative w.r.t $\bm{\alpha}_i$ and setting the derivative to zero, we have
\begin{align}\label{Eq.8}
\hat{\bm{\alpha}}_i = (\bm{X}_m^T\bm{X}_m+\lambda\bm{\Gamma}_i^{gT}\bm{\Gamma}_i^g)^{-1}\bm{X}_m^T\bm{x}_i.
\end{align}
The reconstruction error $r_i$ can be computed by
\begin{align}\label{Eq.9}
  r_i = \|\bm{x}_i-\bm{X}_m\hat{\bm{\alpha}}_i\|_2.
\end{align}
If $r_i$ is greater than a threshold, the pixel $\bm{x}_i$ is referred to as an anomaly. The total computational complexity of the Global PCAroCRD algorithm is $O(n^2+ndm^2+nm^3+ndm+nd)$.

The Local PCAroCRD first selects the surrounding pixels for each pixel $\bm{x}_i$ by a sliding dual window, which is exactly the same as the CRD. Then, the spatial-domain PCA transformation is performed on the surrounding pixels $\bm{X}_s\in\mathbb{R}^{d\times s}$ in the spatial domain:
\begin{align}\label{Eq.10}
   \bar{\bm{X}} = \bm{X}_s\bm{W}_l
\end{align}
where $\check{\bm{X}}=[\check{\bm{x}}_1,\check{\bm{x}}_2,\cdots,\check{\bm{x}}_s]\in\mathbb{R}^{d\times s}$ denotes the transformed data matrix. The projection matrix $\bm{W}_l\in\mathbb{R}^{s\times s}$ is obtained by solving the PCA model:
\begin{align}\label{Eq.11}
   \max_{\bm{W}_l^T\bm{W}_l=\bm{I}}{\rm tr}(\bm{W}_l^T\bm{X}_s^T\bm{X}_s\bm{W}_l).
\end{align}
For the pixel $\bm{x}_i$, the objective function of the Local PCAroCRD can be written as follows~\cite{Su2018}
\begin{align}\label{Eq.12}
\min_{\bm{\alpha}_i}\|\bm{x}_i-\bm{X}_k\bm{\alpha}_i\|_2^2 +\lambda\|\bm{\Gamma}_i^l\bm{\alpha}_i\|_2^2.
\end{align}
where $\bm{X}_k=[\check{\bm{x}}_1,\check{\bm{x}}_2,\cdots,\check{\bm{x}}_k]\in\mathbb{R}^{d\times k}$ denotes the first $k$ principal components of $\bar{\bm{X}}$ and contains the most information of $\bm{X}_s$ in the spatial domain. $\bm{\Gamma}_i^l$ denotes the Tikhonov regularization matrix and is defined as:
\begin{align}\label{Eq.13}
\bm{\Gamma}_{i}^l =
\left[
  \begin{array}{ccc}
    \|\bm{x}_i-\bar{\bm{x}}_1\|_2 & \cdots & 0\\
    \vdots & \ddots & \vdots \\
     0 & \cdots & \|\bm{x}_i-\bar{\bm{x}}_k\|_2
  \end{array}
\right].
\end{align}
Similarly, taking the derivative w.r.t $\bm{\alpha}_i$ and setting the derivative to zero, we have
\begin{align}\label{Eq.14}
\hat{\bm{\alpha}}_i = (\bm{X}_k^T\bm{X}_k+\lambda\bm{\Gamma}_i^{lT}\bm{\Gamma}_i^l)^{-1}\bm{X}_k^T\bm{x}_i.
\end{align}
The reconstruction error $r_i$ can be computed by
\begin{align}\label{Eq.15}
  r_i = \|\bm{x}_i-\bm{X}_k\hat{\bm{\alpha}}_i\|_2.
\end{align}
If $r_i$ is greater than a threshold, then the pixel $\bm{x}_i$ is referred to as an anomaly. The total computational complexity of the Local PCAroCRD algorithm is $O(ns^2+ndk^2+nk^3+ndk+nd)$.

\section{The Proposed ERCRD Method} \label{Sec.3}
\begin{figure}[!htb]
  \centering
  \includegraphics[scale=0.42]{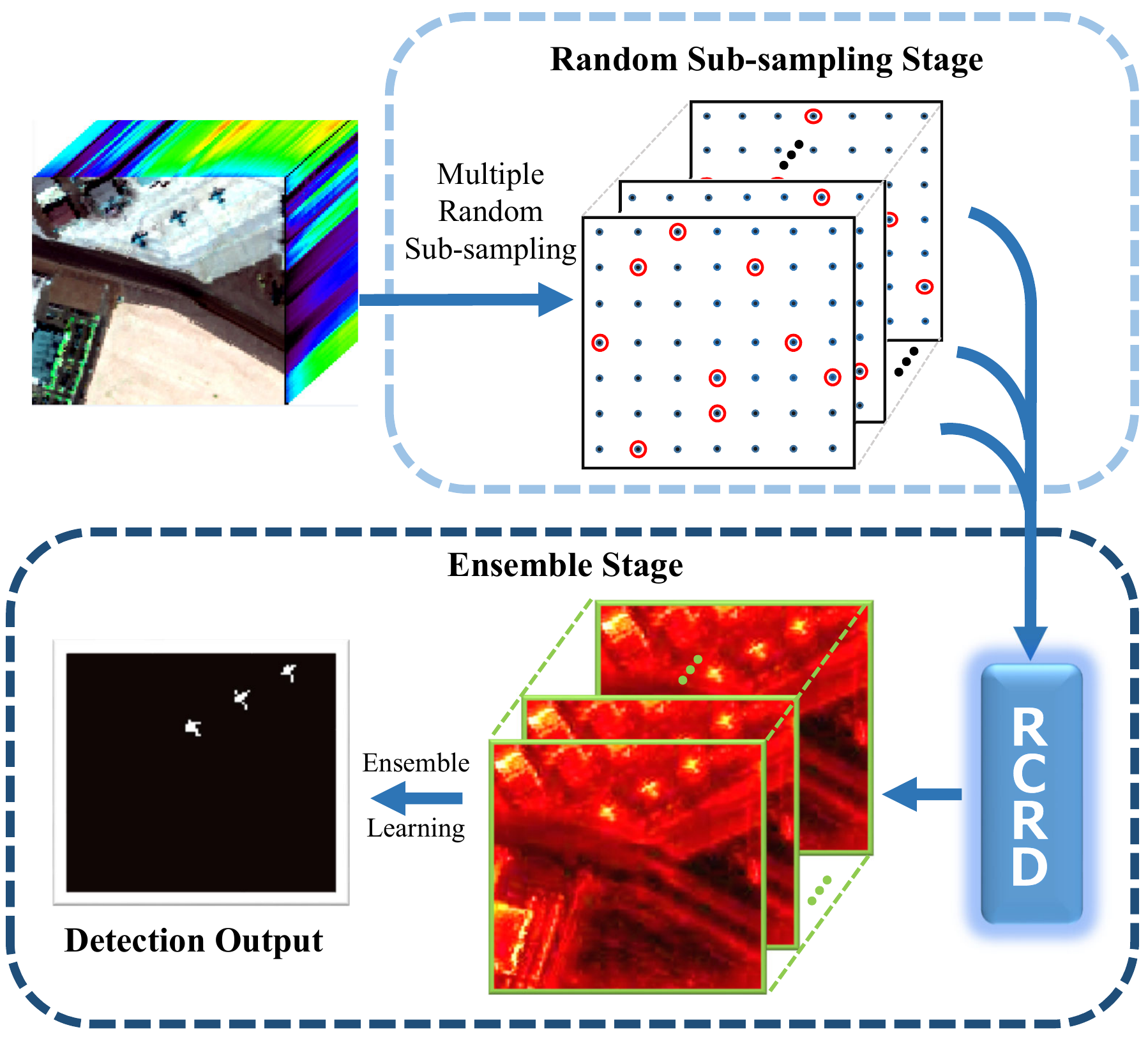}
  \caption{The flowchart of the proposed ERCRD method.}
  \label{Fig.2}
\end{figure}

Although the CRD method performs very well, its computational cost is too high for wide applications and mainly arised from the sliding dual window strategy. 
Moreover, the size of the dual window is determined by multiple repeated tests, needs to be reset once the dataset changes and cannot be identified in advance
with prior knowledge. To alleviate this problem, a novel ensemble and random collaborative representation-based detector (ERCRD) with random sub-sampling and ensemble learning is proposed in this section. Fig. \ref{Fig.2} illustrates the flowchart of the proposed ERCRD method.

\subsection{Random CRD (RCRD)}
The main idea of the CRD hypothesize that the background pixel can be approximated by a linear combination of the background pixels, but the anomalous pixel cannot. In the CRD, the background pixels for each pixel are represented by the surrounding pixels, which are selected by a sliding dual window centered at this pixel.

\begin{figure}[!htb]
  \centering
  \includegraphics[scale=0.26]{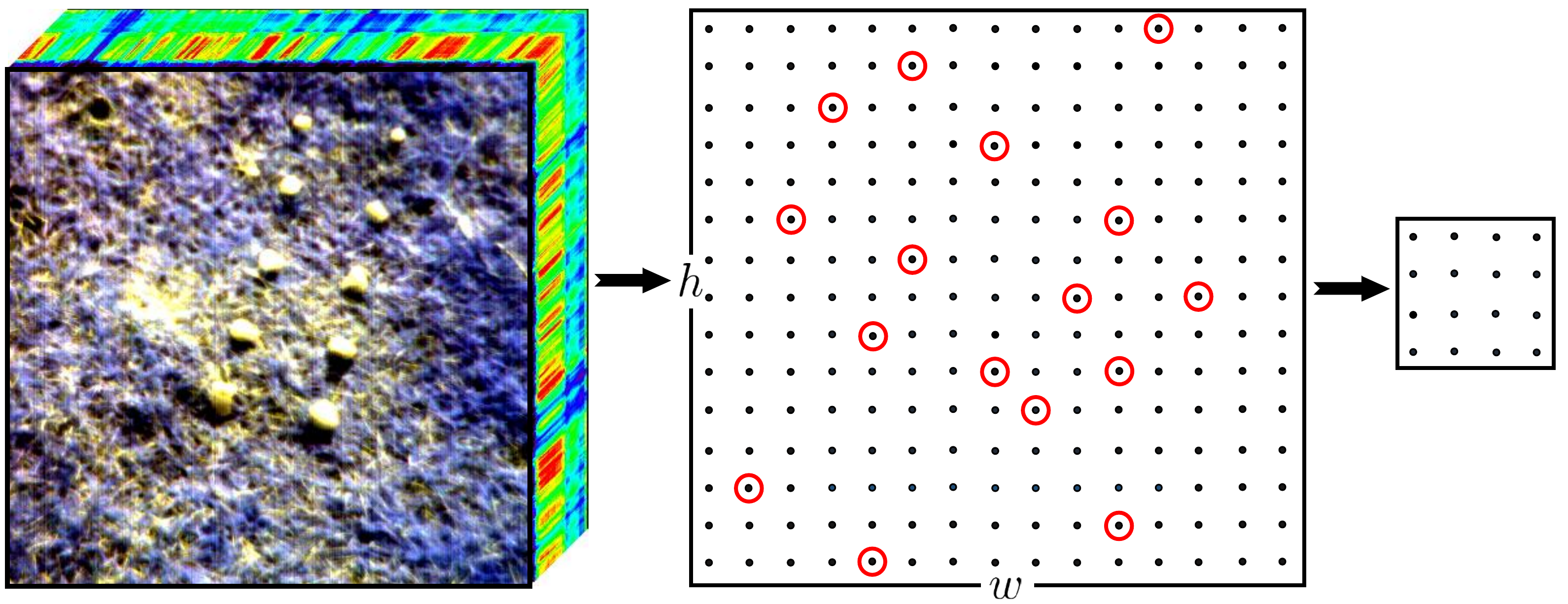}
  \caption{Random sub-sampling.}
  \label{Fig.3}
\end{figure}

Different from the sliding dual window in the CRD, we select the background pixels for each pixel by the random sub-sampling, which is completed from the whole hyperspectral image scene, as shown in Fig. \ref{Fig.3}.

Then, the background pixels obtained by the random sub-sampling are resized into a matrix $\bm{X}_r=[\tilde{\bm{x}}_1,\tilde{\bm{x}}_2,\cdots,\tilde{\bm{x}}_r]\in\mathbb{R}^{d\times r}$. In the CRD, every pixel has its own surrounding pixels, and the surrounding pixels of each pixel are different. Unlike the CRD, we use the same matrix $\bm{X}_r$ for each pixel in our model. Therefore, the objective function of the proposed Random CRD (RCRD) is written as
\begin{align}\label{Eq.16}
   \min _{\bm{A}}\|\bm{X}-\bm{X}_r\bm{A}\|_F^2+\lambda\|\bm{A}\|_F^2,
\end{align}
where $\bm{A}\in\mathbb{R}^{m\times n}$ denotes the weight matrix and $\lambda$ denotes the regularization parameter. Taking the derivative w.r.t $\bm{A}$ and setting the derivative to zero, we have the following equation:
\begin{align}\label{Eq.17}
  \bm{A} =(\bm{X}_r^T\bm{X}_r+\lambda\bm{I})^{-1}\bm{X}_r^T\bm{X}.
\end{align}
Thus, the matrix $\bm{X}$ can be reconstructed by the matrix $\bm{X}_r\bm{A}$. The reconstruction error for the pixel $\bm{x}_i$ is regarded as anomaly score and can be obtained by
\begin{align}\label{Eq.18}
  \delta_i = \|\bm{x}_i-\bm{X}_r\bm{a}_i\|_2.
\end{align}
where $\bm{x}_i$ and $\bm{a}_i$ denote the $i$ column of $\bm{X}$  and $\bm{A}$, respectively.

If $\delta_i$ is larger than a threshold, then the pixel $\bm{x}_i$ is called an anomaly. The computational complexity of the proposed Random CRD is $O(ndr+nd+dr^2+r^3)$. The detailed process can be found in Algorithm \ref{Alg.1}.

\begin{algorithm}[h]
\caption{Random CRD}
\begin{algorithmic}\label{Alg.1}
\STATE {\bf Input:} The two-dimensional HSI matrix $\bm{X}$, the number of random sub-sampling $r$. \\
1. Randomly select $r$ pixels from the matrix $\bm{X}$ and resize these pixels into the matrix $\bm{X}_r$.\\
2. Calculate the weight matrix $\bm{A}$ by Eq. (\ref{Eq.17}).\\
3. Obtain the anomaly score $\delta_i$ for the pixel $\bm{x}_i$ by Eq. (\ref{Eq.18}).\\
\STATE {\bf Output:} The anomaly scores for all pixels.
\end{algorithmic}
\end{algorithm}

\subsection{Ensemble and Random CRD}
As an anomaly detector ensemble that employs Random CRDs, the proposed Ensemble and Random CRD (ERCRD) has multiple Random CRDs acting as `experts' to detect different anomalies. Through multiple Random CRDs, the anomaly score for the pixel $x_i$ is obtained by
\begin{align}\label{Eq.19}
  \gamma_i=\sum_{t=1}^T\delta_i^t,
\end{align}
where $T$ denotes the ensemble size of the Random CRDs. Therefore, the total computational complexity of the proposed ERCRD is $O(ndrT+ndT+dr^2T+r^3T)$. The details of the ERCRD can be found in Algorithm \ref{Alg.2}.

\begin{algorithm}[!htb]
  \caption{Ensemble and Random CRD (ERCRD)}
  \begin{algorithmic}\label{Alg.2}
  \STATE{\bf Input:} The two-dimensional HSI matrix $\bm{X}$, the number of random sub-sampling $r$ and the ensemble size of the Random CRDs $T$.\\
  1. Repeat the Random CRD (Algorithm \ref{Alg.1}) $T$ times. \\
  2. Obtain the anomaly score $\gamma_i$ for the pixel $\bm{x}_i$ by Eq. (\ref{Eq.19}).
  \STATE {\bf Output:} The anomaly scores for all pixels.
  \end{algorithmic}
\end{algorithm}

\section{Experimental Results}\label{Sec.4}
To explore the detection performance of our ERCRD method, we conduct several experiments on a PC with E$5$-$2680$ v$4$ @$2.40$GHz and $256$GB RAM, MATLAB $2016$b. We use four hyperspectral datasets obtained from different scenes, which are described as follows:
\begin{enumerate}
  \item AVIRIS-I Dataset: This dataset was acquired by the Airborne Visible/Infrared Imaging Spectrometer (AVIRIS) from San Diego, CA, USA, with a spatial resolution of $3.5$ m per pixel and a spectral resolution of $10$ nm. This dataset has $224$ spectral bands with wavelengths ranging from $370$ to $2,510$ nm. After removing the bands with water absorption, low signal-to-noise ratio, and poor-quality ($1-6$, $33-35$, $97$, $107-113$, $153-166$, and $221-224$), $189$ bands are retained in this experiment. The size of the entire image scene is $400\times400$ pixels, from which we select a $120\times120$ pixels area in the upper left corner to test and mark it as AVIRIS-I. The three airplanes in the image are considered to be anomalies, which consist of $58$ pixels and should be detected.
  \item AVIRIS-II Dataset: This dataset was derived from \cite{Kang2017}. Same as the above dataset, we select a $100\times100$ pixels area at the center of the San Diego image to test and mark it as AVIRIS-II. The three airplanes consisting of $134$ pixels in the image are considered to be anomalies.
  \item AVIRIS-III Dataset: This dataset was obtained from \cite{Zhang2015}. Again, we select a $200\times240$ pixels area in the upper left of the San Diego image to test and mark it as AVIRIS-III. The six airplanes consisting of $90$ pixels in the image are considered to be anomalies.
  \item Cri Dataset: This dataset was derived from \cite{Zhang2016} and collected by the Nuance Cri hyperspectral sensor, with a spectral resolution of $10$ nm. This dataset has a size of $400\times400$ pixels and $46$ spectral bands with wavelengths ranging from $650$ to $1,100$ nm. The ten rocks consisting of $2,216$ pixels in the image are considered to be anomalies.
\end{enumerate}

\begin{figure}[!ht]
\centering
\subfloat[]{\includegraphics[scale=0.34]{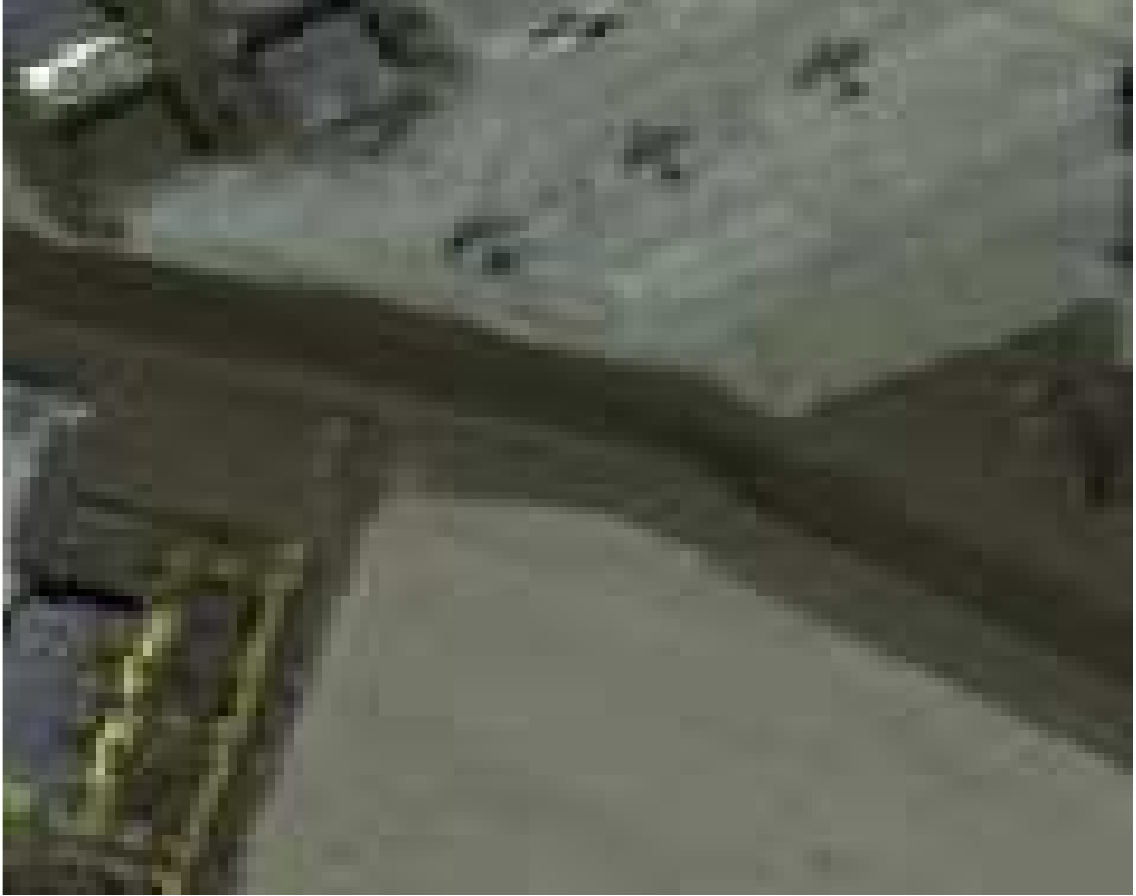}\label{Fig.4a}}~
\subfloat[]{\includegraphics[scale=0.34]{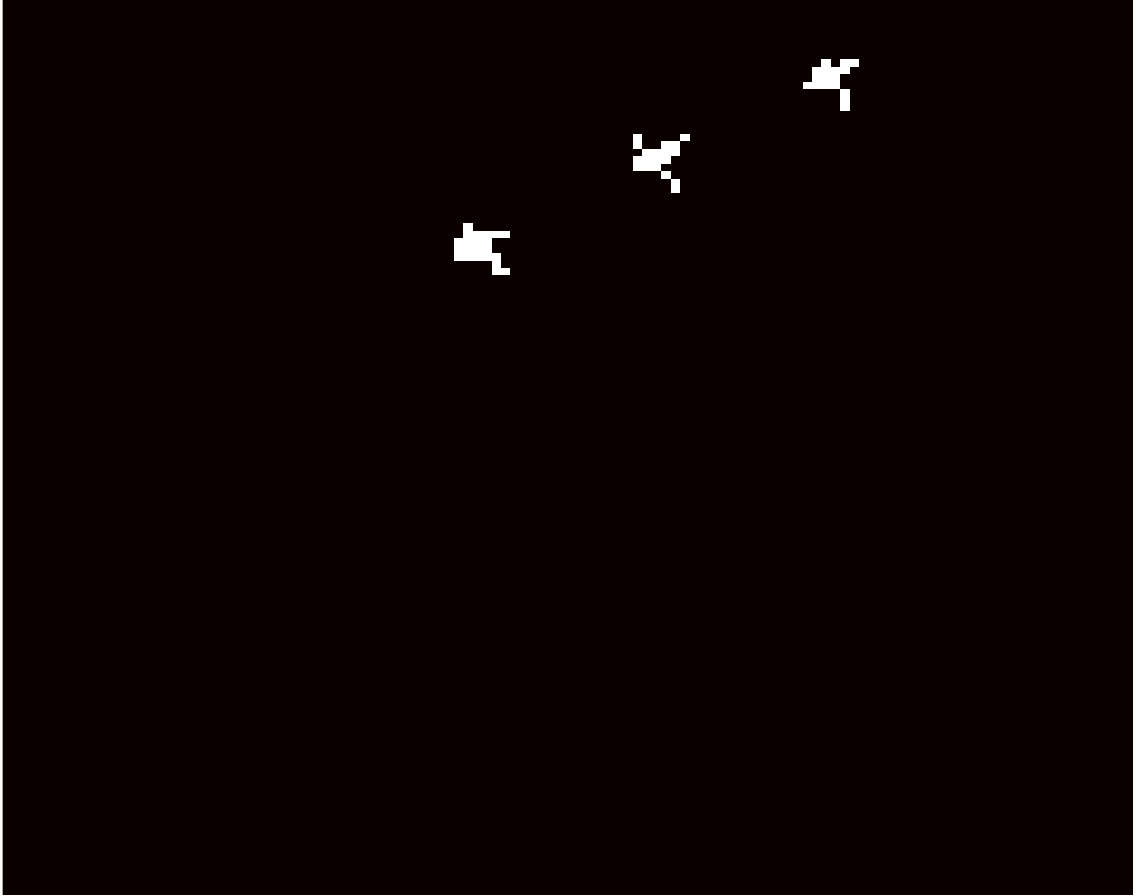}\label{Fig.4b}}\\
\subfloat[]{\includegraphics[scale=0.34]{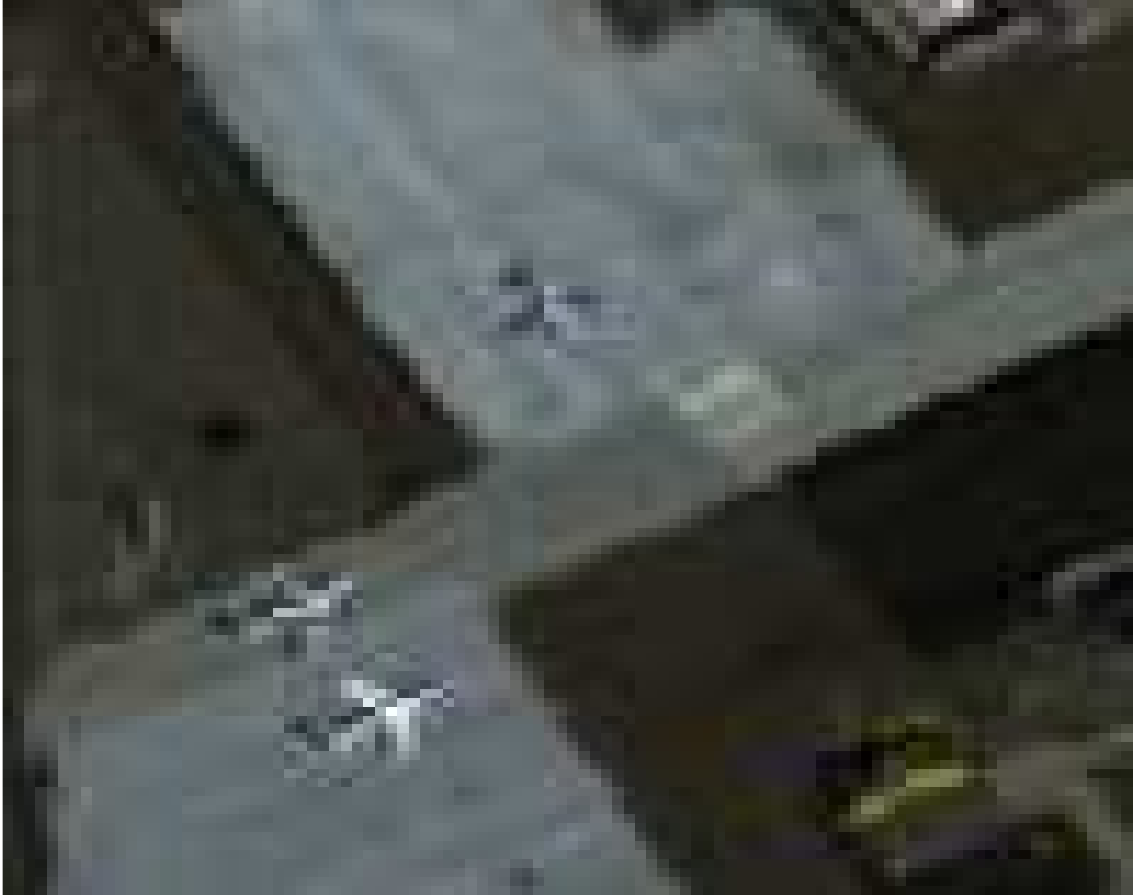}\label{Fig.4c}}~
\subfloat[]{\includegraphics[scale=0.34]{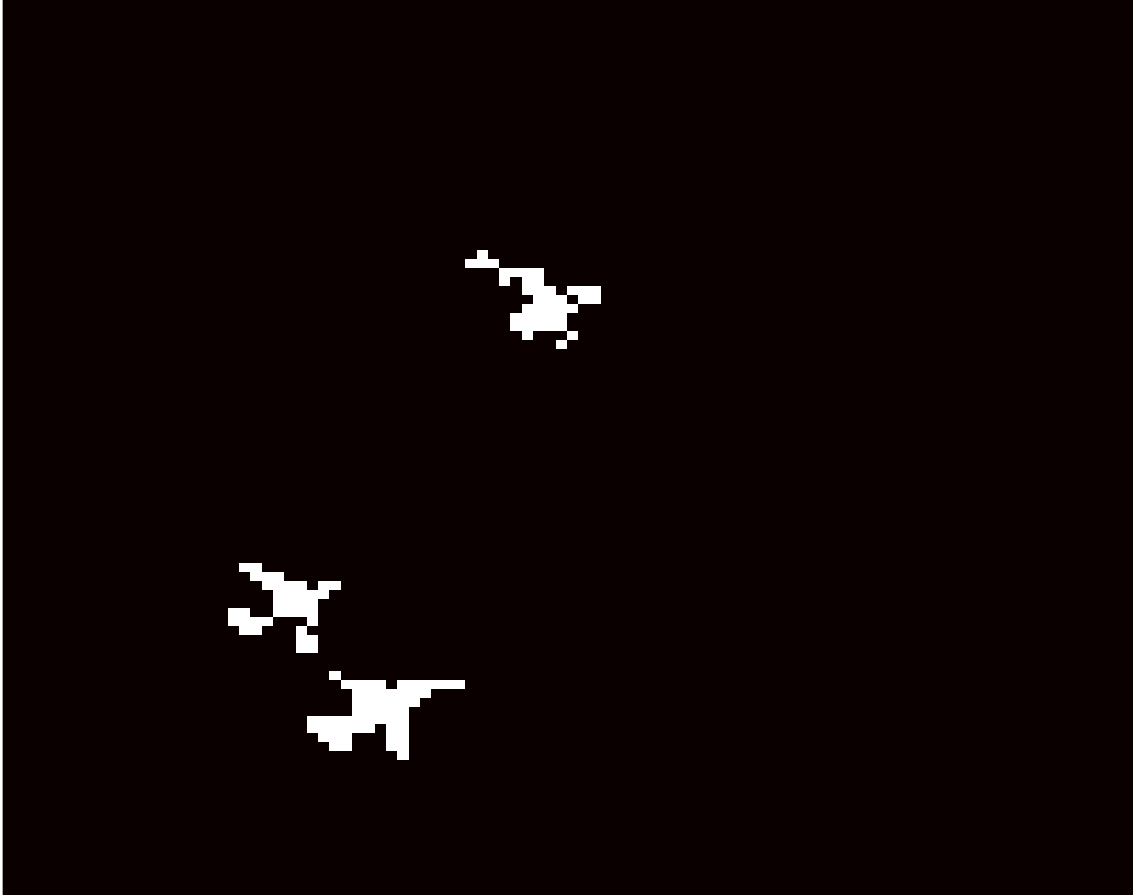}\label{Fig.4d}}\\
\subfloat[]{\includegraphics[scale=0.34]{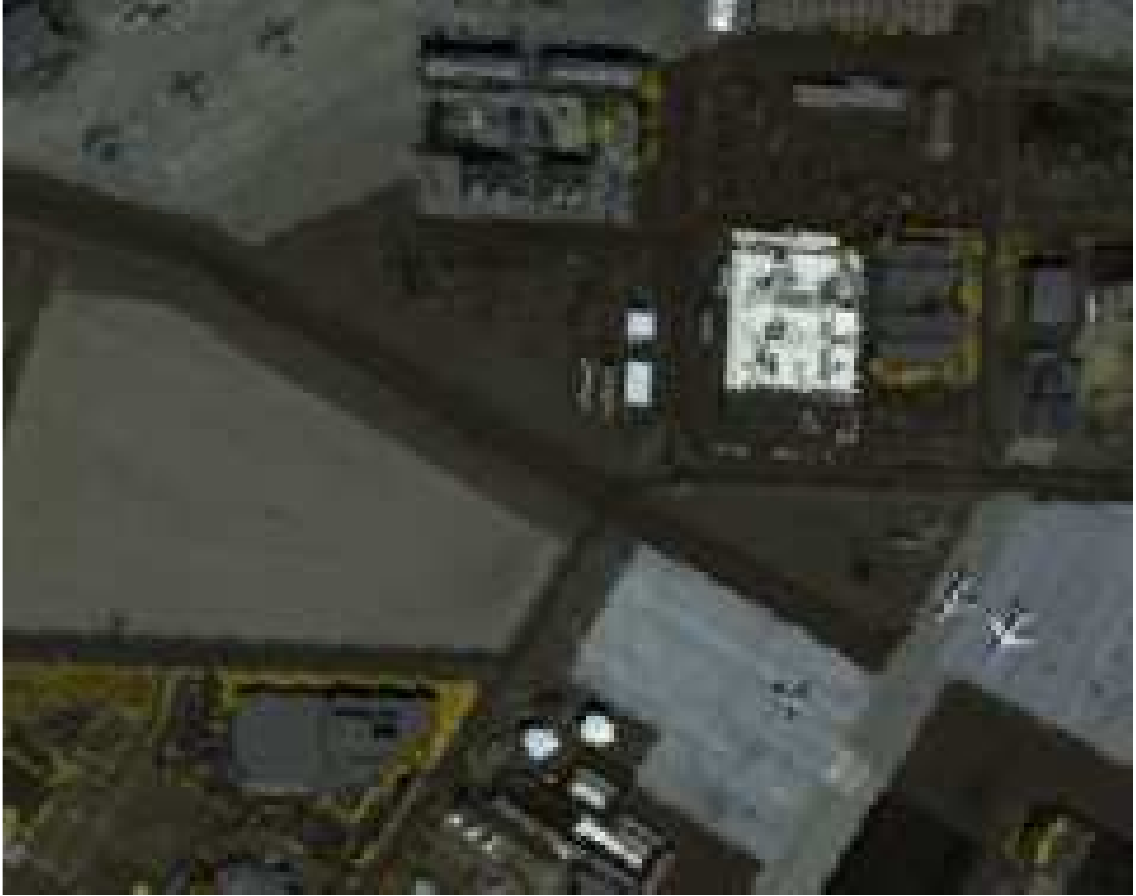}\label{Fig.4e}}~
\subfloat[]{\includegraphics[scale=0.34]{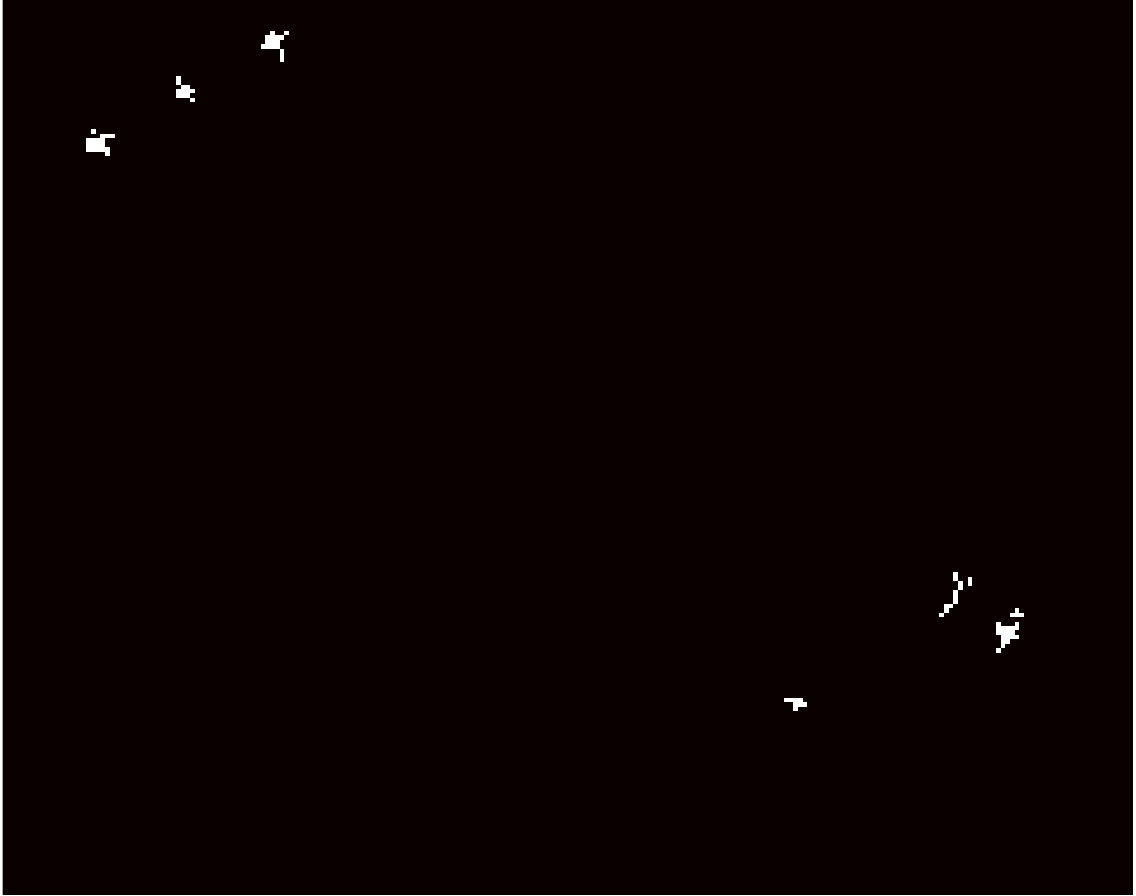}\label{Fig.4f}}\\
\subfloat[]{\includegraphics[scale=0.34]{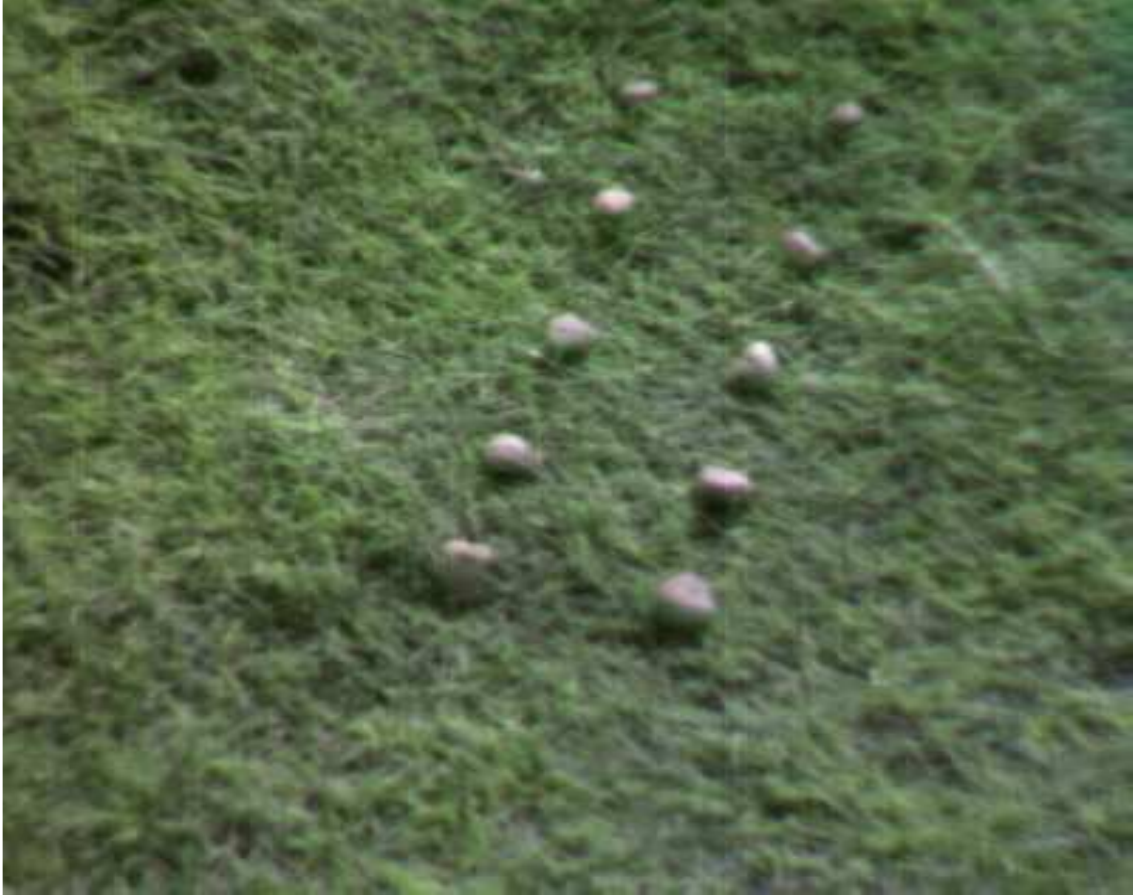}\label{Fig.4g}}~
\subfloat[]{\includegraphics[scale=0.34]{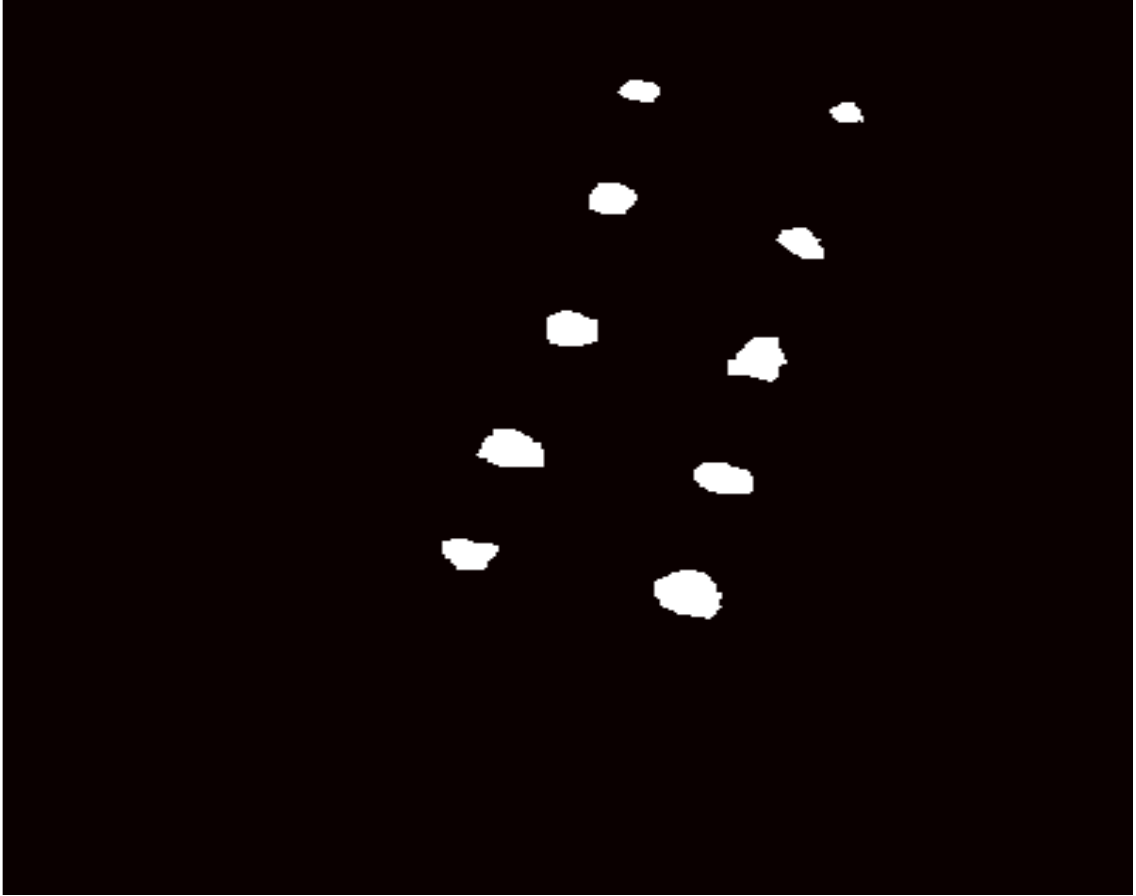}\label{Fig.4h}}
\caption{Image scene descriptions. (a) False color image of AVIRIS-I dataset. (b) Ground truth map of AVIRIS-I dataset. (c) False color image of AVIRIS-II dataset. (d) Ground truth map of AVIRIS-II dataset. (e) False color image of AVIRIS-III dataset. (f) Ground truth map of AVIRIS-III dataset. (g) False color image of Cri dataset. (h) Ground truth map of Cri dataset.}
\label{Fig.4}
\end{figure}

Note that Fig. \ref{Fig.4a}  and \ref{Fig.4b} respectively present the false color image and the corresponding ground truth map of the AVIRIS-I dataset. In the same way, Fig. \ref{Fig.4c}  and \ref{Fig.4d} correspond to the AVIRIS-II dataset, Fig. \ref{Fig.4e}  and \ref{Fig.4f} correspond to the AVIRIS-III dataset, and Fig. \ref{Fig.4g}  and \ref{Fig.4h} correspond to the Cri dataset.

In addition, the color detection map is used as the qualitative evaluation metric in our experiments. The receiver operating characteristic (ROC) curve~\cite{Kerekes2008}, the area value under the ROC curve (AUC), the normalized background-anomaly separation map, and the running time are used as quantitatively evaluation metrics in our experiments. The ROC curve reflects the relations between the detection probability (DP) and the false alarm rate (FAR) at the thresholds ranging from $0$ to $1$ on the strength of ground truth. An excellent detector usually has a high DP value under the same FAR value, which leads to the phenomenon that the corresponding ROC curve is located close to the upper left corner, making the area under the curve larger. The value of the area enclosed by the ROC curve and the false alarm rate axis is called AUC. The normalized background-anomaly separation map describes the normalized anomaly score distributions of the background, and anomalous pixels are represented by a box plot. Generally speaking, a good method should have a high AUC value and a distinct gap between the background box and the anomaly box.

\begin{figure*}[!htb]
  \renewcommand{\thesubfigure}{\Roman{row}-\alph{subfigure}}
  \centering
  \setcounter{row}{1}%
  \subfloat[]{\includegraphics[scale=0.25]{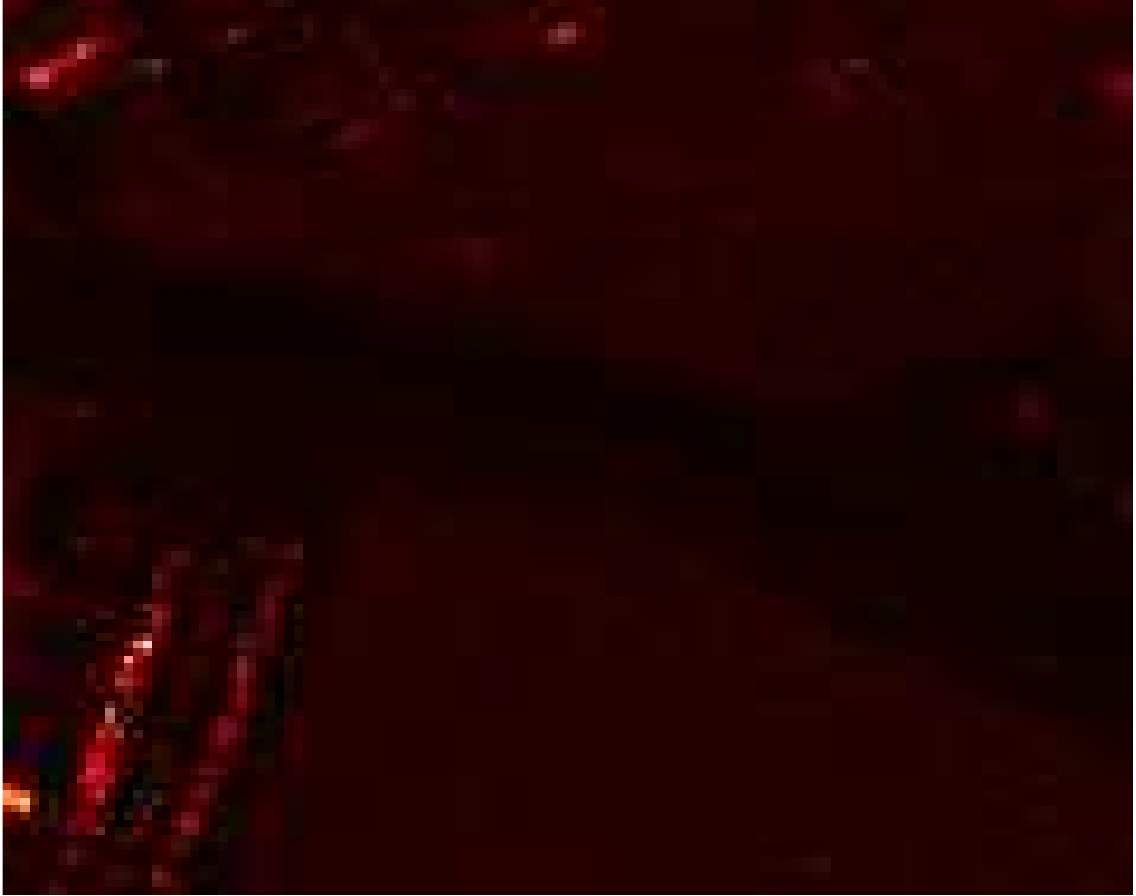}}~
  \subfloat[]{\includegraphics[scale=0.25]{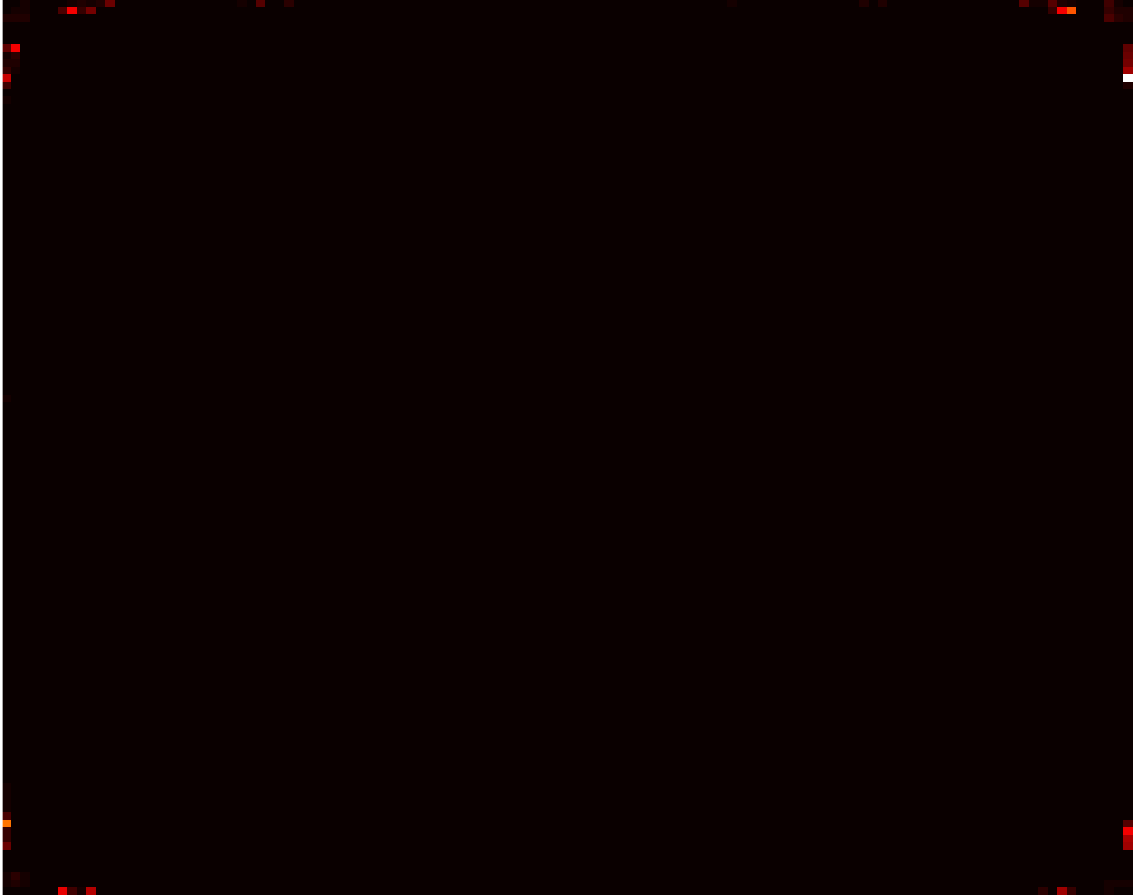}}~
  \subfloat[]{\includegraphics[scale=0.25]{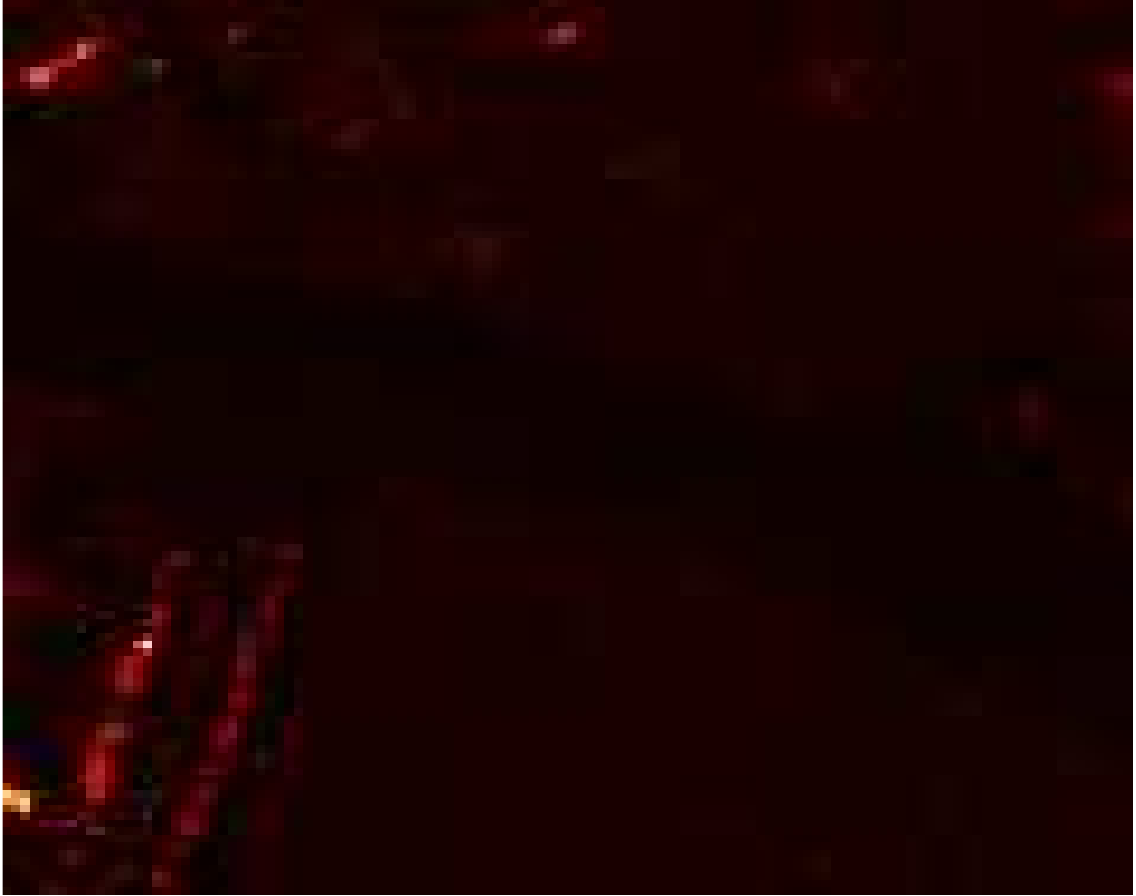}}~
  \subfloat[]{\includegraphics[scale=0.25]{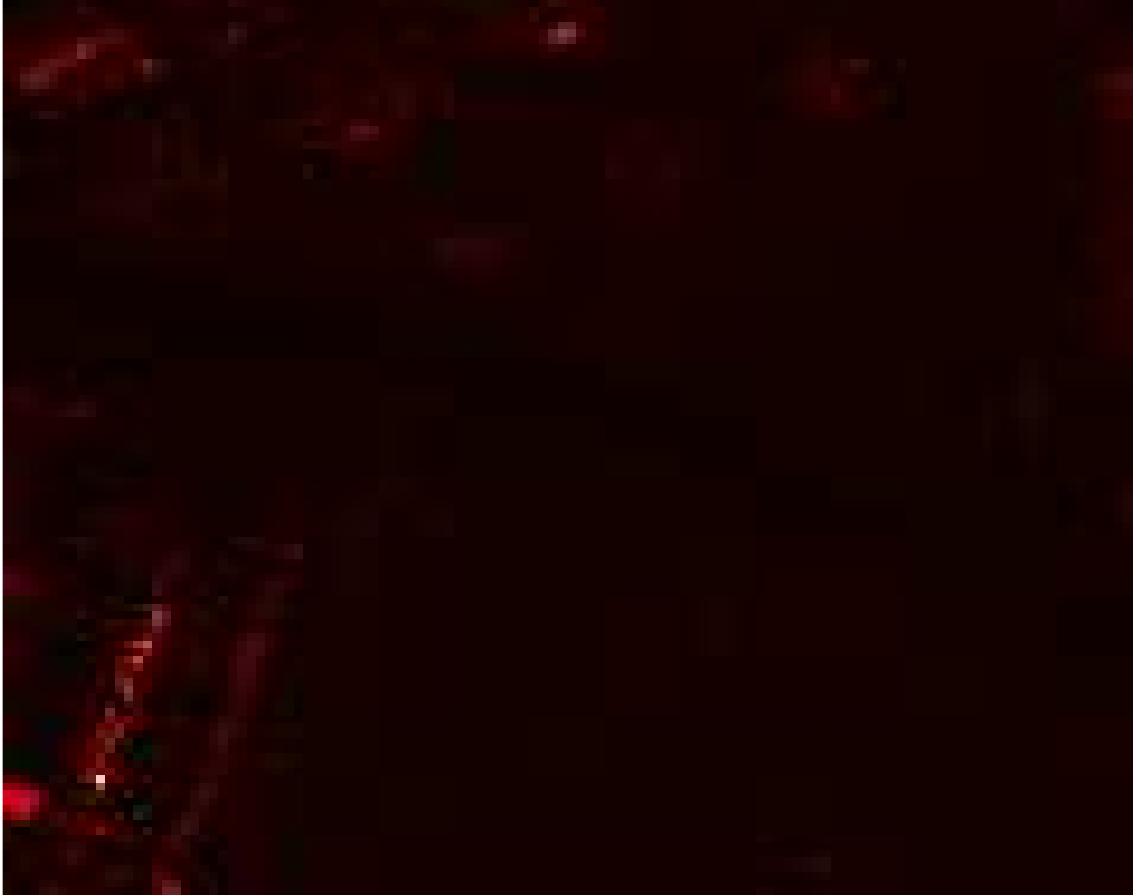}}~
  \subfloat[]{\includegraphics[scale=0.25]{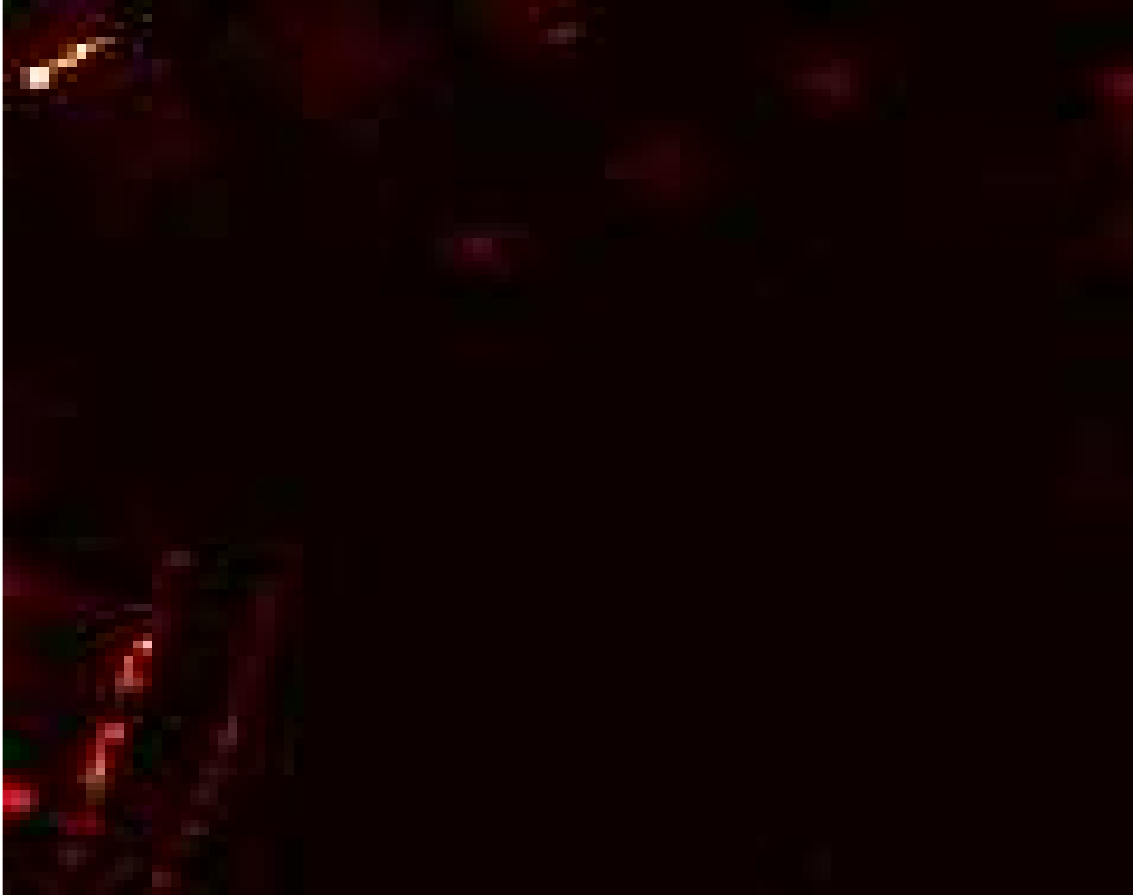}}~
  \subfloat[]{\includegraphics[scale=0.25]{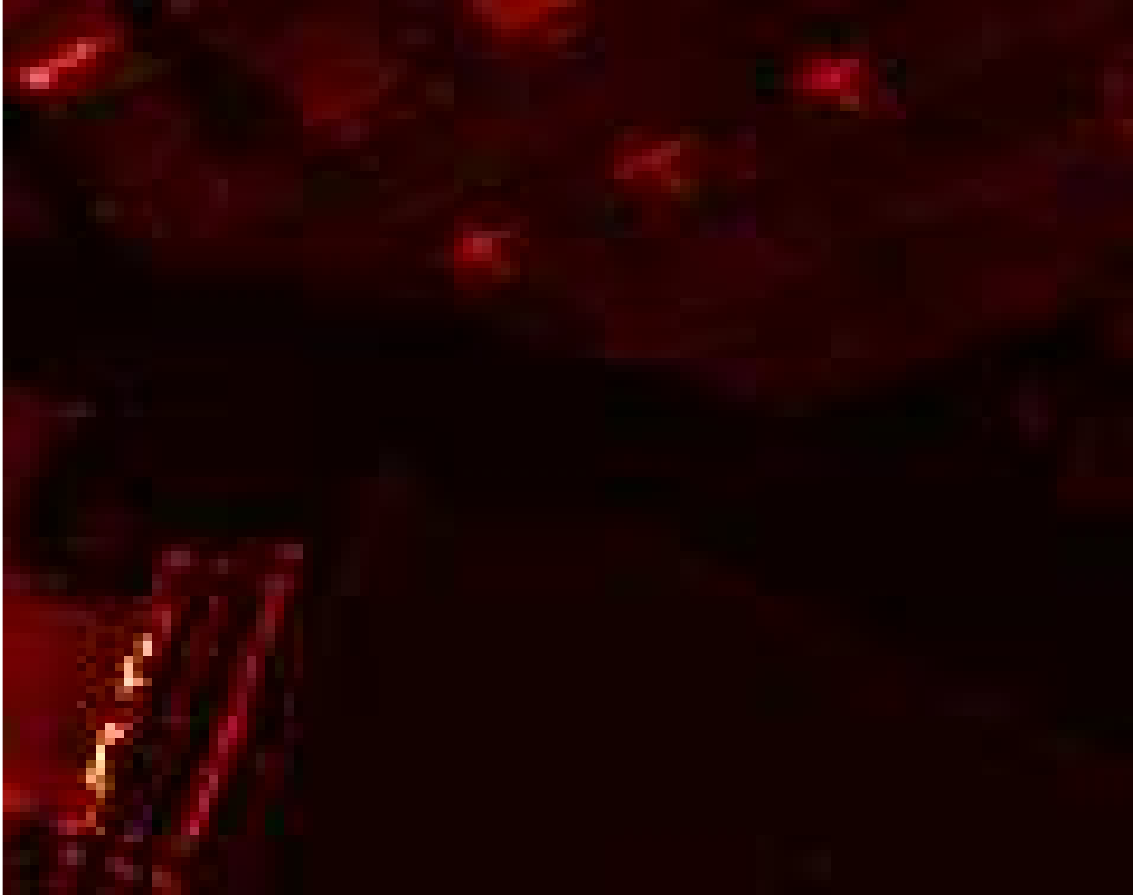}}\\
  \stepcounter{row}%
  \subfloat[]{\includegraphics[scale=0.25]{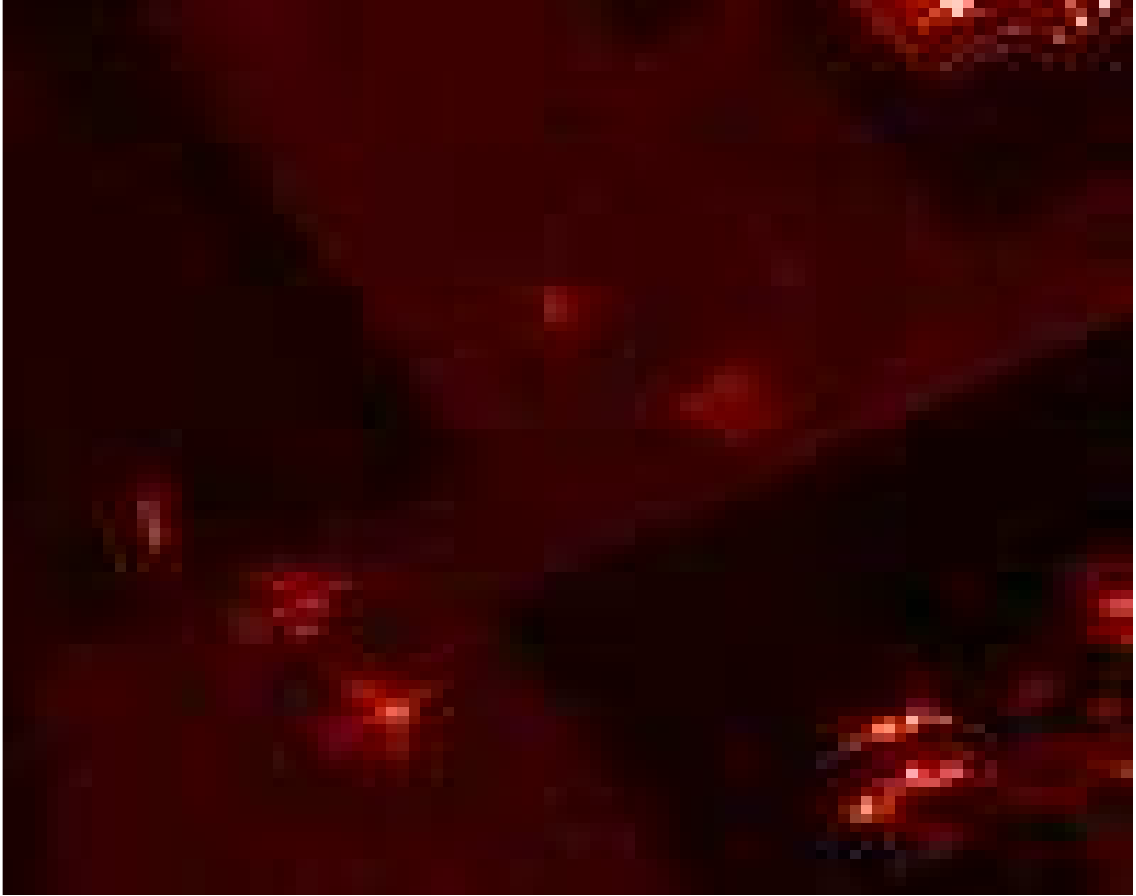}}~
  \subfloat[]{\includegraphics[scale=0.25]{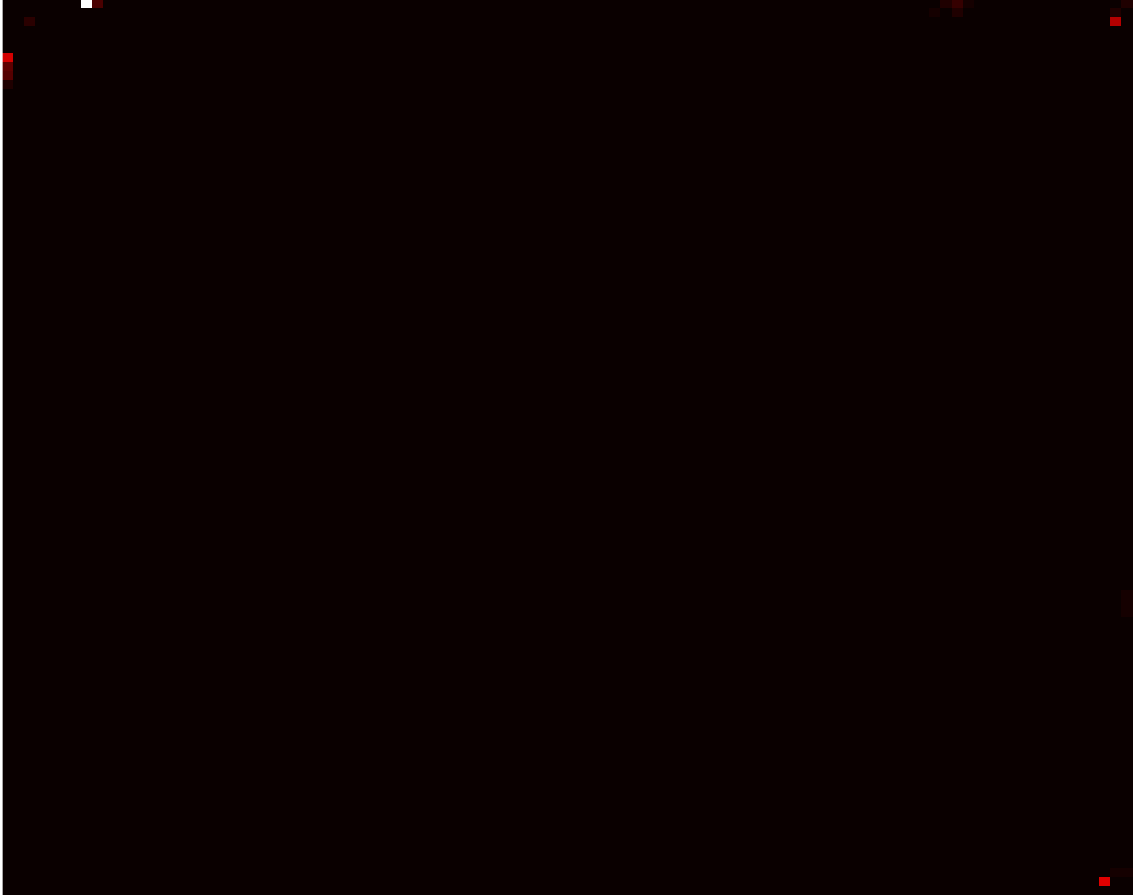}}~
  \subfloat[]{\includegraphics[scale=0.25]{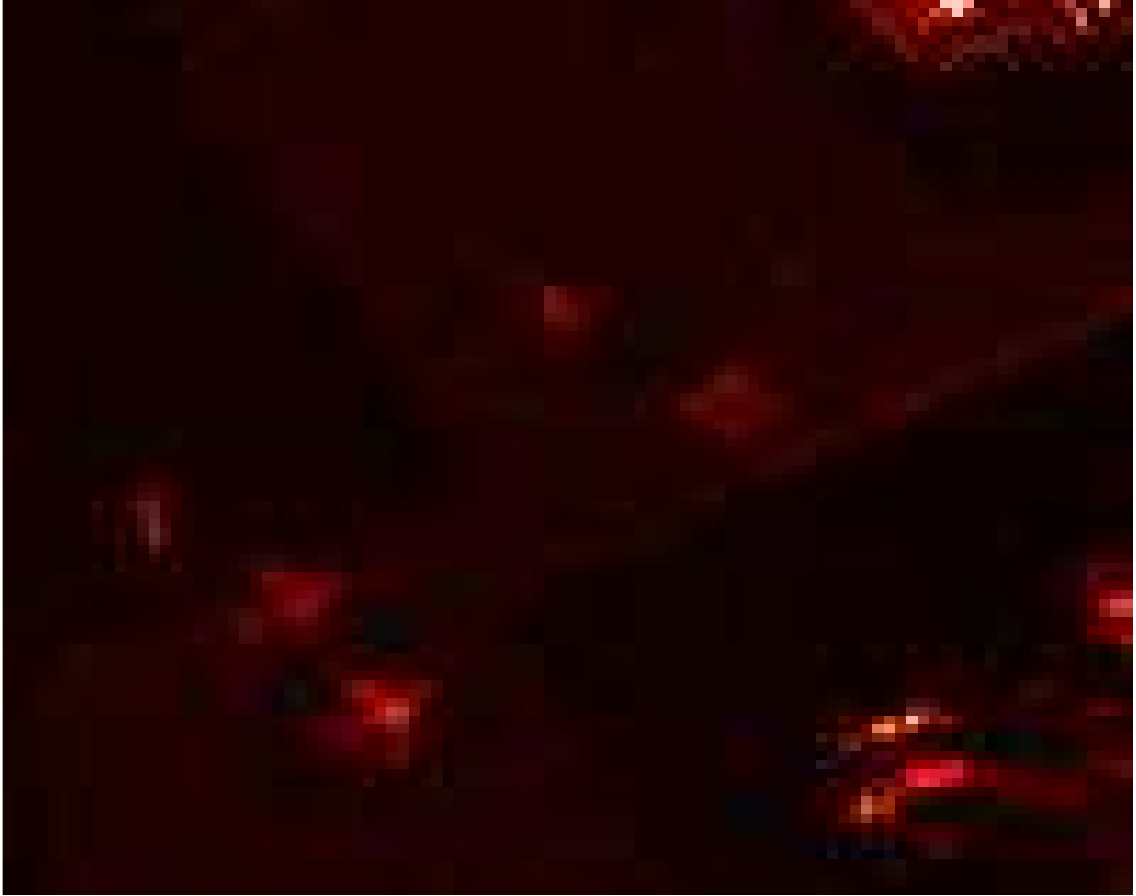}}~
  \subfloat[]{\includegraphics[scale=0.25]{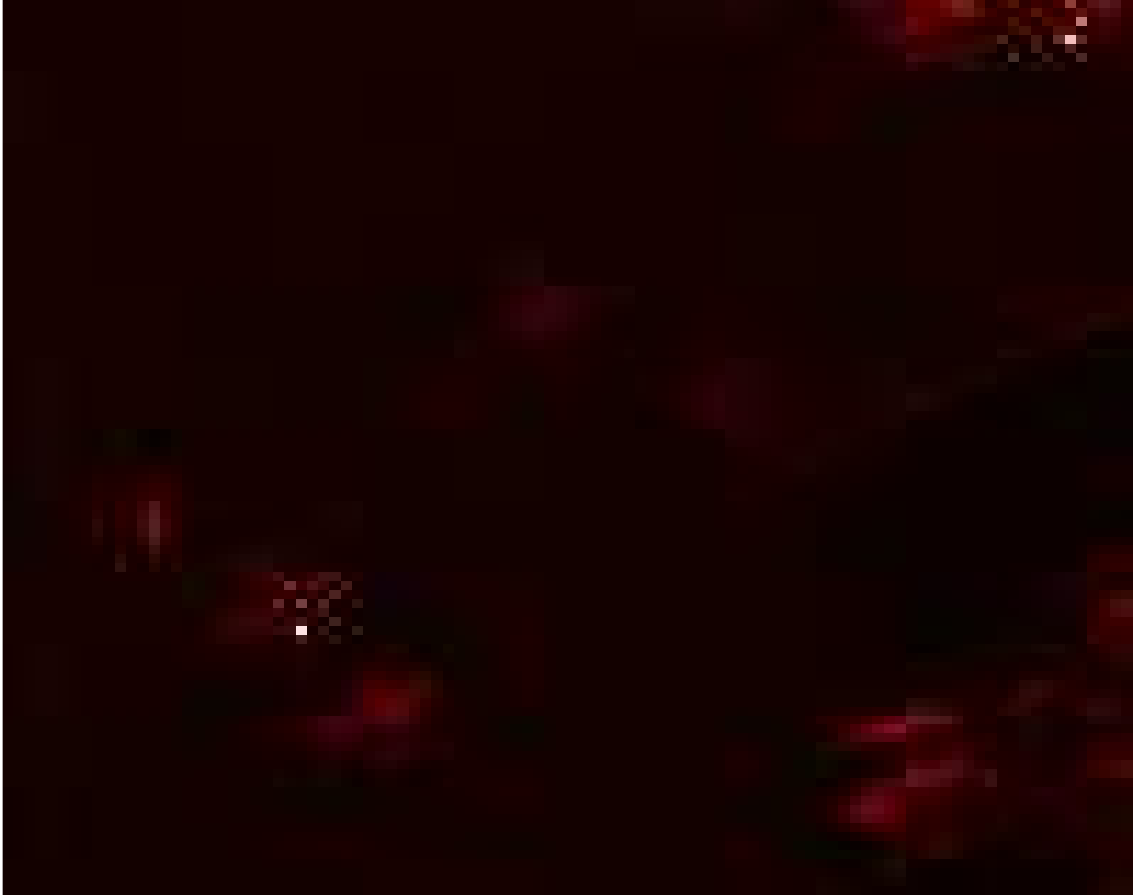}}~
  \subfloat[]{\includegraphics[scale=0.25]{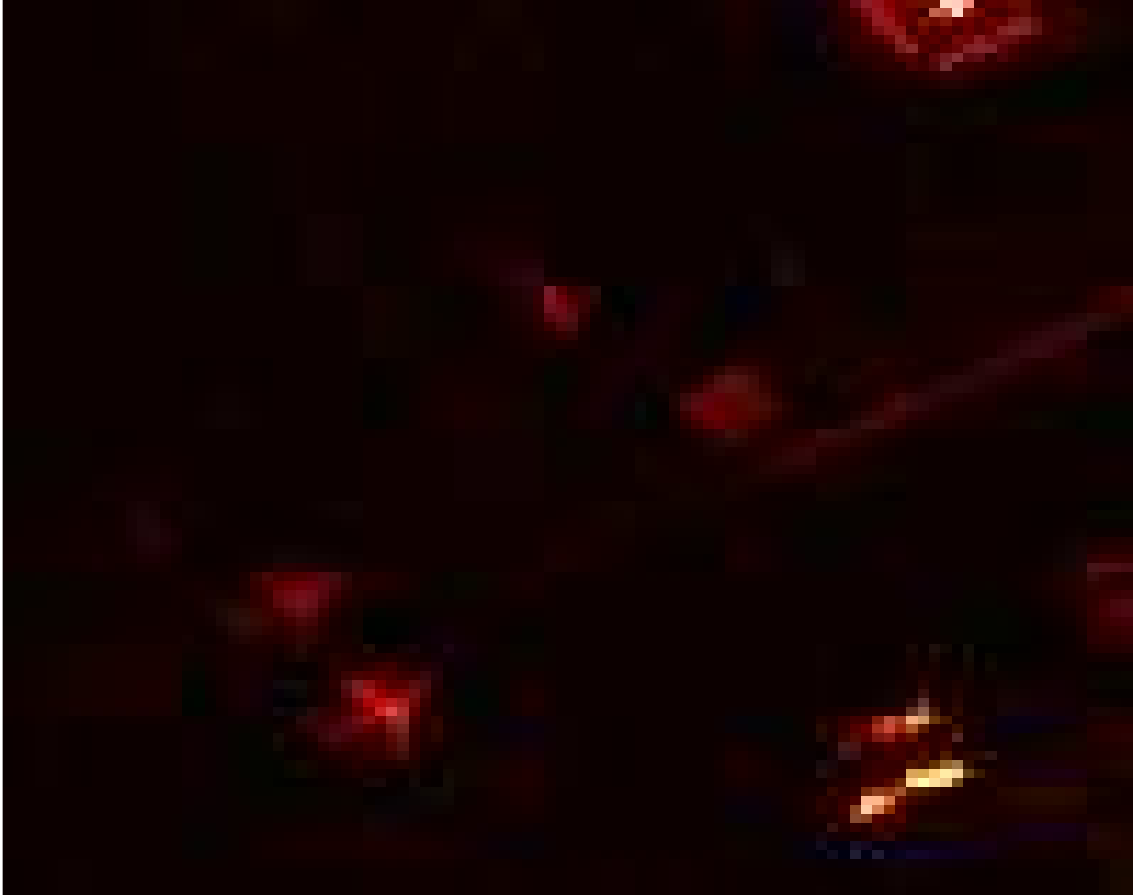}}~
  \subfloat[]{\includegraphics[scale=0.25]{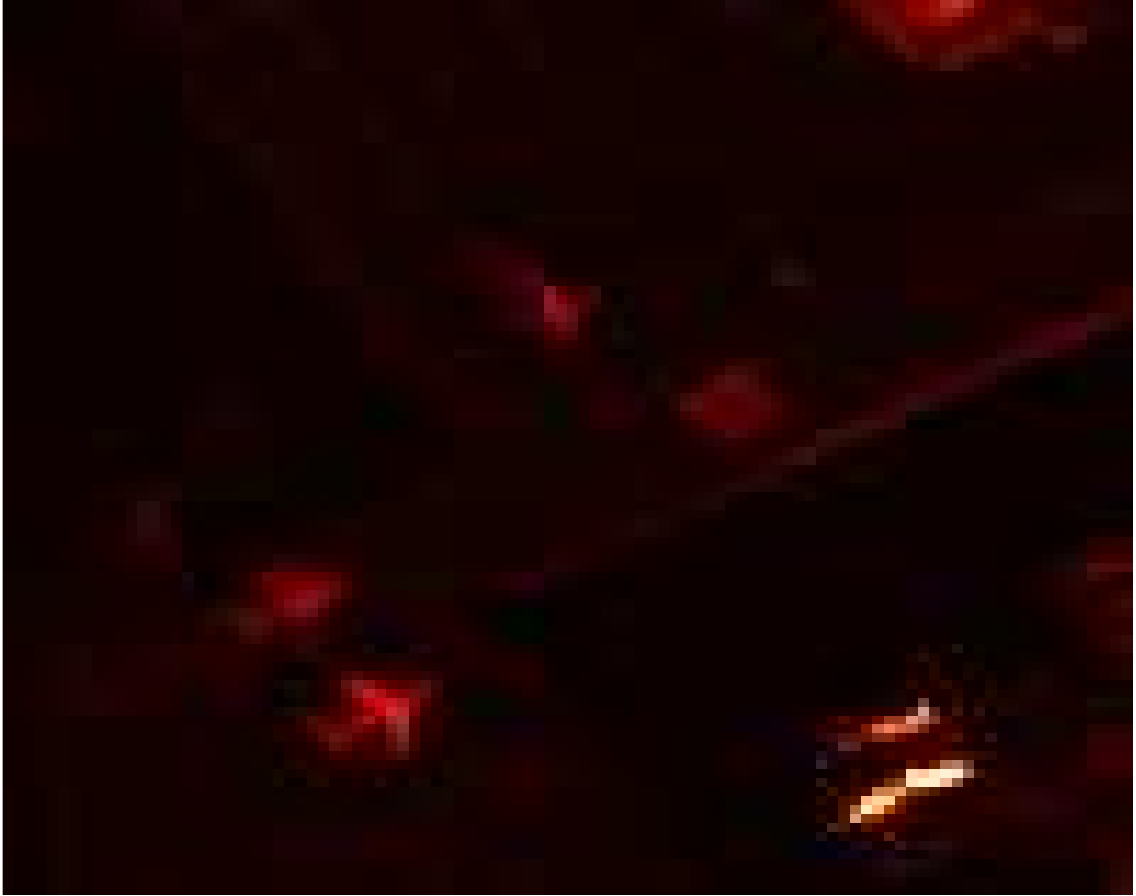}}\\
  \stepcounter{row}%
  \subfloat[]{\includegraphics[scale=0.25]{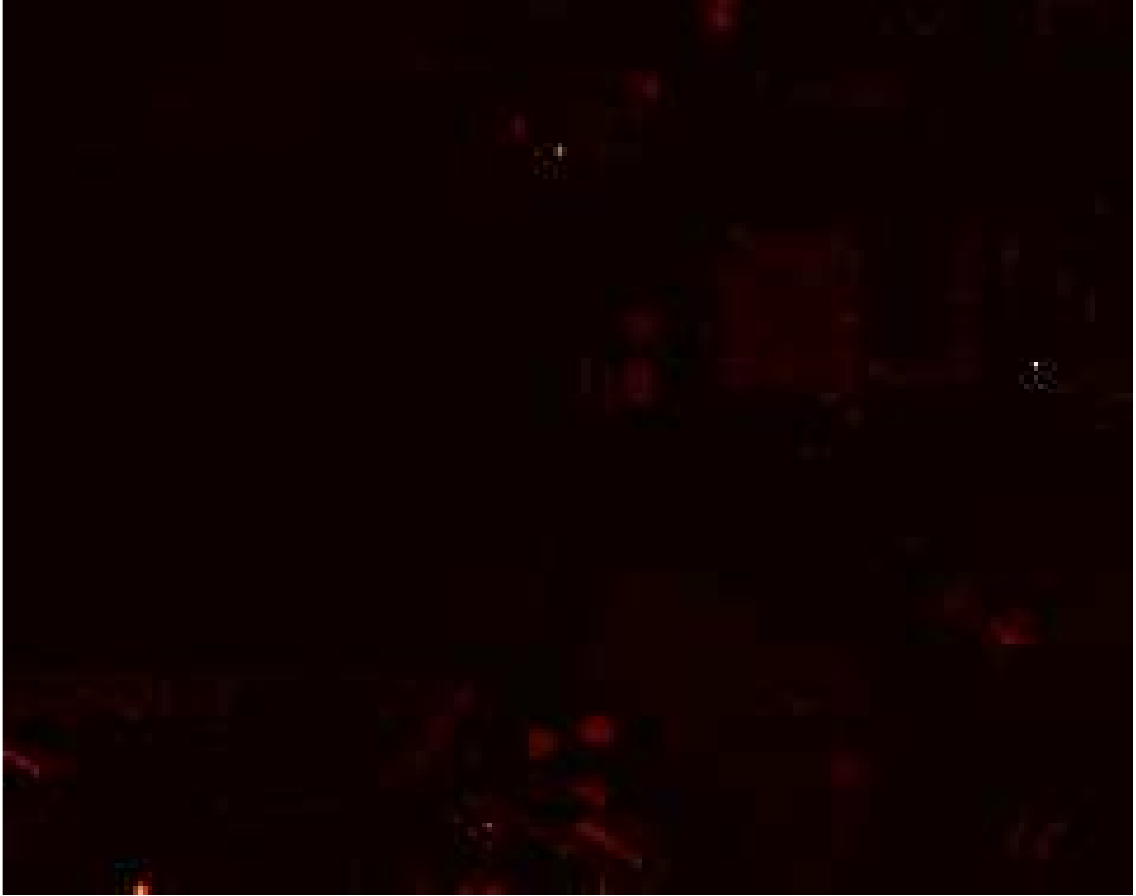}}~
  \subfloat[]{\includegraphics[scale=0.25]{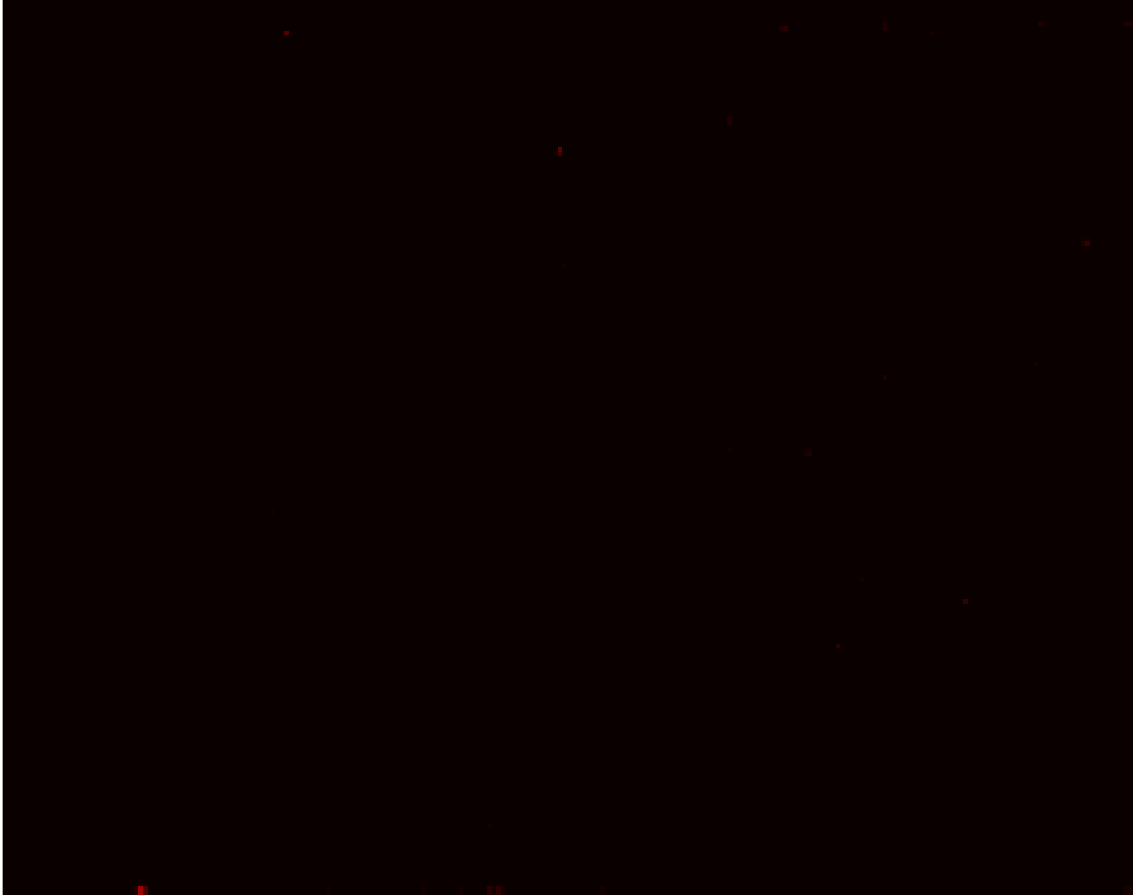}}~
  \subfloat[]{\includegraphics[scale=0.25]{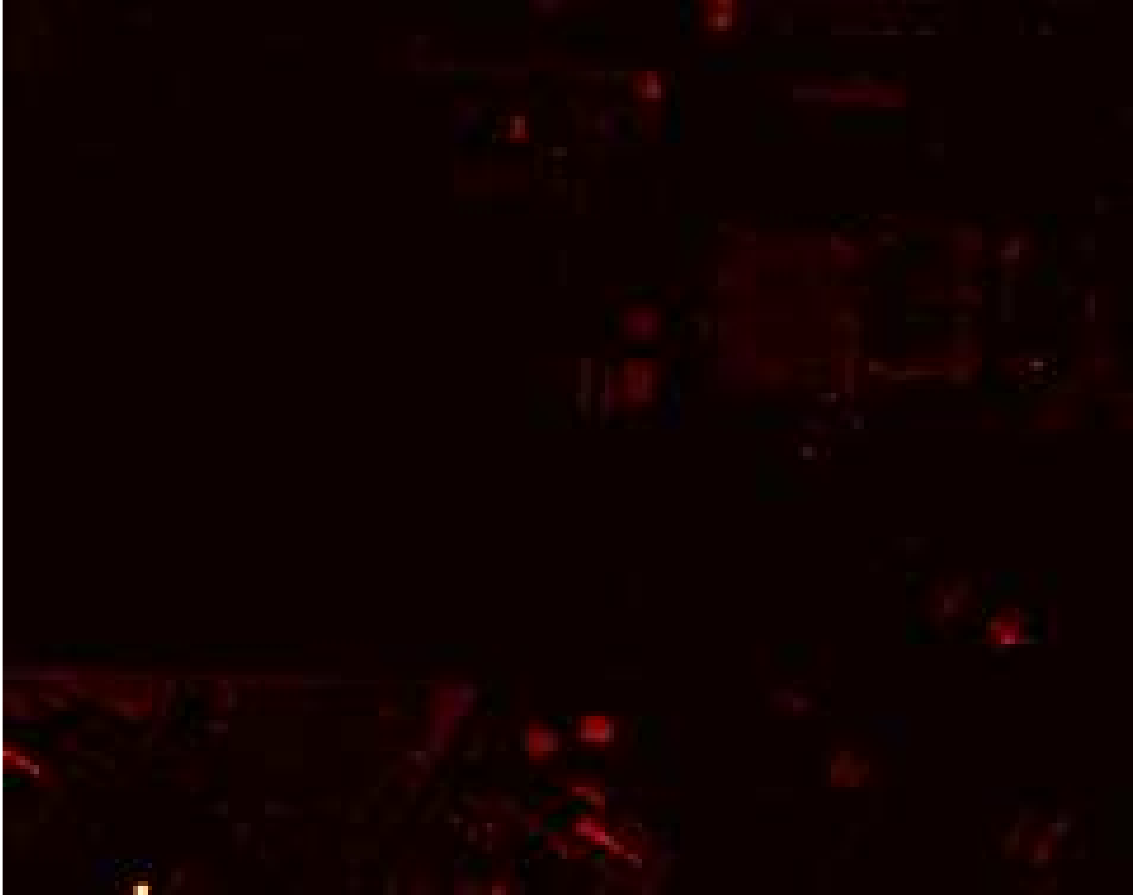}}~
  \subfloat[]{\includegraphics[scale=0.25]{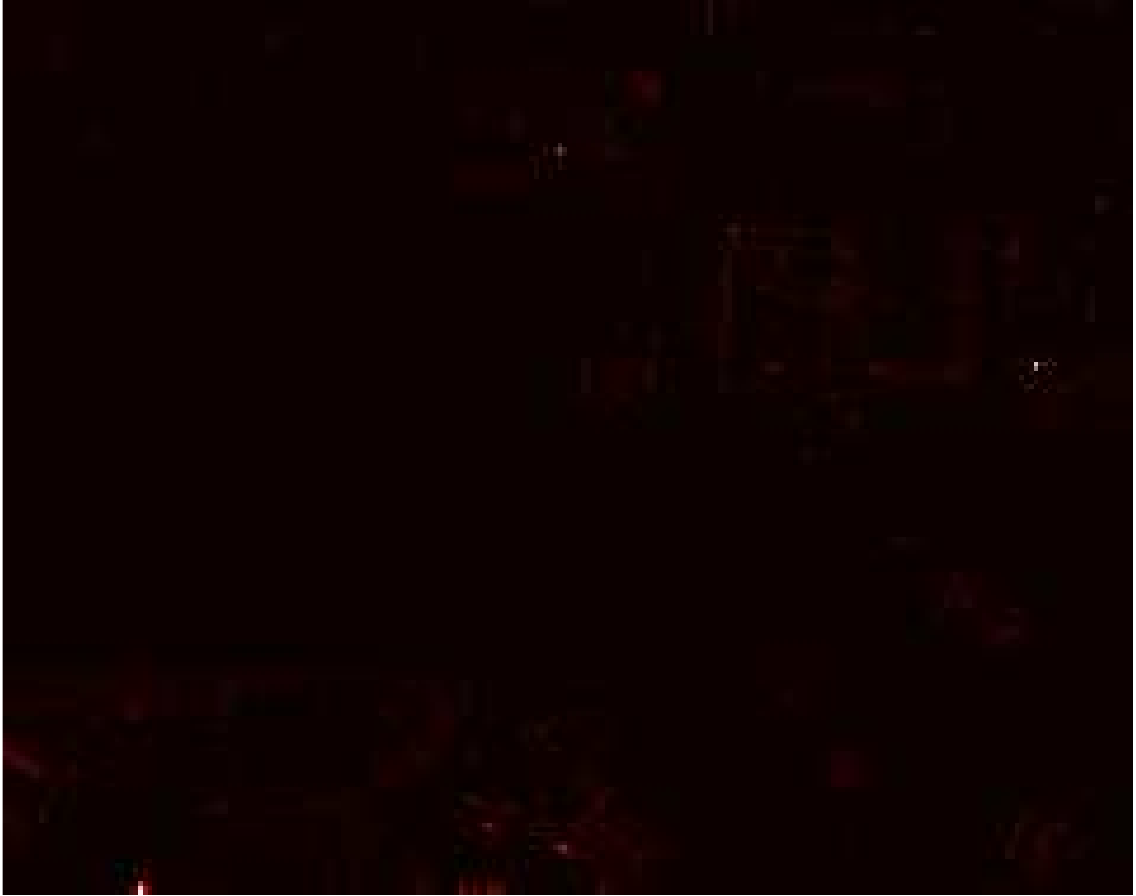}}~
  \subfloat[]{\includegraphics[scale=0.25]{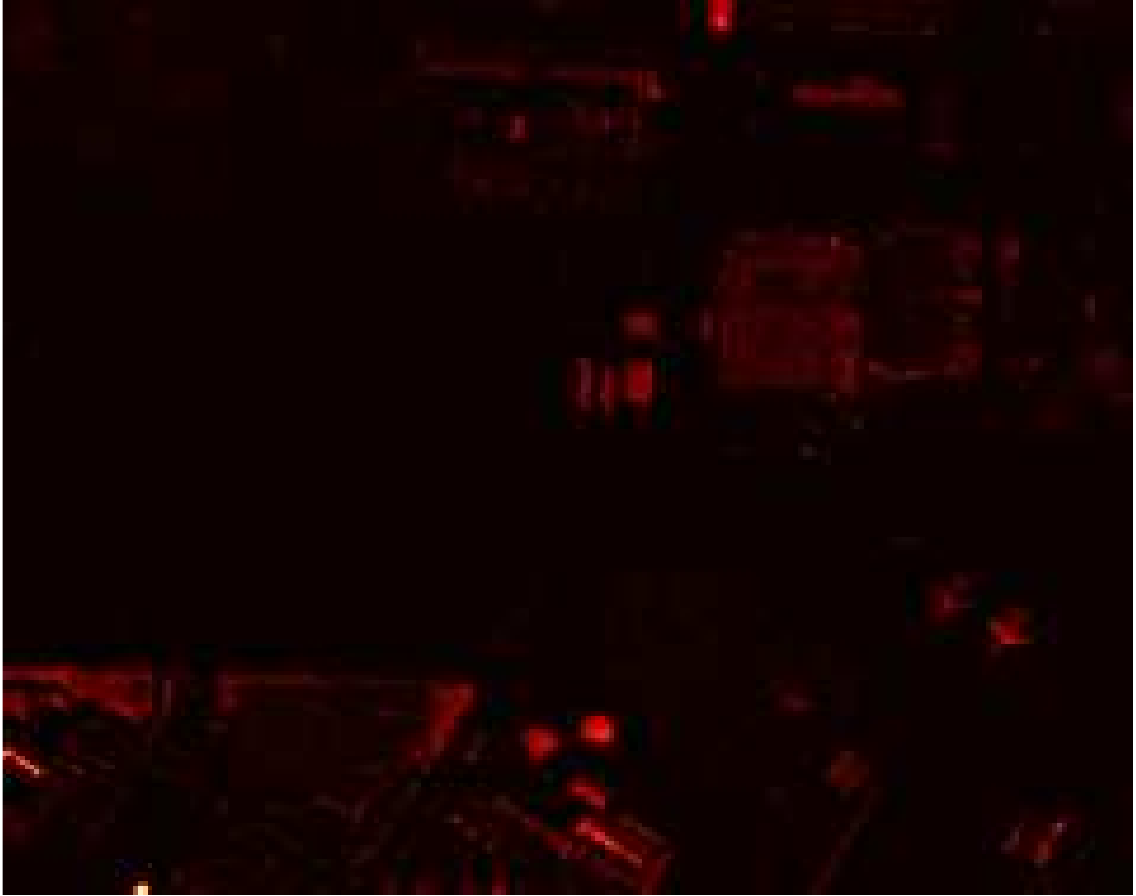}}~
  \subfloat[]{\includegraphics[scale=0.25]{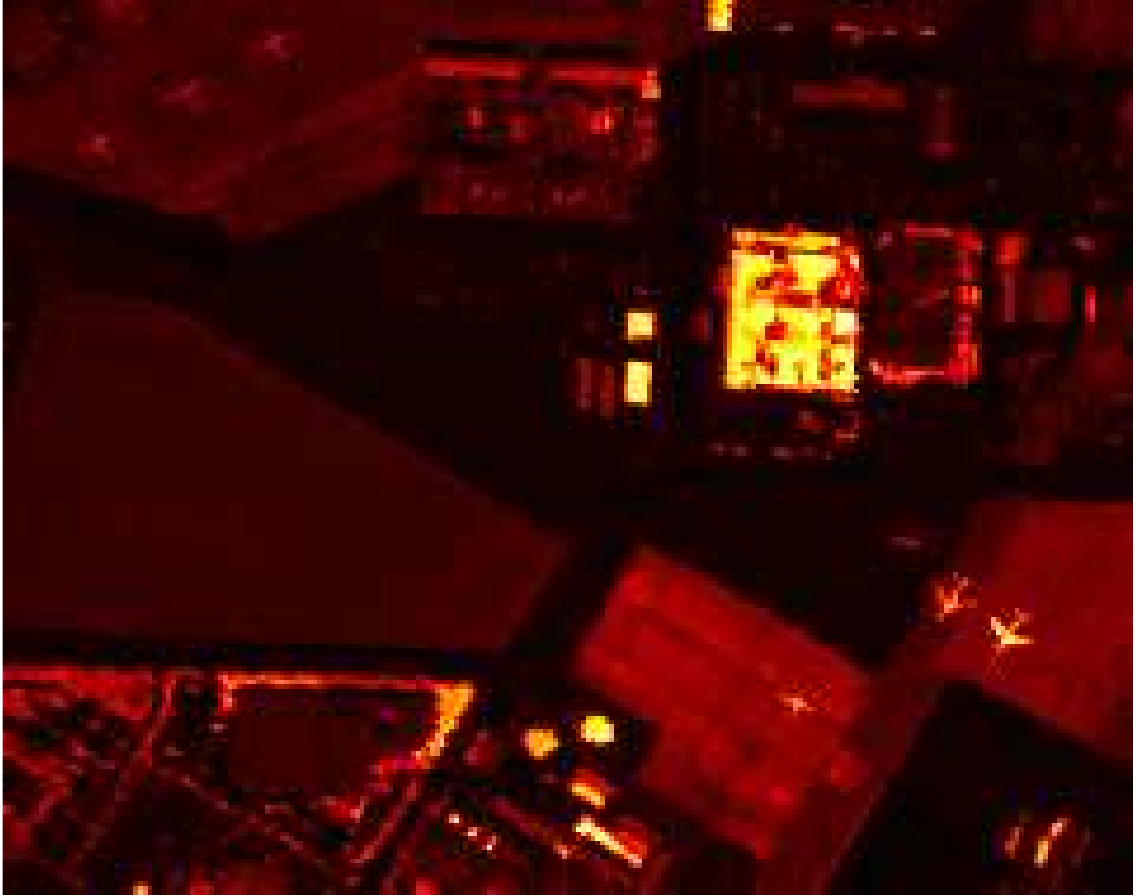}}\\
  \stepcounter{row}%
  \subfloat[]{\includegraphics[scale=0.25]{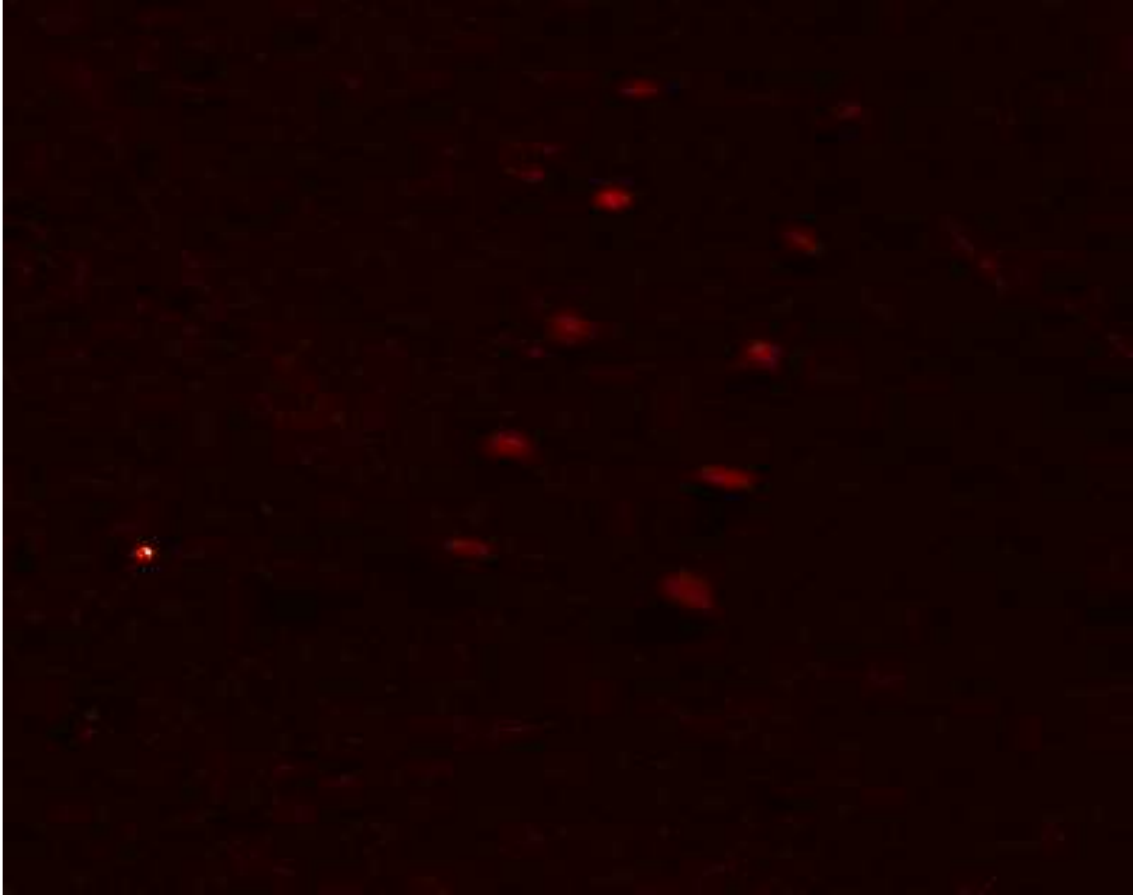}}~
  \subfloat[]{\includegraphics[scale=0.25]{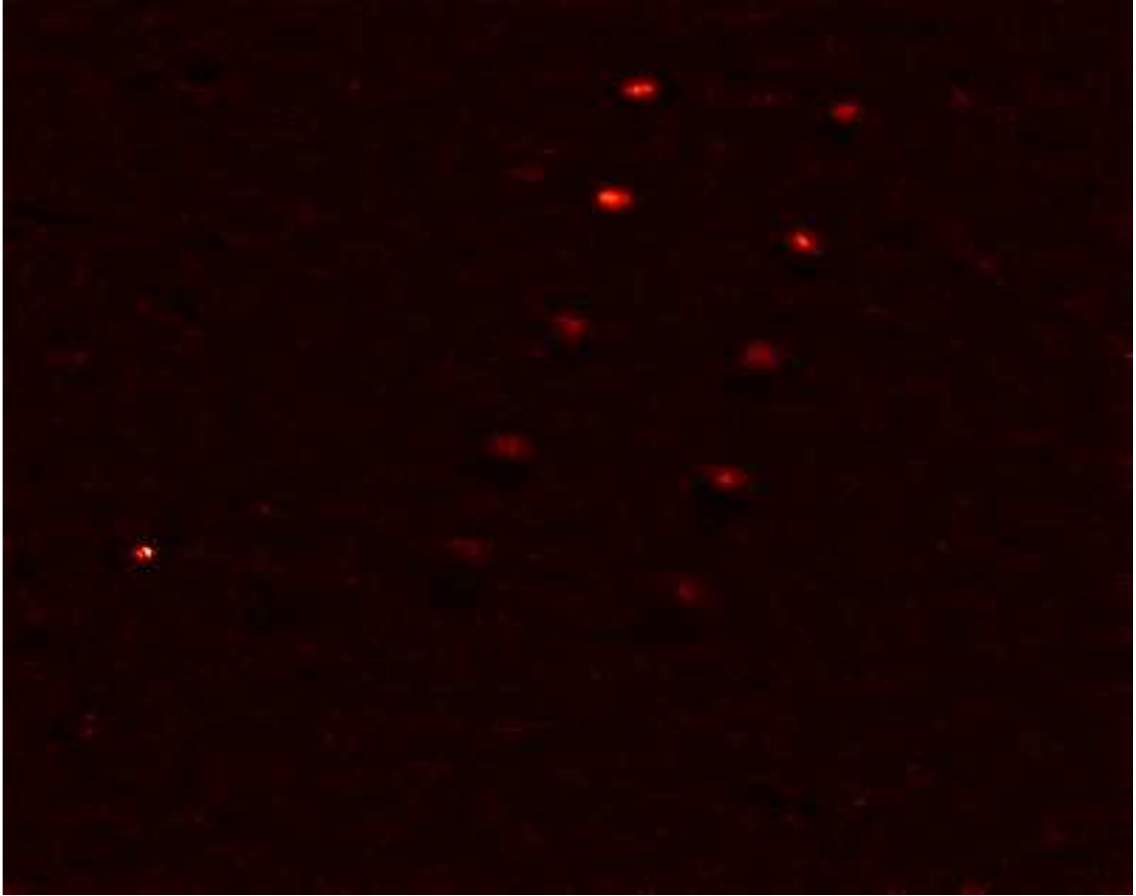}}~
  \subfloat[]{\includegraphics[scale=0.25]{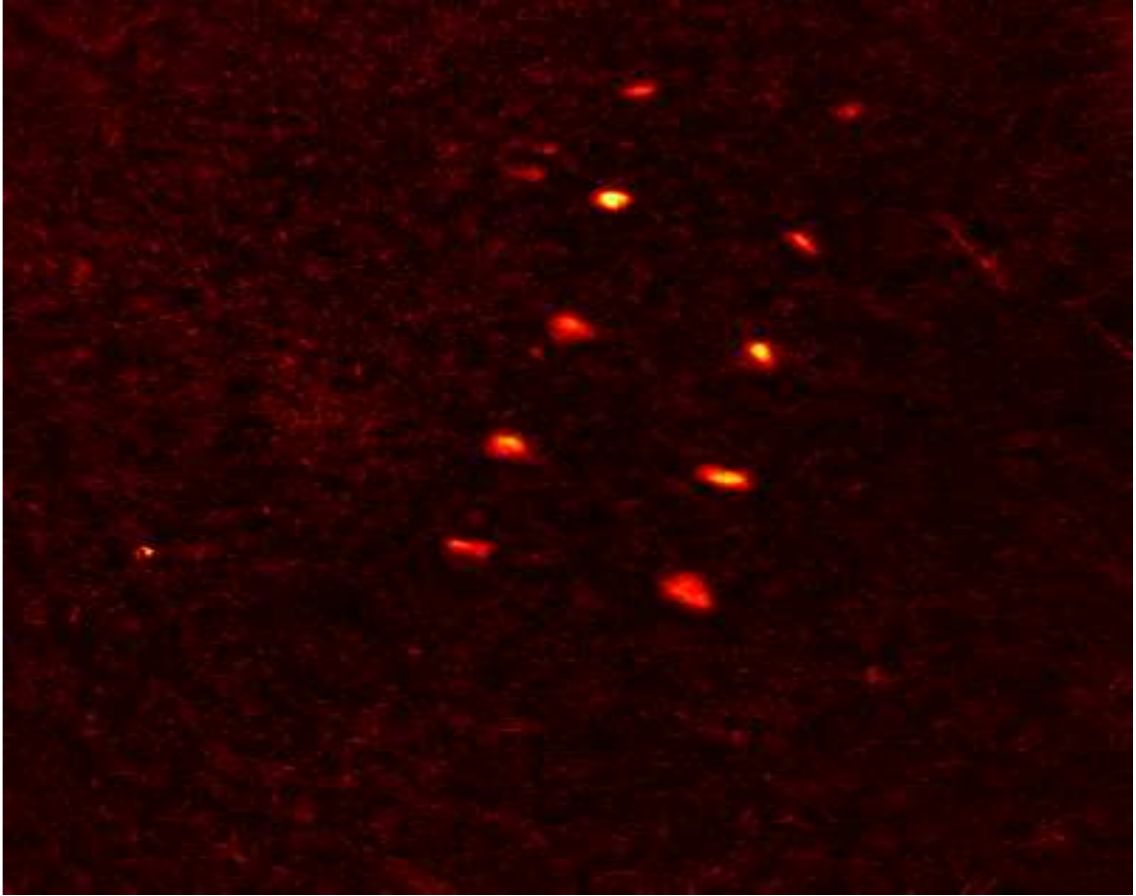}}~
  \subfloat[]{\includegraphics[scale=0.25]{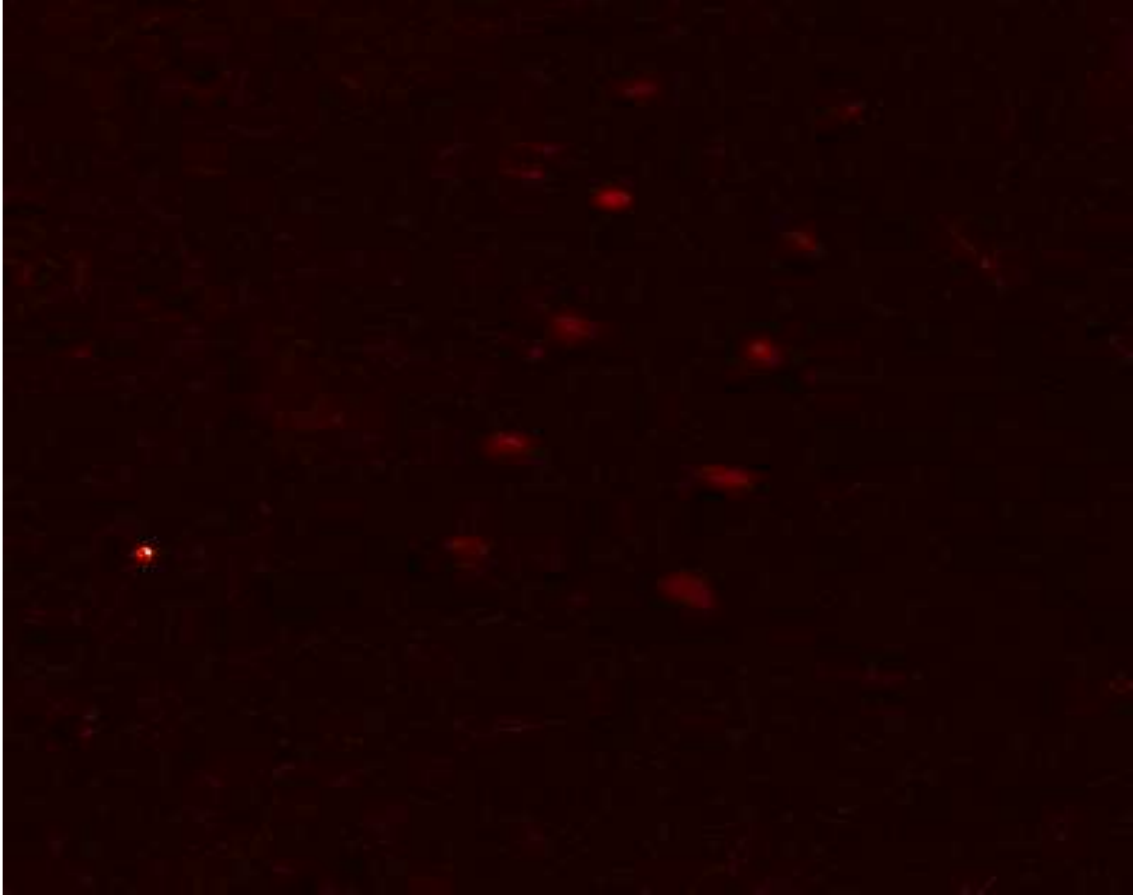}}~
  \subfloat[]{\includegraphics[scale=0.25]{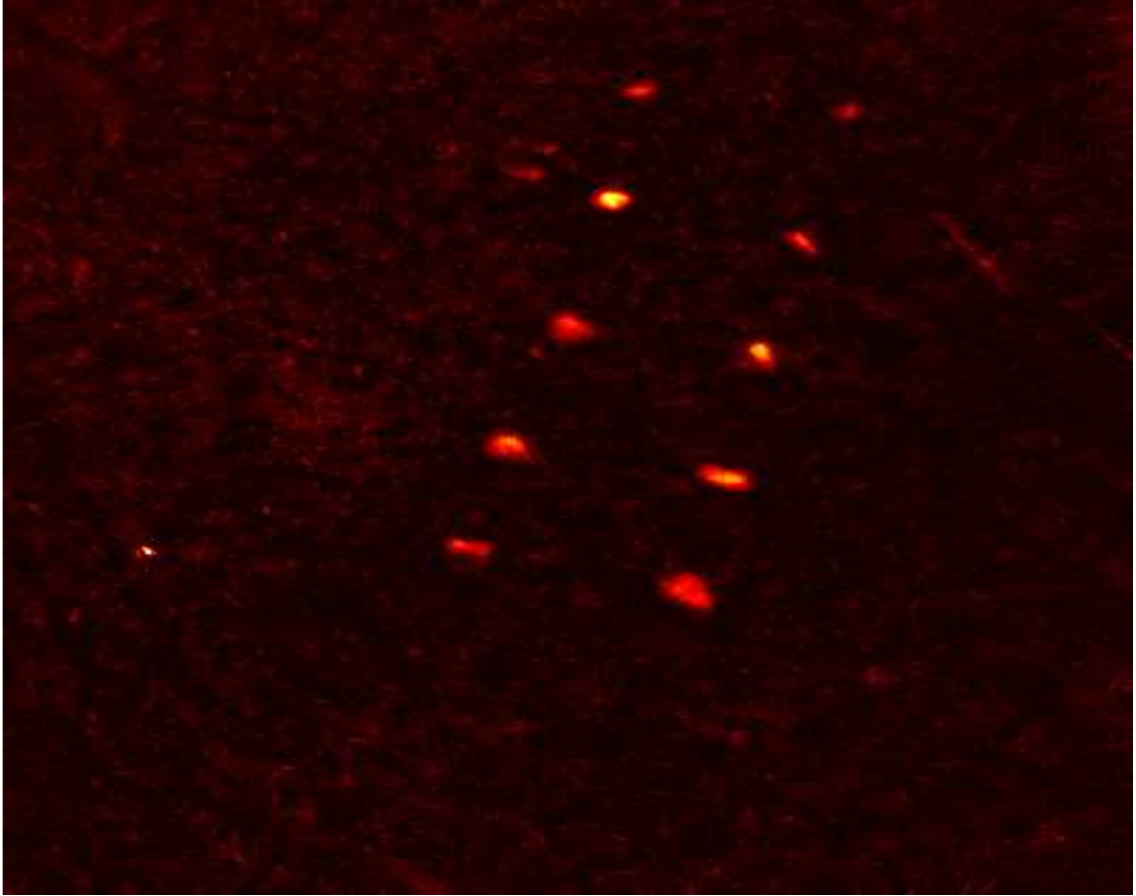}}~
  \subfloat[]{\includegraphics[scale=0.25]{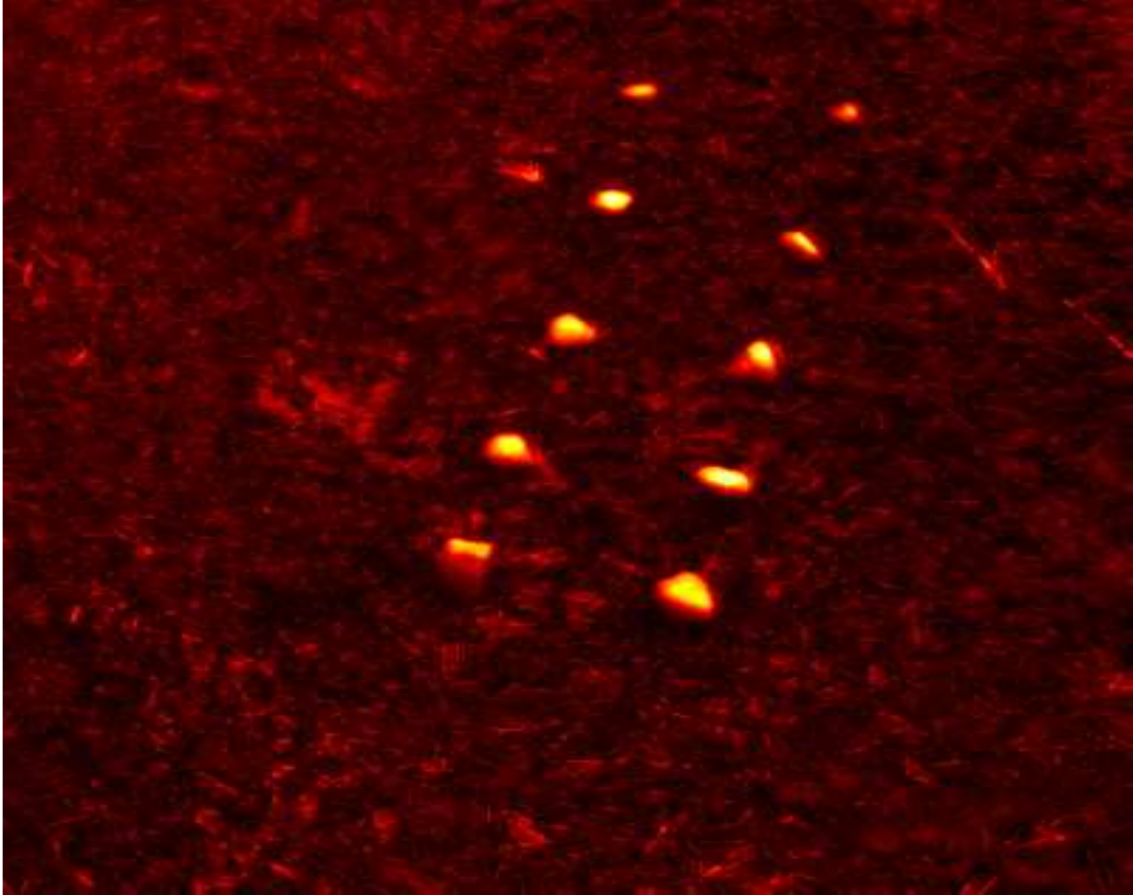}}\\
  \subfloat{\includegraphics[scale=0.35]{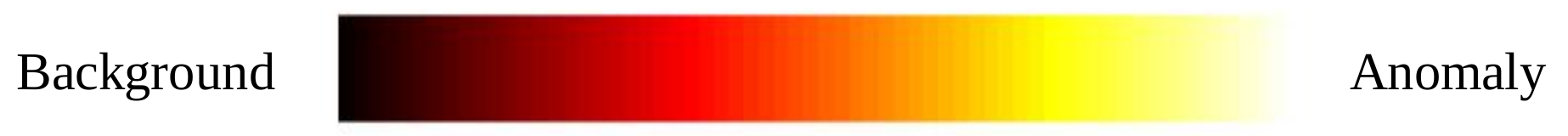}}
  \caption{I is color detection maps obtained by different algorithms for AVIRIS-I dataset. II is color detection maps obtained by different algorithms for AVIRIS-II dataset.
  III is color detection maps obtained by different algorithms for AVIRIS-III dataset. IV is color detection maps obtained by different algorithms for Cri dataset.
  (a) GRX. (b) LRX. (c) SSRX. (d) CBAD. (e) LSMAD. (f) ERCRD.}
  \label{Fig.5}
\end{figure*}

\begin{figure*}
    \renewcommand{\thesubfigure}{\Roman{row}-\alph{subfigure}}
    \centering
    \setcounter{row}{1}%
    \subfloat[]{\includegraphics[scale=0.45]{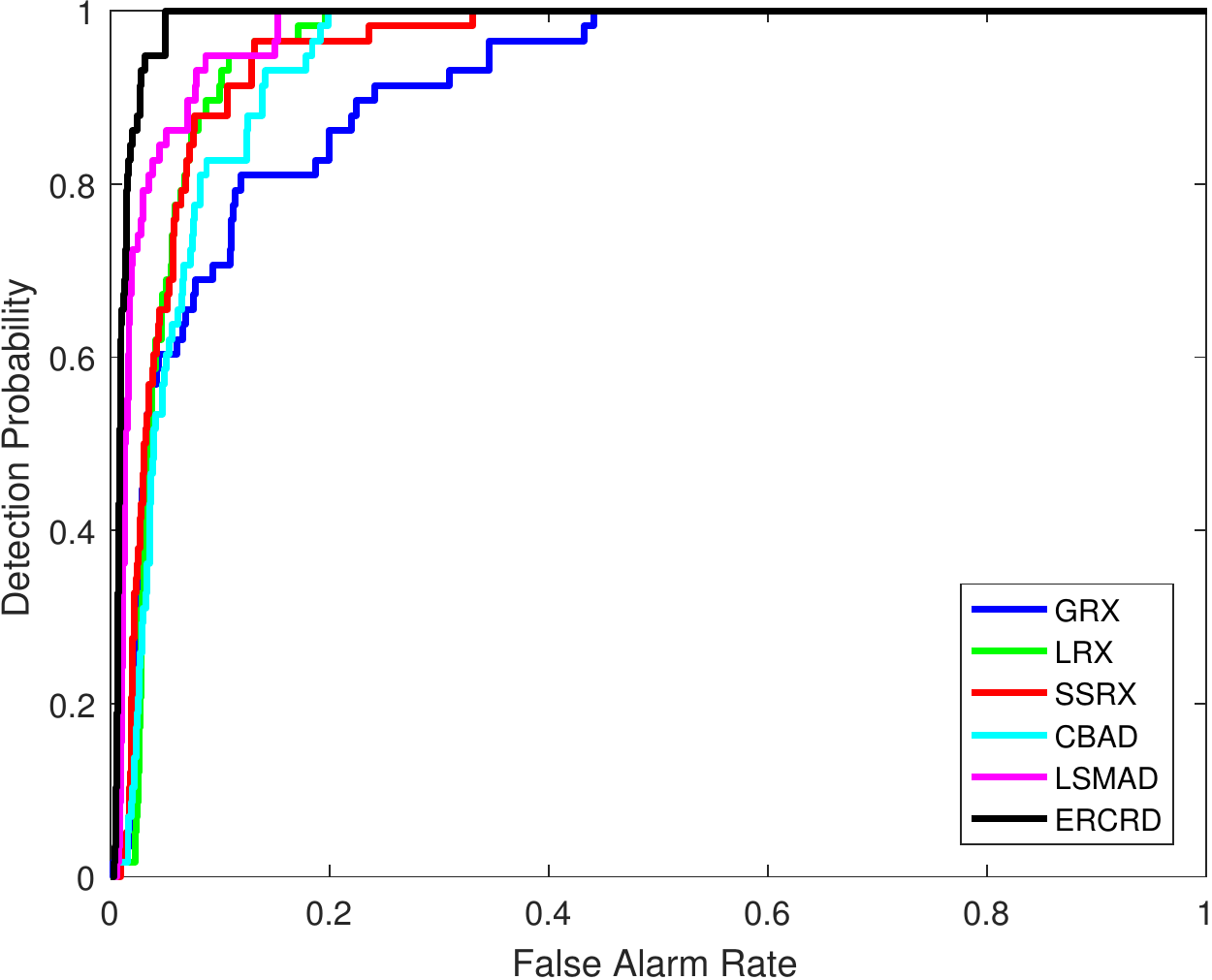}}~
    \subfloat[]{\includegraphics[scale=0.45]{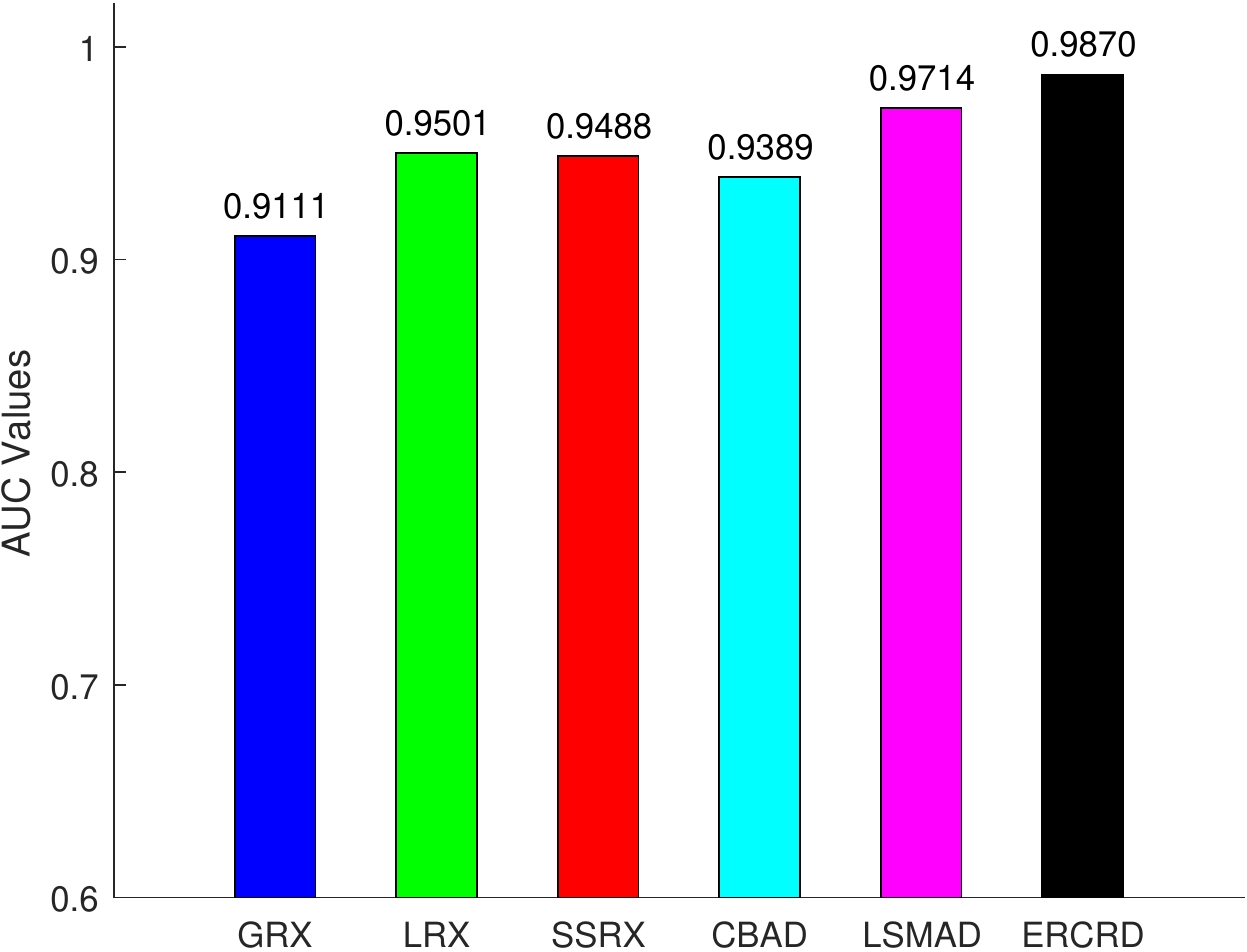}}~
    \subfloat[]{\includegraphics[scale=0.45]{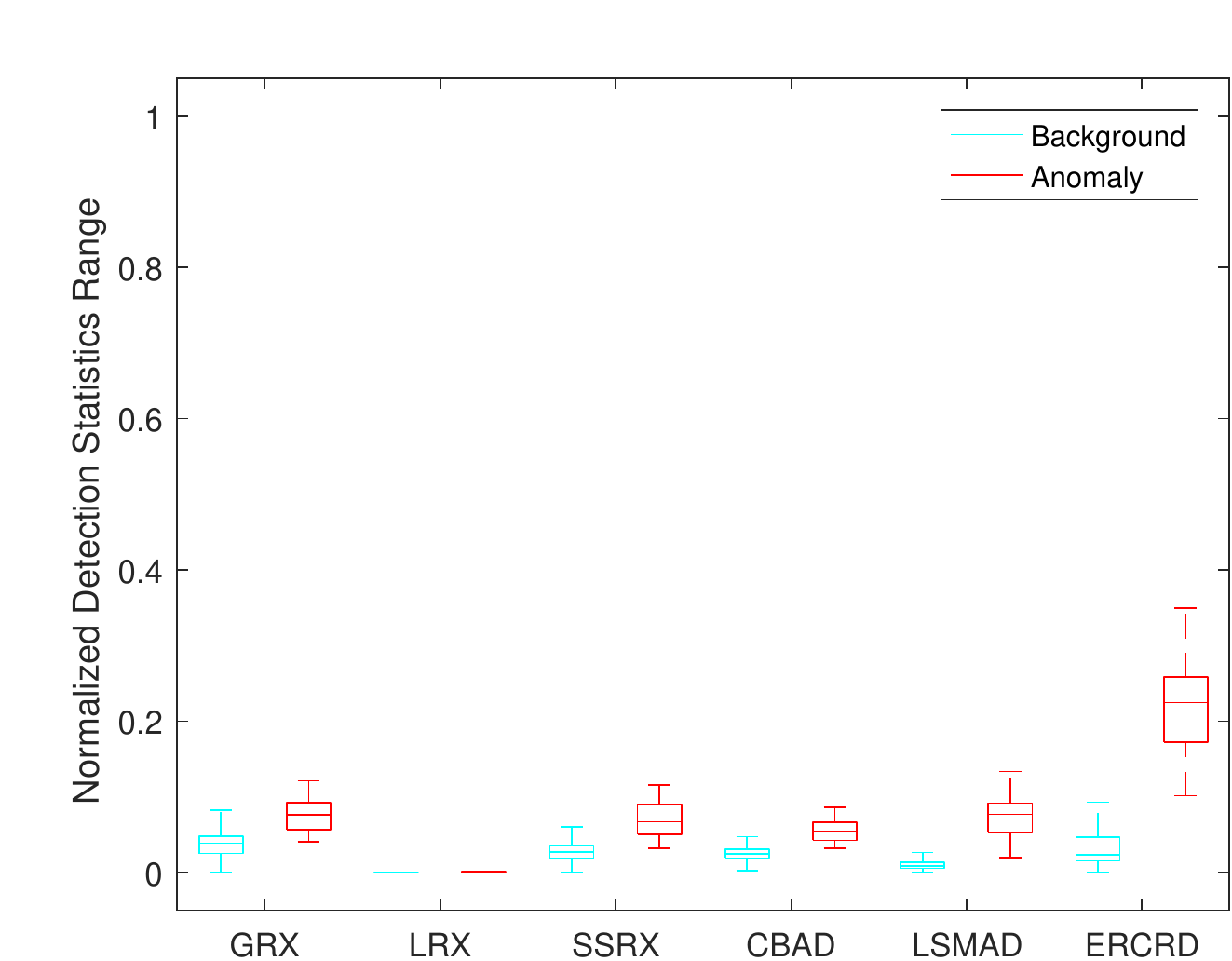}}\\
    \stepcounter{row}%
    \subfloat[]{\includegraphics[scale=0.45]{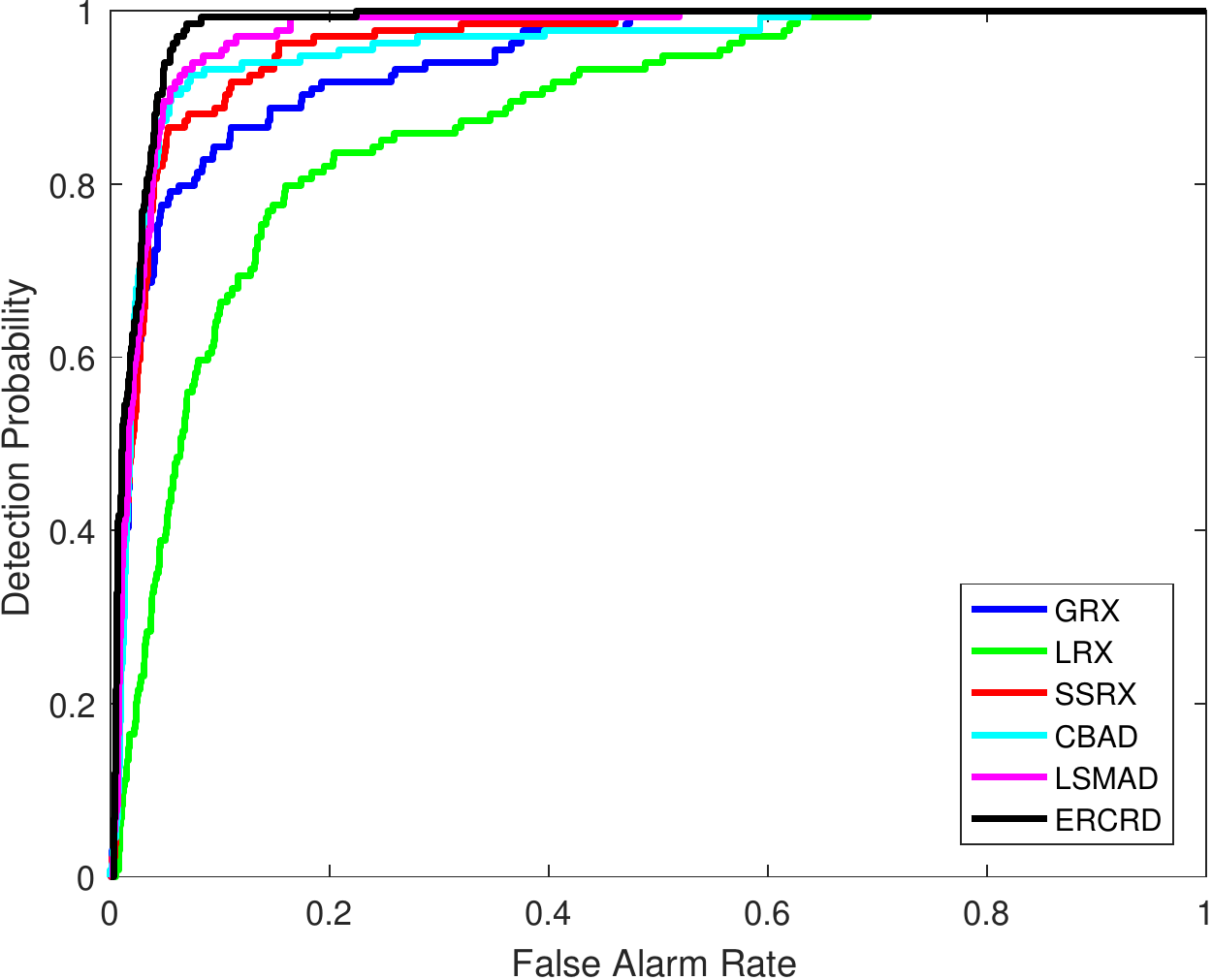}}~
    \subfloat[]{\includegraphics[scale=0.45]{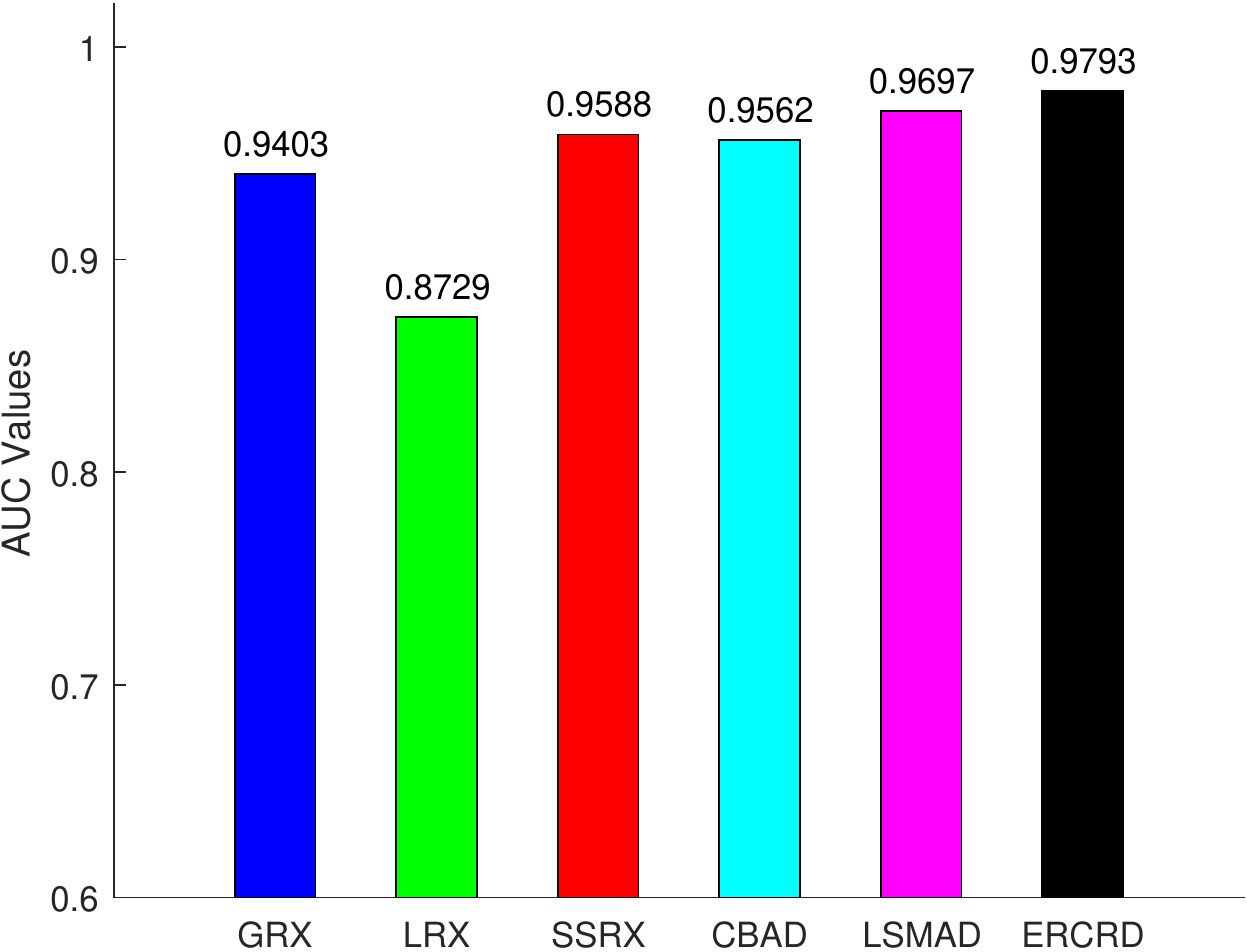}}~
    \subfloat[]{\includegraphics[scale=0.45]{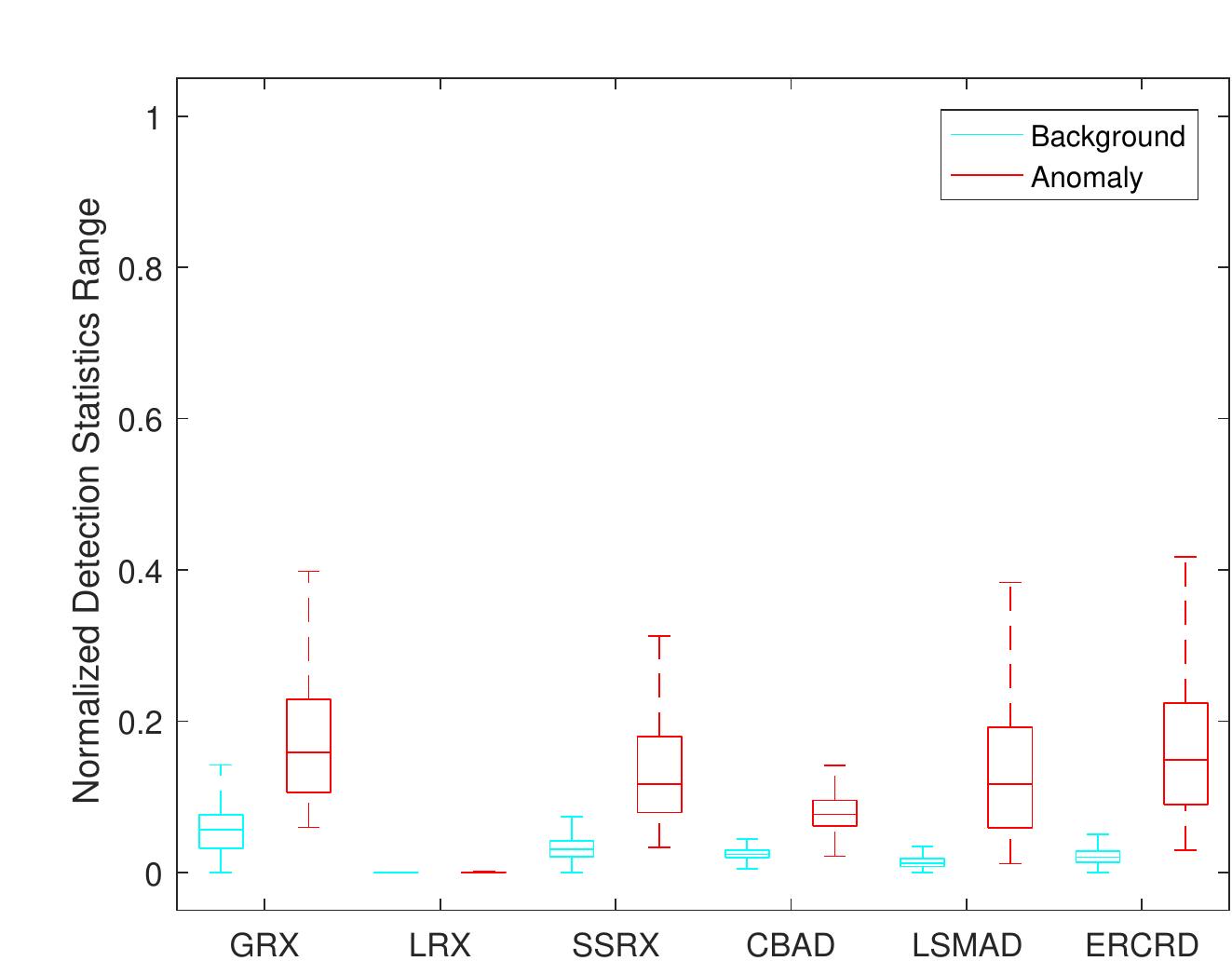}}\\
    \stepcounter{row}%
    \subfloat[]{\includegraphics[scale=0.45]{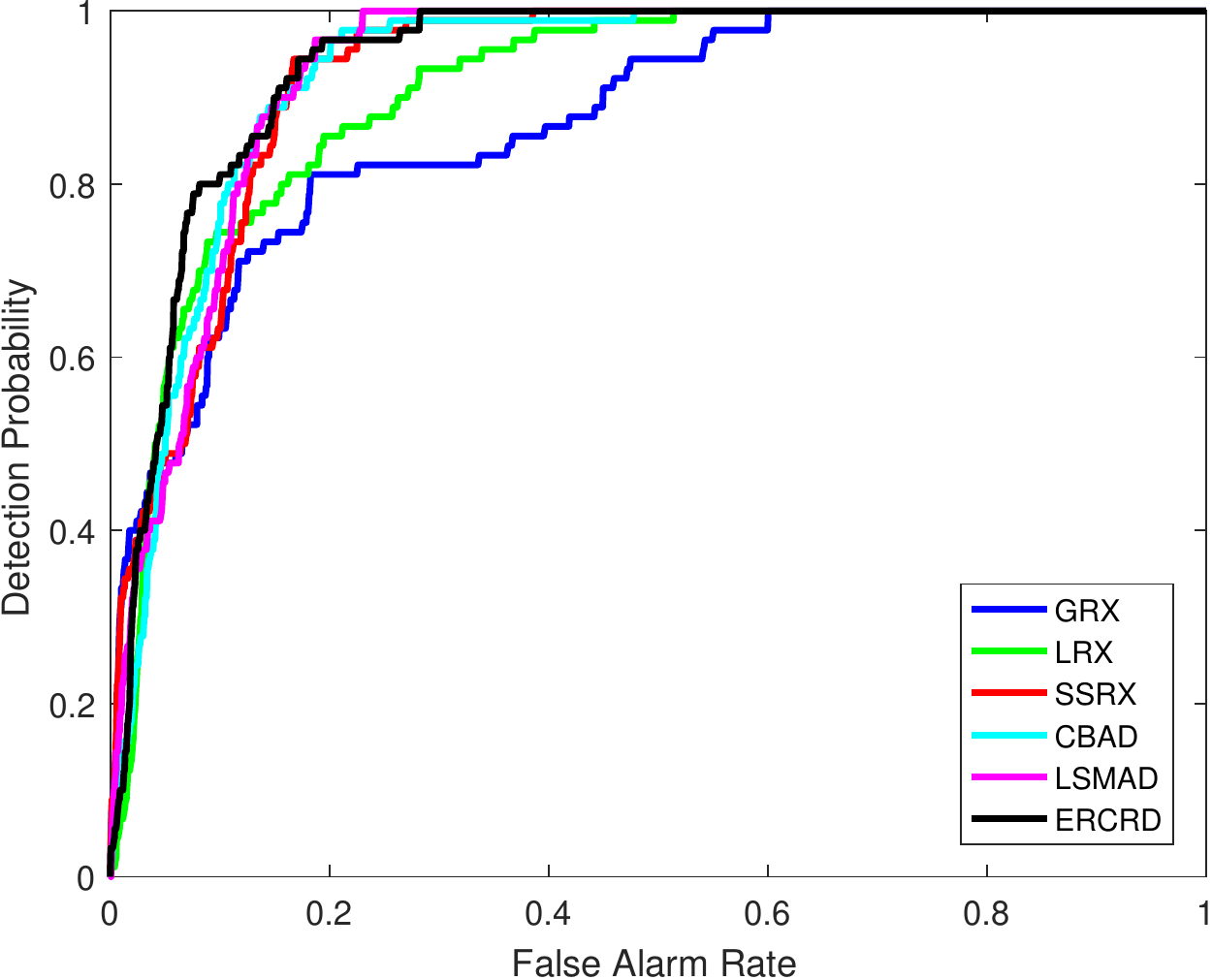}\label{Fig.6.IIIa}}~
    \subfloat[]{\includegraphics[scale=0.45]{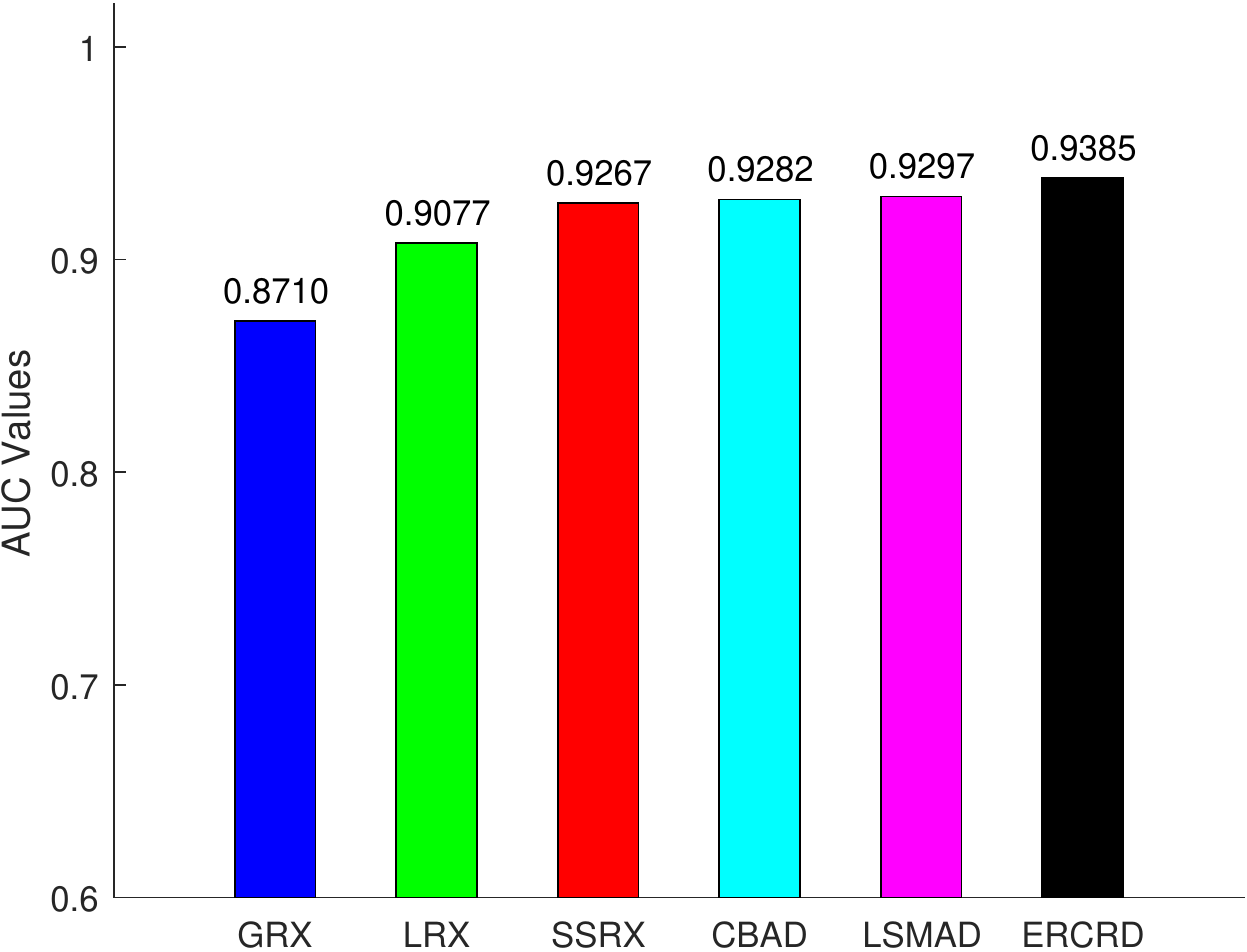}\label{Fig.6.IIIb}}~
    \subfloat[]{\includegraphics[scale=0.45]{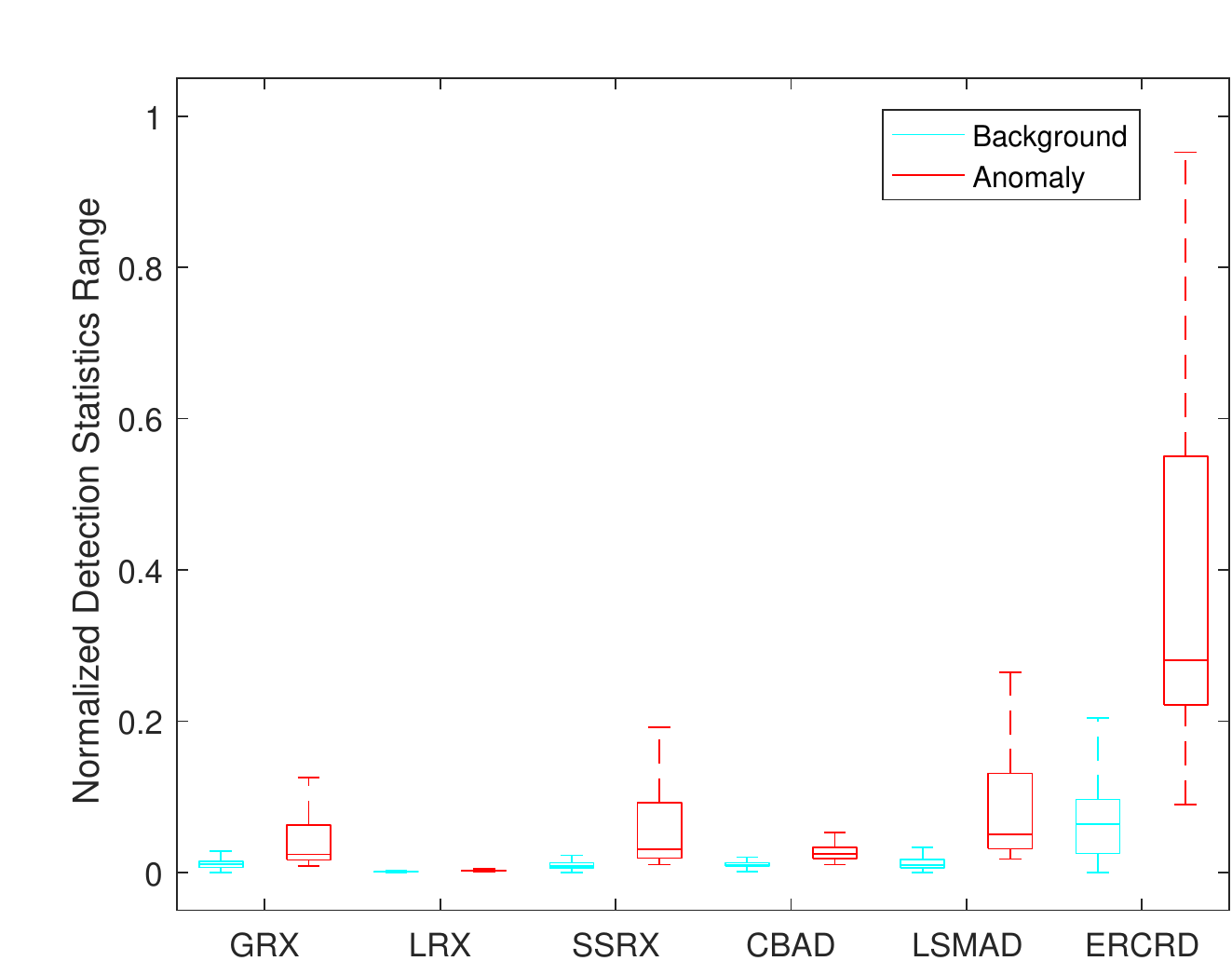}\label{Fig.6.IIIc}}\\
    \stepcounter{row}%
    \subfloat[]{\includegraphics[scale=0.45]{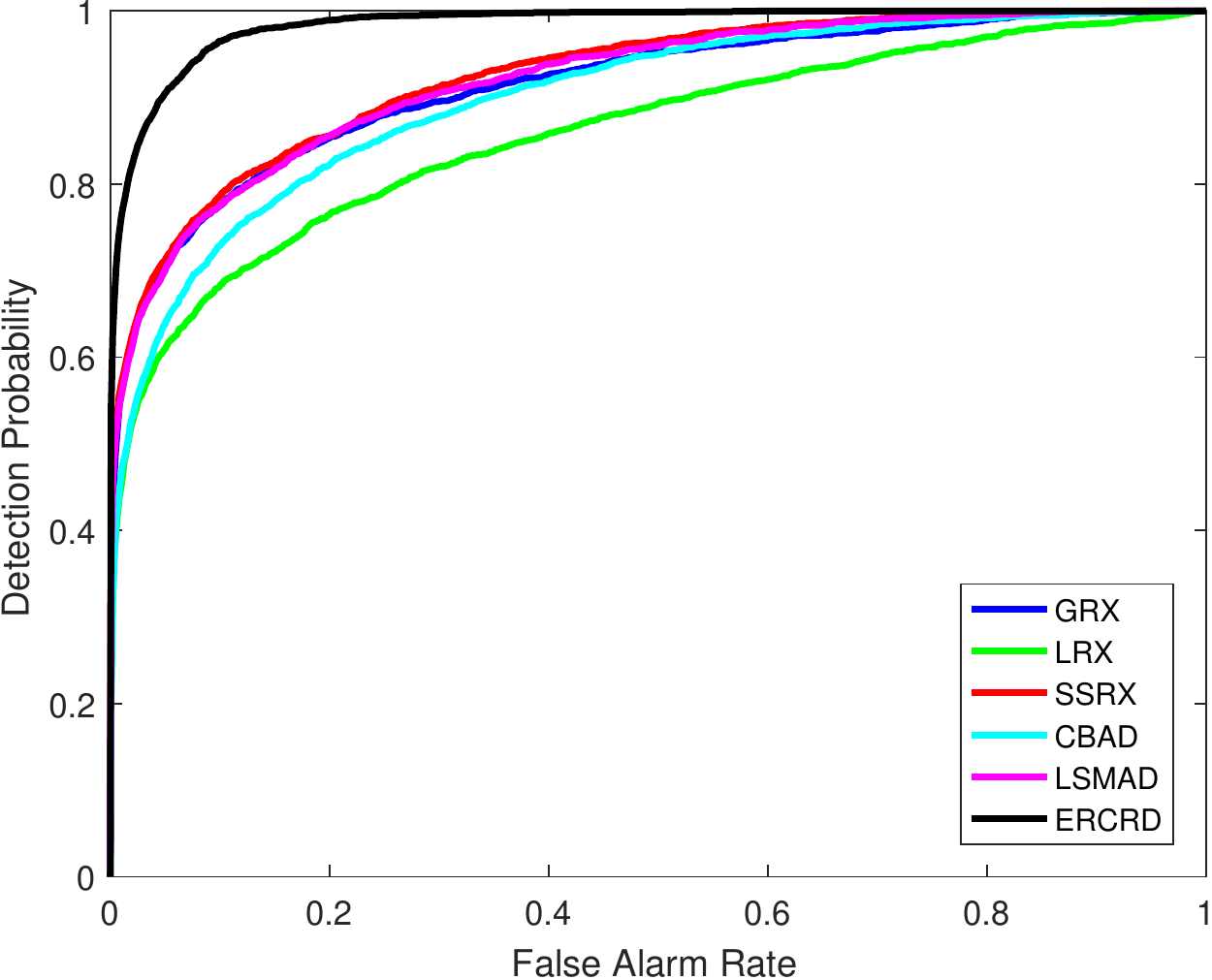}}~
    \subfloat[]{\includegraphics[scale=0.45]{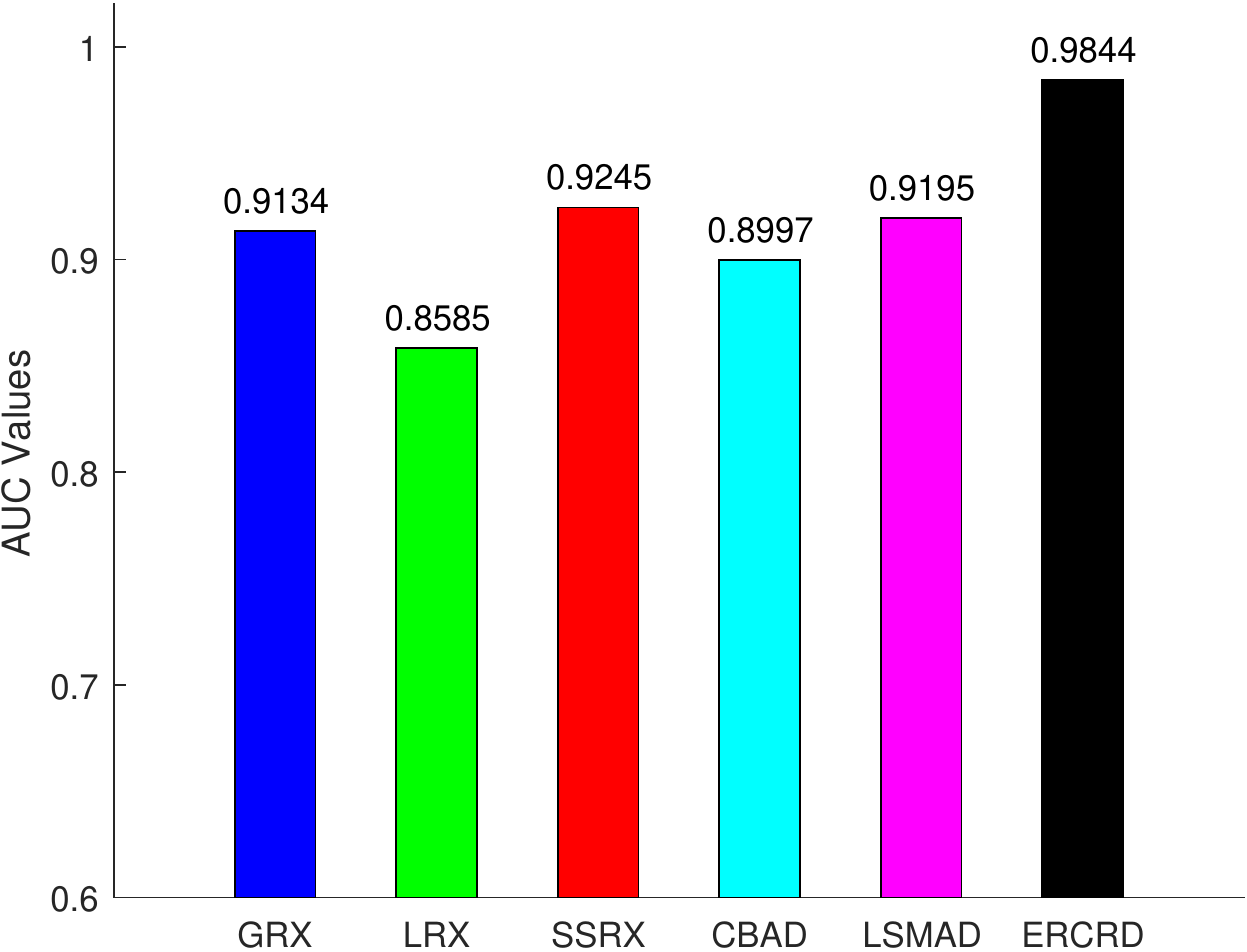}}~
    \subfloat[]{\includegraphics[scale=0.45]{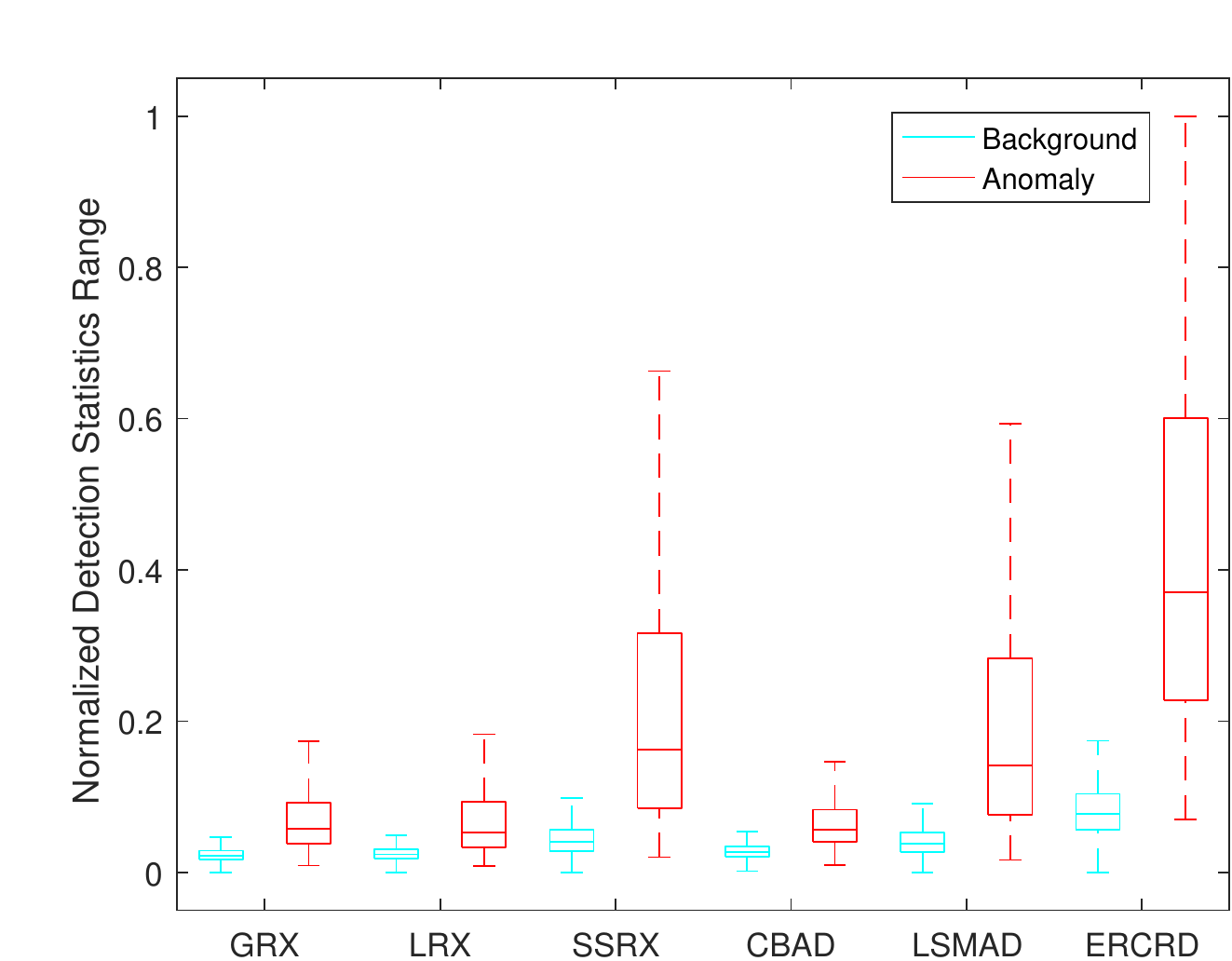}}
    \caption{I is detection accuracy evaluation for AVIRIS-I dataset. II is detection accuracy evaluation for AVIRIS-II dataset.
    III is detection accuracy evaluation for AVIRIS-III dataset. IV is detection accuracy evaluation for Cri dataset.
    (a) ROC curves. (b) AUC values. (c) Normalized background-anomaly separation map.}
    \label{Fig.6}
\end{figure*}
 
\subsection{Detection Performance}
In this subsection, we carry out two experiments to verify the performance of our ERCRD method. Firstly, our ERCRD method is compared with five state-of-the-art methods. Since the proposed ERCRD method is a variant of the CRD, the CRD and four representative variants of the CRD are subsequently compared. The number of random sub-sampling $r$ and the value of ensemble size $T$ are set to be $10$ and $20$, respectively.

In the first experiment, in order to evaluate detection performance, we make a comparison between the proposed ERCRD method and five classic and benchmark methods: GRX~\cite{Reed1990}, LRX~\cite{Reed1990}, SSRX~\cite{Schaum2007}, CBAD~\cite{Carlotto2005}, and LSMAD~\cite{Zhang2016}. Among them, GRX is known as the benchmark anomaly detector for HSI. The LRX, SSRX, and CBAD are three representative improved versions of RX. LSMAD is a typical low-rank and sparse matrix decomposition-based detector with remarkable detection performance. We choose the inner window size $w_{in}$ ranging from $3$ to $23$ and the outer window size $w_{out}$ ranging from $5$ to $25$ for the reason that the detection performance of LRX is sensitive to them. Moreover, we set the parameters of SSRX, CBAD, and LSMAD to be accordant with earlier work~\cite{Schaum2007,Carlotto2005,Zhang2016}.

The color detection maps of different methods based on the AVIRIS-I dataset, AVIRIS-II dataset, AVIRIS-III dataset, and Cri dataset are presented in Fig.~\ref{Fig.5}.
As for the AVIRIS-I dataset, our ERCRD method is able to identify the locations of three airplanes but fails to precisely picture their shapes. The GRX, LRX, SSRX, CBAD, and LRaSMD methods not only fail to detect the anomalies but also misidentify several normal background pixels as anomalies. Moreover, the ROC curves, the corresponding AUC values, and the normalized background-anomaly separation maps are displayed in Fig.~\ref{Fig.6}. It can be observed that the curve of the ERCRD method is closer to the upper left corner than the others and its AUC value is $0.9870$, which is larger than others. We can see that the separation gap for the proposed ERCRD method is larger than those for the other methods. This indicates that the ERCRD method achieves the best separation result. Moreover, the LSMAD, SSRX and CBAD methods obtain relatively better separation capacity, while the GRX and LRX methods perform unsatisfactorily separation capacity.

As for the AVIRIS-II dataset, the LRX and CBAD methods still can not separate the anomalies from the background and even misidentify some normal pixels as anomalies. The GRX, SSRX, LSMAD, and ERCRD methods can identify the locations of three airplanes, but their shapes are fuzzy and some false anomalies are also detected. The ROC curves, corresponding AUC values and normalized background-anomaly separation maps of different methods are given in Fig.~\ref{Fig.6}. It can be concluded that the ROC curve of the ERCRD method is closer to the upper left corner than the others, and its AUC value is $0.9793$, which is also larger than others. Here, the proposed ERCRD method still achieves larger separation gaps while the separation capabilities of the GRX, SSRX, CBAD, and LSMAD methods are slightly poorer. Compared with the above methods, the LRX method performs relatively unsatisfactorily.

As for the AVIRIS-III dataset, the proposed ERCRD method is able to identify the locations of six airplanes but some anomalous pixels are missing and several normal pixels are misidentified. Unfortunately, other methods cannot detect anomalies effectively. In Fig.~\ref{Fig.6.IIIa}, the ROC curves indicate that the proposed ERCRD method obtains a higher detection probability than others. The AUC values of all methods are illustrated in Fig.~\ref{Fig.6.IIIb}; these values indicate that the proposed ERCRD method can achieve the best detection results among all the compared methods. Fig.~\ref{Fig.6.IIIc} presents the separation maps for this dataset. Here, the proposed ERCRD method still achieves larger separation gaps. Moreover, the separation abilities of the other methods are greatly poorer, since their separation gaps are narrower than that of the ERCRD method.

As for the Cri dataset, the SSRX, LSMAD, and proposed ERCRD methods can effectively detect the locations and clear shapes of ten rocks, while the GRX, LRX, and CBAD methods can only detect the positions of several anomalous pixels but the shapes of some are missing. It can be seen from Fig.~\ref{Fig.5} that the proposed ERCRD method performs high detection abilities with a low false alarm rate, high AUC value and achieves larger separation gaps than the other methods.

In addition, the running times for the four datasets are displayed in Table~\ref{Table.1}. It is noteworthy that the running time of the proposed ERCRD and CBAD methods are similar to that of the GRX and SSRX methods; meanwhile, ERCRD method also achieves excellent detection performance. Thus, ERCRD method with lower computational complexity and high efficiency. Furthermore, the LRX and LSMAD methods are more time-consuming than the others.

\begin{table}[!htb]
  \centering
  \caption{Running time (seconds)}
  \begin{tabular}{p{13mm}<{\centering}p{5mm}<{\centering}p{8mm}<{\centering}p{7mm}<{\centering}p{7mm}<{\centering}p{8mm}<{\centering}c}
    \hline\hline
    Dataset & GRX & LRX & SSRX & CBAD & LSMAD & ERCRD \\ \hline\hline
    AVIRIS-I & $0.27$ & $140.75$ & $0.21$ & $0.41$ & $20.28$ & $0.77$ \\
    AVIRIS-II & $0.15$ & $98.04$ & $0.14$ & $0.23$ & $14.50$ & $0.79$ \\
    AVIRIS-III & $0.52$ & $443.55$ & $0.67$ & $3.56$ & $73.99$ & $2.45$ \\
    Cri & $0.79$ & $192.86$ & $0.98$ & $2.76$ & $55.01$ & $1.82$ \\
    \hline \hline
  \end{tabular}
  \label{Table.1}
\end{table}

In the second experiment, the detection performance of the ERCRD method is assessed and compared with the CRD~\cite{Li2015b} and four state-of-the-art variants of CRD: Global PCAroCRD proposed in 2018~\cite{Su2018} , Local PCAroCRD in 2018~\cite{Su2018}, MCRD in 2018~\cite{Imani2018}, and RCRD in 2019~\cite{Ma2019}. It can be seen that the detection performance of the CRD, Local PCAroCRD, MCRD, and RCRD methods are sensitive to the inner window size $w_{in}$ and the outer window size $w_{out}$. Thus, we employed four window sizes: $(5,9)$, $(7,11)$, $(9,13)$ and $(11,15)$. The regularization parameter $\lambda$ of these six methods is set to $10^{-6}$. The AUC values and the corresponding running times of these six methods are displayed in Table~\ref{Table.2} and Table~\ref{Table.3}, respectively.

\begin{table}[!htb]
  \centering
  \caption{AUC values}
  \begin{tabular}{c|p{8mm}<{\centering}|p{11mm}<{\centering}p{12mm}<{\centering}p{13mm}<{\centering}p{7mm}<{\centering}}
  \hline\hline
  \multicolumn{2}{c}{Dataset}  & AVIRIS-I & AVIRIS-II & AVIRIS-III & Cri \\ \hline \hline
  \multirow{4}{*}{CRD}
            & $(5,9)$   & $0.7116$ & $0.9025$ & $0.8393$ & $0.8888$ \\ \cline{2-6}
            & $(7,11)$  & $0.7835$ & $0.8970$ & $0.8880$ & $0.9355$ \\ \cline{2-6}
            & $(9,13)$  & $0.8842$ & $0.9035$ & $0.9388$ & $0.9539$\\ \cline{2-6}
            & $(11,15)$ & $0.9493$ & $0.9179$ & $0.9659$ & $0.9619$ \\ \hline
  \multirow{4}{12mm}{~~~~Local PCAroCRD}
            & $(5,9)$   & $0.8773$ & $0.9517$ & $0.8790$ & $0.6106$ \\ \cline{2-6}
            & $(7,11)$  & $0.9033$ & $0.9455$ & $0.9115$ & $0.6416$ \\ \cline{2-6}
            & $(9,13)$  & $0.9146$ & $0.9468$ & $0.9318$ & $0.6726$ \\ \cline{2-6}
            & $(11,15)$ & $0.9515$ & $0.9590$ & $0.9500$ & $0.6990$ \\ \hline
  \multirow{4}{*}{MCRD}
            & $(5,9)$   & $0.8460$ & $0.9068$ & $0.8206$ & $0.5561$ \\ \cline{2-6}
            & $(7,11)$  & $0.9138$ & $0.9162$ & $0.8964$ & $0.5463$ \\ \cline{2-6}
            & $(9,13)$  & $0.9603$ & $0.9262$ & $0.9415$ & $0.5393$ \\ \cline{2-6}
            & $(11,15)$ & $0.9904$ & $0.9495$ & $0.9696$ & $0.5396$ \\ \hline
  \multirow{4}{*}{RCRD}
            & $(5,9)$   & $0.7030$ & $0.9016$ & $0.8324$ & $0.6361$ \\ \cline{2-6}
            & $(7,11)$  & $0.7846$ & $0.8963$ & $0.8824$ & $0.6261$ \\ \cline{2-6}
            & $(9,13)$  & $0.8804$ & $0.9051$ & $0.9341$ & $0.6113$ \\ \cline{2-6}
            & $(11,15)$ & $0.9446$ & $0.9197$ & $0.9588$ & $0.6322$ \\ \hline
  \multicolumn{2}{c|}{Global PCAroCRD} & $0.9403$ & $0.9259$ & $0.9180$ & $0.8183$ \\ \hline
  \multicolumn{2}{c|}{ERCRD} & $0.9870$ & $0.9793$ & $0.9385$ & $0.9844$ \\ \hline \hline
  \end{tabular}
  \label{Table.2}
\end{table}

\begin{table}[!htb]
  \centering
  \caption{Running time (seconds)}
  \begin{tabular}{c|p{8mm}<{\centering}|p{11mm}<{\centering}p{12mm}<{\centering}p{13mm}<{\centering}p{8mm}<{\centering}}
  \hline\hline
  \multicolumn{2}{c}{Dataset} & AVIRIS-I & AVIRIS-II & AVIRIS-III & Cri \\ \hline \hline
  \multirow{4}{*}{CRD}
            & $(5,9)$  & $19.31$ & $13.11$ & $63.92$ & $186.66$ \\ \cline{2-6}
            & $(7,11)$  & $25.12$ & $18.53$ & $88.05$ & $264.28$ \\ \cline{2-6}
            & $(9,13)$  & $33.21$ & $23.43$ & $119.41$ & $360.93$ \\ \cline{2-6}
            & $(11,15)$  & $44.66$ & $31.01$ & $153.33$ & $434.43$ \\ \hline
  \multirow{4}{12mm}{~~~~Local PCAroCRD}
            & $(5,9)$  & $11.39$ & $7.72$ & $39.79$ & $101.31$ \\ \cline{2-6}
            & $(7,11)$  & $15.61$ & $10.48$ & $52.32$ & $131.17$ \\ \cline{2-6}
            & $(9,13)$  & $20.97$ & $13.89$ & $70.89$ & $186.62$ \\ \cline{2-6}
            & $(11,15)$  & $26.82$ & $18.12$ & $91.24$ & $214.92$ \\ \hline
  \multirow{4}{*}{MCRD}
            & $(5,9)$  & $58.45$ & $52.14$ & $249.23$ & $816.61$ \\ \cline{2-6}
            & $(7,11)$  & $82.64$ & $72.11$ & $311.20$ & $1195.25$ \\ \cline{2-6}
            & $(9,13)$  & $115.99$ & $91.41$ & $463.19$ & $1667.27$ \\ \cline{2-6}
            & $(11,15)$  & $167.34$ & $114.62$ & $568.85$ & $2051.54$ \\ \hline
  \multirow{4}{*}{RCRD}
            & $(5,9)$  & $12.55$ & $8.43$ & $42.65$ & $99.89$ \\ \cline{2-6}
            & $(7,11)$  & $15.88$ & $10.56$ & $51.18$ & $123.59$ \\ \cline{2-6}
            & $(9,13)$  & $17.47$ & $11.65$ & $56.84$ & $145.64$ \\ \cline{2-6}
            & $(11,15)$  & $23.72$ & $16.15$ & $79.23$ & $224.16$ \\ \hline
  \multicolumn{2}{c|}{Global PCAroCRD} & $52.90$ & $37.37$ & $115.89$ & $809.39$ \\ \hline
  \multicolumn{2}{c|}{ERCRD} & $0.77$ & $0.79$ & $2.45$ & $1.82$ \\ \hline \hline
  \end{tabular}
  \label{Table.3}
\end{table}

In AVIRIS-I, the AUC value of the ERCRD is $0.9870$, only smaller than that of MCRD with window sizes $(11,15)$ and larger than that of the others. The running time of ERCRD is $0.77$s, which is much lower than the others. The running time of MCRD with window sizes $(11,15)$ is $167.34$s, which is much higher than that of the ERCRD. In AVIRIS-II, the AUC value of ERCRD is $0.9793$, higher than the others. The running time of ERCRD is $0.79$s, which is much lower than that of the others. In AVIRIS-III, the AUC value of ERCRD is $0.9385$, smaller than that of CRD with window sizes $(9,13)$ and $(11,15)$, Local PCAroCRD with window sizes $(11,15)$, MCRD with window sizes $(9,13)$ and $(11,15)$, RCRD with window sizes $(11,15)$. Note that the running time of the ERCRD is $2.45$s, which is much lower than others. In Cri dataset, the AUC value of the ERCRD is $0.9844$, higher than the others. The running time of ERCRD is $1.82$s, which is much lower than the others.

\begin{figure}[!htb]
  \centering
  \subfloat[]{\includegraphics[scale=0.5]{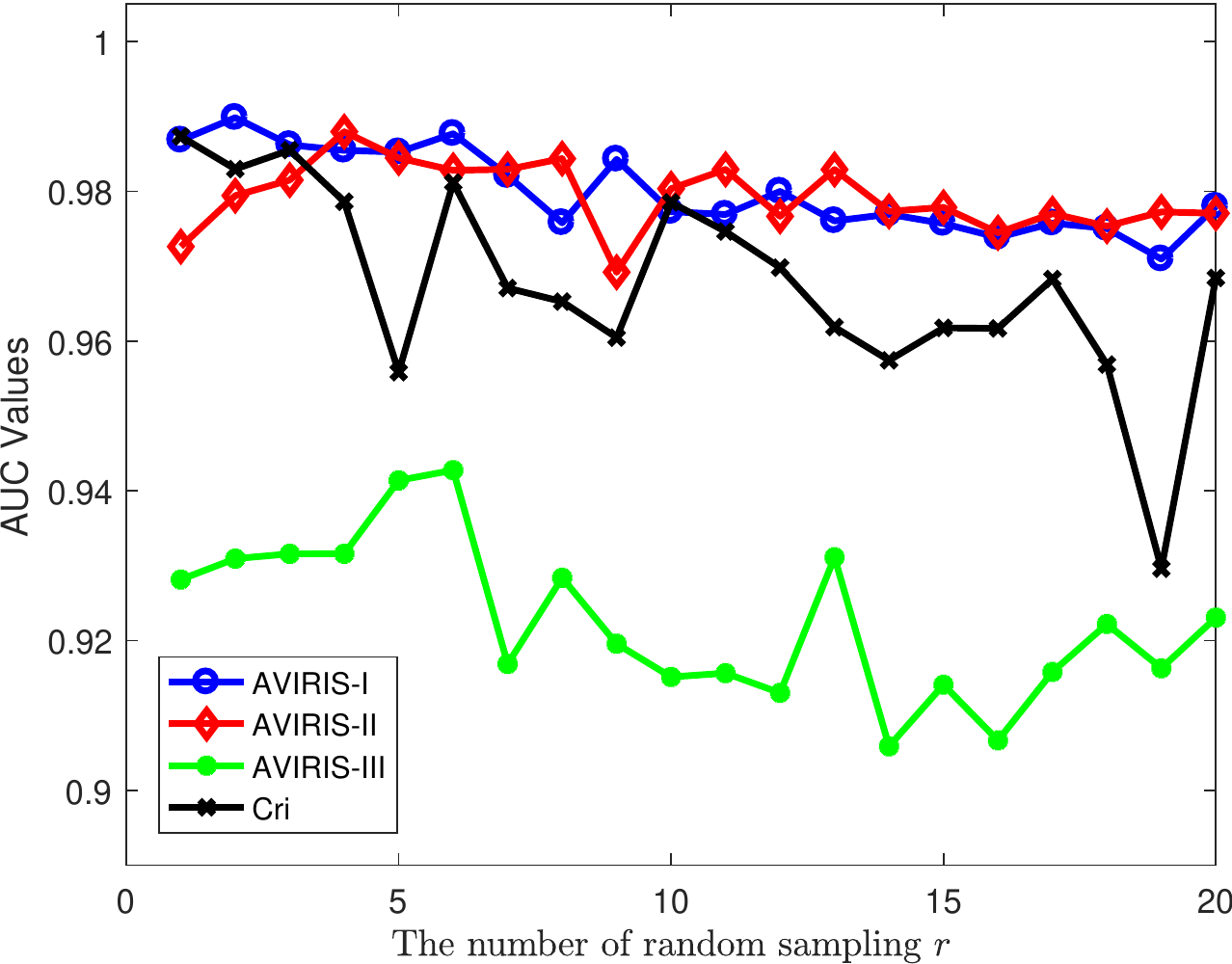}\label{Fig.7a}}\\
  \subfloat[]{\includegraphics[scale=0.5]{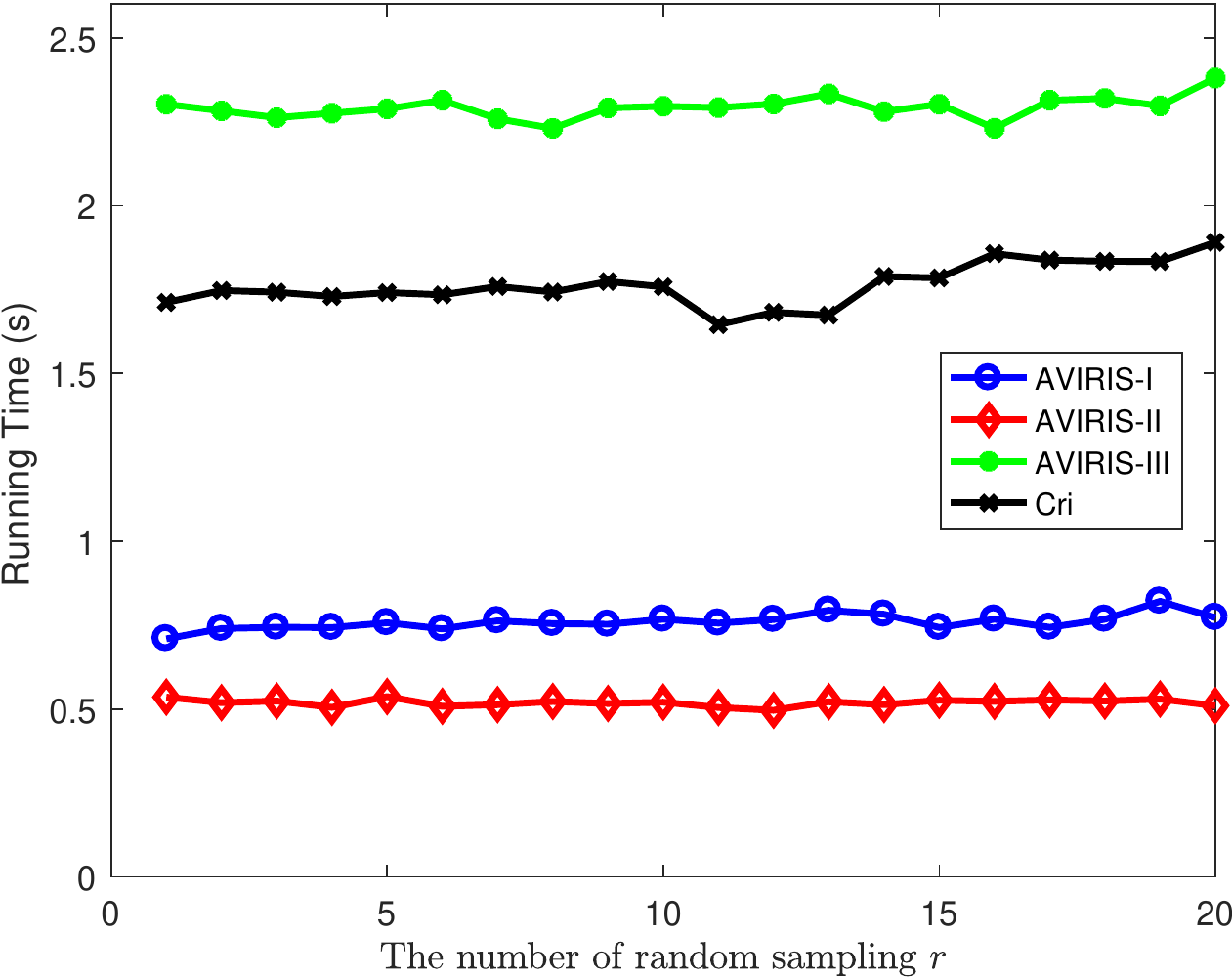}\label{Fig.7b}}
  \caption{Effect of the number of random sub-sampling $r$ on each dataset. (a) AUC values. (b) Running time.}
  \label{Fig.7}
\end{figure}
  
\begin{figure}[!htb]
  \centering
  \subfloat[]{\includegraphics[scale=0.5]{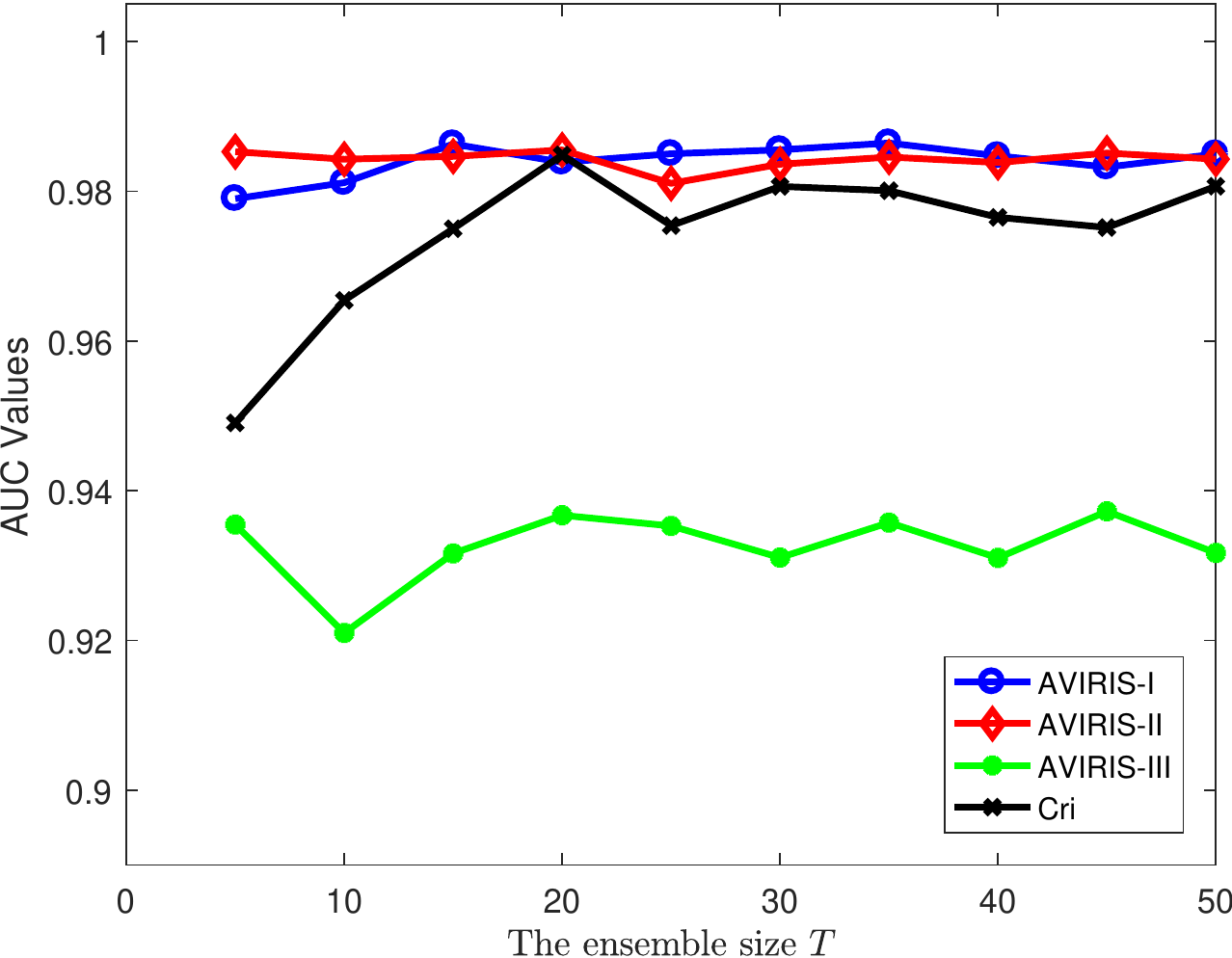}\label{Fig.8a}}\\
  \subfloat[]{\includegraphics[scale=0.5]{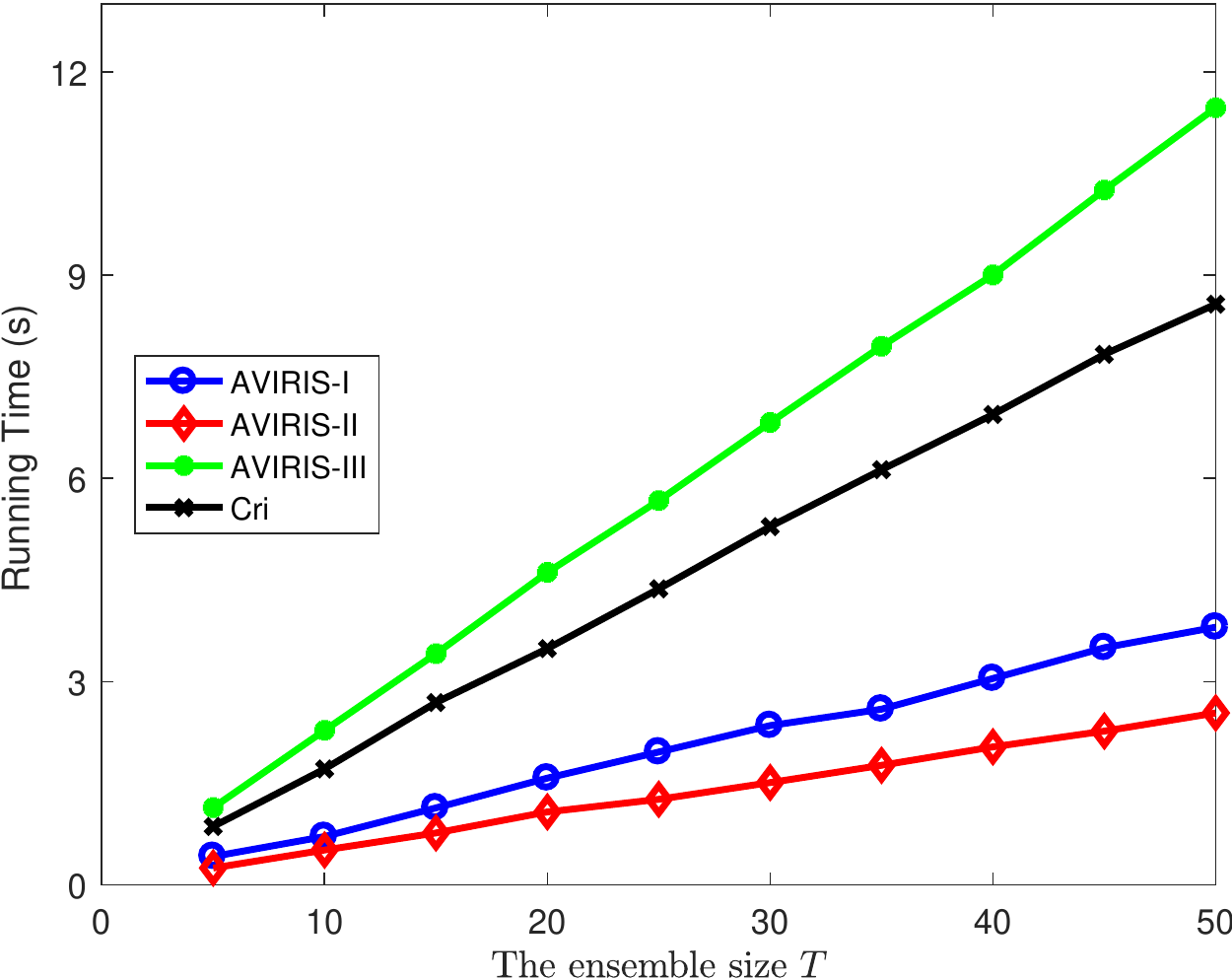}\label{Fig.8b}}
  \caption{Effect of the ensemble size $T$ on each dataset. (a) AUC values. (b) Running time.}
  \label{Fig.8}
\end{figure}

\subsection{Parameter Analysis and Discussion}
There are two parameters in the proposed ERCRD method: the number of random sub-sampling $r$ and the ensemble size $T$. We will conduct parameter analysis by comparing the AUC value and running time on four datasets.

Fig.~\ref{Fig.7} presents the impact of different random sub-sampling numbers $r$ on the detection performance and running time for each dataset. The value of the random sub-sampling number $r$ is varied in the range $[1,2,\cdots,20]$ and the ensemble size $T$ is set to $10$. It can be seen from Fig.~\ref{Fig.7a} that the AUC values on AVIRIS-I and AVIRIS-II datasets fluctuate in a small range from $0.97$ to $0.99$, while on the AVIRIS-III dataset it fluctuates in a slightly larger range from $0.91$ to $0.95$, and within a relatively larger range from $0.93$ to $0.99$ in Cri dataset. It is known from Fig.~\ref{Fig.7b} that the running time on each dataset is nearly stable while the random sub-sampling number $r$ increases.

Then the impact of different ensemble size $T$ on the four datasets is analyzed. The value of ensemble size $T$ is changed within the range of $[5,10,\cdots,50]$ and the number of random sampling $r$ is set to $5$. In Fig.~\ref{Fig.8a}, we see that the AUC values on AVIRIS-I and AVIRIS-II datasets are nearly stable, while on AVIRIS-III and Cri datasets are increasing at first, and then fluctuate on a small scale. Fig.~\ref{Fig.8b} illustrates that with the ensemble size $T$ increasing, the running time of each dataset increases almost linearly. 

\section{Conclusion}\label{Sec.5}
Hyperspectral anomaly detection (HAD) has attracted more and more interest and in-depth research. More recently, the collaboration representation-based detector (CRD) has become an active topic for HAD methods. Although CRD achieved good performance, its computational cost mainly arising from the sliding dual window strategy is too high. Besides, it takes multiple repeated tests to determine the size of the dual window, which needs to be reset once the dataset changes and cannot be identified in advance. In this paper, we propose a novel ensemble and random collaborative representation-based detector (ERCRD) for HAD, which comprises two closely related stages: 1) we process a random sub-sampling on CRD (RCRD) to obtain multiple detection results without using sliding dual window, which reduces the computational complexity and makes it more feasible in practice; 2) we adopt ensemble learning to refine the multiple results of RCRD, which act as various ``experts" providing abundant complementary information to better target different anomalies. These two stages perfectly form an organic and theoretical detector, which can not only improve the accuracy and stability of HAD methods but also enhance its generalization ability. Experimental results on four real hyperspectral datasets validate that our method outperforms its counterparts in the aspect of detection accuracy and running time.

\bibliographystyle{IEEEtran}
\bibliography{IEEEabrv,refs}

\begin{thebibliography}{10}
\providecommand{\url}[1]{#1}
\csname url@samestyle\endcsname
\providecommand{\newblock}{\relax}
\providecommand{\bibinfo}[2]{#2}
\providecommand{\BIBentrySTDinterwordspacing}{\spaceskip=0pt\relax}
\providecommand{\BIBentryALTinterwordstretchfactor}{4}
\providecommand{\BIBentryALTinterwordspacing}{\spaceskip=\fontdimen2\font plus
\BIBentryALTinterwordstretchfactor\fontdimen3\font minus
  \fontdimen4\font\relax}
\providecommand{\BIBforeignlanguage}[2]{{%
\expandafter\ifx\csname l@#1\endcsname\relax
\typeout{** WARNING: IEEEtran.bst: No hyphenation pattern has been}%
\typeout{** loaded for the language `#1'. Using the pattern for}%
\typeout{** the default language instead.}%
\else
\language=\csname l@#1\endcsname
\fi
#2}}
\providecommand{\BIBdecl}{\relax}
\BIBdecl

\bibitem{Shaw2002}
G.~{Shaw} and D.~{Manolakis}, ``Signal processing for hyperspectral image
  exploitation,'' \emph{IEEE Signal Process. Mag.}, vol.~19, no.~1, pp. 12--16,
  Jan. 2002.

\bibitem{Stein2002}
D.~W.~J. {Stein}, S.~G. {Beaven}, L.~E. {Hoff}, E.~M. {Winter}, A.~P. {Schaum},
  and A.~D. {Stocker}, ``Anomaly detection from hyperspectral imagery,''
  \emph{IEEE Signal Process. Mag.}, vol.~19, no.~1, pp. 58--69, Jan. 2002.

\bibitem{Chang2002}
C.-I. Chang and S.-S. Chiang, ``Anomaly detection and classification for
  hyperspectral imagery,'' \emph{IEEE Trans. Geosci. Remote Sens.}, vol.~40,
  no.~6, pp. 1314--1325, Jun. 2002.

\bibitem{Zhang2018a}
L.~{Zhang}, Q.~{Zhang}, B.~{Du}, X.~{Huang}, Y.~Y. {Tang}, and D.~{Tao},
  ``Simultaneous spectral-spatial feature selection and extraction for
  hyperspectral images,'' \emph{IEEE Trans. Cybern.}, vol.~48, no.~1, pp.
  16--28, 2018.

\bibitem{Luo2019}
F.~{Luo}, B.~{Du}, L.~{Zhang}, L.~{Zhang}, and D.~{Tao}, ``Feature learning
  using spatial-spectral hypergraph discriminant analysis for hyperspectral
  image,'' \emph{IEEE Trans. Cybern.}, vol.~49, no.~7, pp. 2406--2419, 2019.

\bibitem{Bioucas-Dias2012}
J.~M. {Bioucas-Dias}, A.~{Plaza}, N.~{Dobigeon}, M.~{Parente}, Q.~{Du},
  P.~{Gader}, and J.~{Chanussot}, ``Hyperspectral unmixing overview:
  Geometrical, statistical, and sparse regression-based approaches,''
  \emph{IEEE J. Sel. Topics Appl. Earth Observ. Remote Sens.}, vol.~5, no.~2,
  pp. 354--379, Apr. 2012.

\bibitem{Wang2017a}
R.~Wang, F.~Nie, and W.~Yu, ``Fast spectral clustering with anchor graph for
  large hyperspectral images,'' \emph{IEEE Geosci. Remote Sens. Lett.},
  vol.~14, no.~11, pp. 2003--2007, Nov. 2017.

\bibitem{Wang2019}
R.~{Wang}, F.~{Nie}, Z.~{Wang}, F.~{He}, and X.~{Li}, ``Scalable graph-based
  clustering with nonnegative relaxation for large hyperspectral image,''
  \emph{IEEE Trans. Geosci. Remote Sens.}, vol.~57, no.~10, pp. 7352--7364,
  Oct. 2019.

\bibitem{Liu2019}
S.~{Liu}, D.~{Marinelli}, L.~{Bruzzone}, and F.~{Bovolo}, ``A review of change
  detection in multitemporal hyperspectral images: Current techniques,
  applications, and challenges,'' \emph{IEEE Geosci. Remote Sens. Mag.},
  vol.~7, no.~2, pp. 140--158, Jun. 2019.

\bibitem{Yuan2016}
Y.~{Yuan}, D.~{Ma}, and Q.~{Wang}, ``Hyperspectral anomaly detection by graph
  pixel selection,'' \emph{IEEE Trans. Cybern.}, vol.~46, no.~12, pp.
  3123--3134, 2016.

\bibitem{Li2021}
L.~Li, W.~Li, Q.~Du, and R.~Tao, ``Low-rank and sparse decomposition with
  mixture of gaussian for hyperspectral anomaly detection,'' \emph{IEEE
  Transactions on Cybernetics}, vol.~51, no.~9, pp. 4363--4372, 2021.

\bibitem{Nasrabadi2014}
N.~M. {Nasrabadi}, ``Hyperspectral target detection: An overview of current and
  future challenges,'' \emph{IEEE Signal Process. Mag.}, vol.~31, no.~1, pp.
  34--44, Jan. 2014.

\bibitem{Huber-Lerner2016}
M.~{Huber-Lerner}, O.~{Hadar}, S.~R. {Rotman}, and R.~{Huber-Shalem},
  ``Hyperspectral band selection for anomaly detection: The role of data
  gaussianity,'' \emph{IEEE J. Sel. Topics Appl. Earth Observ. Remote Sens.},
  vol.~9, no.~2, pp. 732--743, Feb. 2016.

\bibitem{Wang2017}
L.~{Wang}, C.~{Chang}, L.~{Lee}, Y.~{Wang}, B.~{Xue}, M.~{Song}, C.~{Yu}, and
  S.~{Li}, ``Band subset selection for anomaly detection in hyperspectral
  imagery,'' \emph{IEEE Trans. Geosci. Remote Sens.}, vol.~55, no.~9, pp.
  4887--4898, Sep. 2017.

\bibitem{Banerjee2006}
A.~{Banerjee}, P.~{Burlina}, and C.~{Diehl}, ``A support vector method for
  anomaly detection in hyperspectral imagery,'' \emph{IEEE Trans. Geosci.
  Remote Sens.}, vol.~44, no.~8, pp. 2282--2291, Aug. 2006.

\bibitem{Sakla2011}
W.~{Sakla}, A.~{Chan}, J.~{Ji}, and A.~{Sakla}, ``An {SVDD}-based algorithm for
  target detection in hyperspectral imagery,'' \emph{IEEE Geosci. Remote Sens.
  Lett.}, vol.~8, no.~2, pp. 384--388, Mar. 2011.

\bibitem{Kang2017}
X.~{Kang}, X.~{Zhang}, S.~{Li}, K.~{Li}, J.~{Li}, and J.~A. {Benediktsson},
  ``Hyperspectral anomaly detection with attribute and edge-preserving
  filters,'' \emph{IEEE Trans. Geosci. Remote Sens.}, vol.~55, no.~10, pp.
  5600--5611, Oct. 2017.

\bibitem{Li2018}
S.~{Li}, K.~{Zhang}, Q.~{Hao}, P.~{Duan}, and X.~{Kang}, ``Hyperspectral
  anomaly detection with multiscale attribute and edge-preserving filters,''
  \emph{IEEE Geosci. Remote Sens. Lett.}, vol.~15, no.~10, pp. 1605--1609, Oct.
  2018.

\bibitem{Taghipour2017}
A.~{Taghipour} and H.~{Ghassemian}, ``Hyperspectral anomaly detection using
  attribute profiles,'' \emph{IEEE Geosci. Remote Sens. Lett.}, vol.~14, no.~7,
  pp. 1136--1140, Jul. 2017.

\bibitem{Zhang2016a}
X.~{Zhang}, G.~{Wen}, and W.~{Dai}, ``A tensor decomposition-based anomaly
  detection algorithm for hyperspectral image,'' \emph{IEEE Trans. Geosci.
  Remote Sens.}, vol.~54, no.~10, pp. 5801--5820, Oct. 2016.

\bibitem{Xu2018}
Y.~{Xu}, Z.~{Wu}, J.~{Chanussot}, and Z.~{Wei}, ``Joint reconstruction and
  anomaly detection from compressive hyperspectral images using mahalanobis
  distance-regularized tensor rpca,'' \emph{IEEE Trans. Geosci. Remote Sens.},
  vol.~56, no.~5, pp. 2919--2930, May. 2018.

\bibitem{Zhang2018}
X.~{Zhang} and G.~{Wen}, ``A fast and adaptive method for determining $k_{1}$ ,
  $k_{2}$ , and $k_{3}$ in the tensor decomposition-based anomaly detection
  algorithm,'' \emph{IEEE Geosci. Remote Sens. Lett.}, vol.~15, no.~1, pp.
  3--7, Jan. 2018.

\bibitem{Xie2019}
W.~{Xie}, T.~{Jiang}, Y.~{Li}, X.~{Jia}, and J.~{Lei}, ``Structure tensor and
  guided filtering-based algorithm for hyperspectral anomaly detection,''
  \emph{IEEE Trans. Geosci. Remote Sens.}, vol.~57, no.~7, pp. 4218--4230, Jul.
  2019.

\bibitem{Li2017}
W.~{Li}, G.~{Wu}, and Q.~{Du}, ``Transferred deep learning for anomaly
  detection in hyperspectral imagery,'' \emph{IEEE Geosci. Remote Sens. Lett.},
  vol.~14, no.~5, pp. 597--601, May. 2017.

\bibitem{Zhao2017b}
C.~Zhao, X.~Li, and H.~Zhu, ``Hyperspectral anomaly detection based on stacked
  denoising autoencoders,'' \emph{J. Appl. Remote Sens.}, vol.~11, no.~4, p.
  042605, 2017.

\bibitem{Ma2018}
N.~Ma, Y.~Peng, S.~Wang, and L.~Phw, ``An unsupervised deep hyperspectral
  anomaly detector,'' \emph{Sensors}, vol.~18, no.~3, p. 693, 2018.

\bibitem{Reed1990}
I.~S. {Reed} and X.~{Yu}, ``Adaptive multiple-band {CFAR} detection of an
  optical pattern with unknown spectral distribution,'' \emph{IEEE Trans.
  Acoust., Speech, Signal Process.}, vol.~38, no.~10, pp. 1760--1770, Oct.
  1990.

\bibitem{Kwon2005}
H.~{Kwon} and N.~M. {Nasrabadi}, ``Kernel {RX}-algorithm: A nonlinear anomaly
  detector for hyperspectral imagery,'' \emph{IEEE Trans. Geosci. Remote
  Sens.}, vol.~43, no.~2, pp. 388--397, Feb. 2005.

\bibitem{Carlotto2005}
M.~J. {Carlotto}, ``A cluster-based approach for detecting man-made objects and
  changes in imagery,'' \emph{IEEE Trans. Geosci. Remote Sens.}, vol.~43,
  no.~2, pp. 374--387, Feb. 2005.

\bibitem{Schaum2007}
A.~P. Schaum, ``Hyperspectral anomaly detection beyond {RX},'' in \emph{Proc.
  SPIE}, vol. 6565, May 2007.

\bibitem{Molero2013}
J.~M. {Molero}, E.~M. {Garzn}, I.~{Garca}, and A.~{Plaza}, ``Analysis and
  optimizations of global and local versions of the {RX} algorithm for anomaly
  detection in hyperspectral data,'' \emph{IEEE J. Sel. Topics Appl. Earth
  Observ. Remote Sens.}, vol.~6, no.~2, pp. 801--814, Apr. 2013.

\bibitem{Liu2013}
W.~{Liu} and C.~{Chang}, ``Multiple-window anomaly detection for hyperspectral
  imagery,'' \emph{IEEE J. Sel. Topics Appl. Earth Observ. Remote Sens.},
  vol.~6, no.~2, pp. 644--658, Apr. 2013.

\bibitem{Guo2014}
Q.~{Guo}, B.~{Zhang}, Q.~{Ran}, L.~{Gao}, J.~{Li}, and A.~{Plaza},
  ``Weighted-{RXD} and linear filter-based {RXD}: Improving background
  statistics estimation for anomaly detection in hyperspectral imagery,''
  \emph{IEEE J. Sel. Topics Appl. Earth Observ. Remote Sens.}, vol.~7, no.~6,
  pp. 2351--2366, Jun. 2014.

\bibitem{Zhou2016}
J.~{Zhou}, C.~{Kwan}, B.~{Ayhan}, and M.~T. {Eismann}, ``A novel cluster kernel
  {RX} algorithm for anomaly and change detection using hyperspectral images,''
  \emph{IEEE Trans. Geosci. Remote Sens.}, vol.~54, no.~11, pp. 6497--6504,
  Nov. 2016.

\bibitem{Chang2018}
S.~{Chang}, B.~{Du}, and L.~{Zhang}, ``{BASO}: A background-anomaly component
  projection and separation optimized filter for anomaly detection in
  hyperspectral images,'' \emph{IEEE Trans. Geosci. Remote Sens.}, vol.~56,
  no.~7, pp. 3747--3761, Jul. 2018.

\bibitem{Imani2017}
M.~{Imani}, ``{RX} anomaly detector with rectified background,'' \emph{IEEE
  Geosci. Remote Sens. Lett.}, vol.~14, no.~8, pp. 1313--1317, Aug. 2017.

\bibitem{Zhao2017a}
C.~{Zhao}, X.~{Yao}, and Y.~{Yan}, ``Modified kernel {RX} algorithm based on
  background purification and inverse-of-matrix-free calculation,'' \emph{IEEE
  Geosci. Remote Sens. Lett.}, vol.~14, no.~4, pp. 544--548, Apr. 2017.

\bibitem{Zhao2018}
C.~{Zhao} and Y.~{Xi-Feng}, ``Fast real-time kernel {RX} algorithm based on
  cholesky decomposition,'' \emph{IEEE Geosci. Remote Sens. Lett.}, vol.~15,
  no.~11, pp. 1760--1764, Nov. 2018.

\bibitem{Chen2011}
Y.~{Chen}, N.~M. {Nasrabadi}, and T.~D. {Tran}, ``Sparse representation for
  target detection in hyperspectral imagery,'' \emph{IEEE J. Sel. Topics Signal
  Process.}, vol.~5, no.~3, pp. 629--640, Jun. 2011.

\bibitem{Li2015}
J.~{Li}, H.~{Zhang}, L.~{Zhang}, and L.~{Ma}, ``Hyperspectral anomaly detection
  by the use of background joint sparse representation,'' \emph{IEEE J. Sel.
  Topics Appl. Earth Observ. Remote Sens.}, vol.~8, no.~6, pp. 2523--2533, Jun.
  2015.

\bibitem{Zhang2017a}
Y.~{Zhang}, B.~{Du}, Y.~{Zhang}, and L.~{Zhang}, ``Spatially adaptive sparse
  representation for target detection in hyperspectral images,'' \emph{IEEE
  Geosci. Remote Sens. Lett.}, vol.~14, no.~11, pp. 1923--1927, Nov. 2017.

\bibitem{Zhao2017}
R.~{Zhao}, B.~{Du}, and L.~{Zhang}, ``Hyperspectral anomaly detection via a
  sparsity score estimation framework,'' \emph{IEEE Trans. Geosci. Remote
  Sens.}, vol.~55, no.~6, pp. 3208--3222, Jun. 2017.

\bibitem{Li2018a}
F.~{Li}, X.~{Zhang}, L.~{Zhang}, D.~{Jiang}, and Y.~{Zhang}, ``Exploiting
  structured sparsity for hyperspectral anomaly detection,'' \emph{IEEE Trans.
  Geosci. Remote Sens.}, vol.~56, no.~7, pp. 4050--4064, Jul. 2018.

\bibitem{Ling2019}
Q.~{Ling}, Y.~{Guo}, Z.~{Lin}, and W.~{An}, ``A constrained sparse
  representation model for hyperspectral anomaly detection,'' \emph{IEEE Trans.
  Geosci. Remote Sens.}, vol.~57, no.~4, pp. 2358--2371, Apr. 2019.

\bibitem{Sun2014}
W.~Sun, C.~Liu, J.~Li, Y.~M. Lai, and W.~Li, ``Low-rank and sparse matrix
  decomposition-based anomaly detection for hyperspectral imagery,'' \emph{J.
  Appl. Remote Sens.}, vol.~8, no.~1, pp. 1--18, 2014.

\bibitem{Zhang2016}
Y.~{Zhang}, B.~{Du}, L.~{Zhang}, and S.~{Wang}, ``A low-rank and sparse matrix
  decomposition-based mahalanobis distance method for hyperspectral anomaly
  detection,'' \emph{IEEE Trans. Geosci. Remote Sens.}, vol.~54, no.~3, pp.
  1376--1389, Mar. 2016.

\bibitem{Xu2016}
Y.~{Xu}, Z.~{Wu}, J.~{Li}, A.~{Plaza}, and Z.~{Wei}, ``Anomaly detection in
  hyperspectral images based on low-rank and sparse representation,''
  \emph{IEEE Trans. Geosci. Remote Sens.}, vol.~54, no.~4, pp. 1990--2000, Apr.
  2016.

\bibitem{Qu2018}
Y.~{Qu}, W.~{Wang}, R.~{Guo}, B.~{Ayhan}, C.~{Kwan}, S.~{Vance}, and H.~{Qi},
  ``Hyperspectral anomaly detection through spectral unmixing and
  dictionary-based low-rank decomposition,'' \emph{IEEE Trans. Geosci. Remote
  Sens.}, vol.~56, no.~8, pp. 4391--4405, Aug. 2018.

\bibitem{Madathil2019}
B.~{Madathil} and S.~N. {George}, ``Simultaneous reconstruction and anomaly
  detection of subsampled hyperspectral images using $l_{1/2}$ regularized
  joint sparse and low-rank recovery,'' \emph{IEEE Trans. Geosci. Remote
  Sens.}, vol.~57, no.~7, pp. 5190--5197, Jul. 2019.

\bibitem{Cheng2020}
T.~{Cheng} and B.~{Wang}, ``Graph and total variation regularized low-rank
  representation for hyperspectral anomaly detection,'' \emph{IEEE Trans.
  Geosci. Remote Sens.}, vol.~58, no.~1, pp. 391--406, Jan. 2020.

\bibitem{Li2015b}
W.~{Li} and Q.~{Du}, ``Collaborative representation for hyperspectral anomaly
  detection,'' \emph{IEEE Trans. Geosci. Remote Sens.}, vol.~53, no.~3, pp.
  1463--1474, Mar. 2015.

\bibitem{Imani2018}
M.~Imani, ``Anomaly detection using morphology-based collaborative
  representation in hyperspectral imagery,'' \emph{European Journal of Remote
  Sensing}, vol.~51, no.~1, pp. 457--471, 2018.

\bibitem{Vafadar2018}
M.~{Vafadar} and H.~{Ghassemian}, ``Anomaly detection of hyperspectral imagery
  using modified collaborative representation,'' \emph{IEEE Geosci. Remote
  Sens. Lett.}, vol.~15, no.~4, pp. 577--581, Apr. 2018.

\bibitem{Su2018}
H.~{Su}, Z.~{Wu}, Q.~{Du}, and P.~{Du}, ``Hyperspectral anomaly detection using
  collaborative representation with outlier removal,'' \emph{IEEE J. Sel.
  Topics Appl. Earth Observ. Remote Sens.}, vol.~11, no.~12, pp. 5029--5038,
  Dec. 2018.

\bibitem{Ma2019}
N.~{Ma}, Y.~{Peng}, and S.~{Wang}, ``A fast recursive collaboration
  representation anomaly detector for hyperspectral image,'' \emph{IEEE Geosci.
  Remote Sens. Lett.}, vol.~16, no.~4, pp. 588--592, Apr. 2019.

\bibitem{Liu2008}
F.~T. Liu, K.~M. Ting, and Z.-H. Zhou, ``Isolation forest,'' in \emph{Proc.
  ICDM}, 2008, pp. 413--422.

\bibitem{Liu2012}
------, ``Isolation-based anomaly detection,'' \emph{ACM Trans. Knowl. Discov.
  Data}, vol.~6, no.~1, pp. 3:1--3:39, Mar. 2012.

\bibitem{Li2020}
S.~{Li}, K.~{Zhang}, P.~{Duan}, and X.~{Kang}, ``Hyperspectral anomaly
  detection with kernel isolation forest,'' \emph{IEEE Trans. Geosci. Remote
  Sens.}, vol.~58, no.~1, pp. 319--329, Jan. 2020.

\bibitem{wang2020}
R.~Wang, F.~Nie, Z.~Wang, F.~He, and X.~Li, ``Multiple features and isolation
  forest-based fast anomaly detector for hyperspectral imagery,'' \emph{IEEE
  Trans. Geosci. Remote Sens.}, vol.~58, no.~9, pp. 6664--6676, 2020.

\bibitem{Chang2020}
S.~Chang, B.~Du, and L.~Zhang, ``A subspace selection-based discriminative
  forest method for hyperspectral anomaly detection,'' \emph{IEEE Trans.
  Geosci. Remote Sens.}, vol.~58, no.~6, pp. 4033--4046, 2020.

\bibitem{Song2021}
X.~Song, S.~Aryal, K.~M. Ting, Z.~Liu, and B.~He, ``Spectral-spatial anomaly
  detection of hyperspectral data based on improved isolation forest,''
  \emph{IEEE Trans. Geosci. Remote Sens.}, pp. 1--16, 2021.

\bibitem{Zhang2021}
Y.~Zhang, Y.~Dong, K.~Wu, and T.~Chen, ``Hyperspectral anomaly detection with
  otsu-based isolation forest,'' \emph{IEEE J. Sel. Topics Appl. Earth Observ.
  Remote Sens.}, vol.~14, pp. 9079--9088, 2021.

\bibitem{Zhang2015}
Y.~{Zhang}, B.~{Du}, and L.~{Zhang}, ``A sparse representation-based binary
  hypothesis model for target detection in hyperspectral images,'' \emph{IEEE
  Trans. Geosci. Remote Sens.}, vol.~53, no.~3, pp. 1346--1354, Mar. 2015.

\bibitem{Kerekes2008}
J.~{Kerekes}, ``Receiver operating characteristic curve confidence intervals
  and regions,'' \emph{IEEE Geosci. Remote Sens. Lett.}, vol.~5, no.~2, pp.
  251--255, Apr. 2008.

\end{thebibliography}

\end{document}